\documentclass[aps,showpacs,twocolumn,prd,preprintnumbers,amsmath,amssymb,letterpaper]{revtex4-1}

\newcommand{\beqn}{\begin{eqnarray}}
\newcommand{\eeqn}{\end{eqnarray}}
\def\half{\frac{1}{2}}

\def\araa{Ann. Rev. Astron. Astrop.}
\def\pasp{Publications of the Astronomical Society of the Pacific}
\def\aap{Astronomy and Astrophysics}
\def\apjs{Astrophysical Journal, Supplement}
\def\beqar{\begin{eqnarray}}
\def\eeqar{\end{eqnarray}}
\newcommand{\llabel}[1]{\label{#1}}              

\newcommand{\labeq}[2]{ \begin{equation} \llabel{#1}
{#2}
\end{equation}}

\newcommand{\beq}{\begin{equation}}
\newcommand{\eeq}{\end{equation}}
\usepackage{graphicx}
\usepackage{epsf}
\usepackage{amsmath}
\usepackage{graphics,epsfig,placeins,subfigure,wrapfig}

\begin{document}
\title{The merger of binary white dwarf--neutron stars: \\ Simulations in full general relativity}
\author{Vasileios Paschalidis${}^1$, Yuk Tung Liu${}^1$, Zachariah Etienne${}^1$, and Stuart L. Shapiro${}^{1,2}$
}                           
%
\affiliation{
${}^1$ Department of Physics, University of Illinois at Urbana-Champaign,
Urbana, IL 61801 \\
${}^2$ Department of Astronomy and NCSA, University of Illinois at Urbana-Champaign,
Urbana, IL 61801 
}
%
\begin{abstract}

We perform fully general relativistic (GR) simulations to address the
inspiral and merger of binary white dwarf-neutron stars.
The initial binary is in a circular orbit at the Roche critical separation. 
The goal is to determine the ultimate fate of such systems. 
We focus on binaries whose total mass exceeds the maximum mass ($M_{\rm max}$) a cold, 
degenerate EOS can support against gravitational collapse. 
The time and length scales span many orders of magnitude, making fully general
relativistic hydrodynamic (GRHD) simulations computationally
prohibitive.  For this reason, we model the WD as a ``pseudo-white
dwarf'' (pWD) as in our binary WDNS head-on collisions study \cite{WDNS_PAPERII}. 
Our GRHD simulations of a pWDNS system with a $0.98 M_\odot$  WD and a $1.4 M_\odot$ NS 
show that the merger remnant is a spinning Thorne-Zytkow-like Object (TZlO) surrounded by a massive disk.
The final total rest mass exceeds $M_{\rm max}$, but the remnant does not collapse promptly.
To assess whether the object will ultimately collapse after cooling, we introduce 
radiative thermal cooling. We first apply our cooling algorithm to TZlOs formed in 
pWDNS head-on collisions, and show that these 
objects collapse and form black holes on the cooling time scale, as expected. 
However, when we cool the spinning TZlO formed in 
the merger of a circular-orbit pWDNS binary, the remnant does not collapse,
demonstrating that differential rotational support is sufficient to prevent
collapse. Given that the final total mass exceeds $M_{\rm max}$ for our cold EOS,
magnetic fields and/or viscosity may redistribute angular momentum,
ultimately leading to delayed collapse to a BH. We
infer that the merger of realistic massive WDNS binaries likely will 
lead to the formation of spinning TZlOs that undergo delayed collapse. 
\end{abstract}

\pacs{04.25.D-,04.25.dk,04.40.Dg}

\maketitle
%

\section{Introduction}
\label{sec:introduction}

During inspiral and merger, compact binaries emit a large flux of
gravitational waves (GWs), making them among the most promising
sources for GWs detectable by ground-based laser interferometers like
LIGO \cite{LIGO1,LIGO2}, VIRGO \cite{VIRGO1,VIRGO2}, GEO \cite{GEO},
TAMA \cite{TAMA1,TAMA2} and AIGO \cite{AIGO}, as well as by proposed
space-based interferometers such as LISA \cite{LISA} and DECIGO
\cite{DECIGO}.  Extracting physical information about these binaries
from their GWs may shed light on determining their ultimate fate, but
requires careful modeling of these systems in full general relativity
(see \cite{BSBook} for review and references therein).
Most effort to date has 
focused on modeling black hole--black hole (BHBH) binaries  (see \cite{2010CQGra..27k4004H} and references therein),
and neutron star--neutron star (NSNS) binaries (see \cite{DuezNSNSReview} for a review),
with some recent work on black hole--neutron star (BHNS) binaries in
full GR
\cite{Rantsiou08, Loffler06, Faber, Faber06, Shibata06, Shibata07, Shibata08,Yamamoto08, 
Etienne08a, Etienne08, Duez08,2009PhRvD..79d4030S,2009PhRvD..79l4018K,2010AAS...21530001M,
2010arXiv1006.2839C,2010CQGra..27k4106D,2010arXiv1007.4160P,2010arXiv1007.4203F,2010arXiv1008.1460K}. 

As a follow-up to our investigation of binary WDNS {\it head-on} collisions \cite{WDNS_PAPERII}, 
in this work we perform fully general relativistic simulations of
{\it circular-orbit} WDNS binaries through inspiral and merger.  Throughout we call this the ``inspiral
case'' to distinguish it from the ``head-on'' collision case. 
WDNS binaries are promising sources of low-frequency GWs (for LISA and
DECIGO) and, as we argued in \cite{WDNS_PAPERI}, possibly also
high-frequency GWs (for LIGO, VIRGO, GEO, TAMA and AIGO), if the remnant
ultimately collapses to a black hole. 

Like NSNS binaries, WDNS binaries are known to exist. In \cite{WDNS_PAPERI} 
we compiled tables of 20 observed WDNS binaries and their measured
orbital properties.  The NS masses in these systems range between
$1.26\ M_\odot$ and $2.08\ M_\odot$, and their distribution is centered
around $1.5\ M_\odot$.  On the other hand, the WD masses in these
systems range between $0.125\ M_\odot$ and $1.3\ M_\odot$, and their
distribution is centered around $0.6\ M_\odot$. Eighteen of these observed
WDNS binaries have total mass greater than $1.65\ M_\odot$, 8 of which 
have a WD component with mass greater than $0.8\ M_\odot$,  and 5 have
total mass greater than $2.2\ M_\odot$.  This is interesting because the
expected Tolman-Oppenheimer-Volkoff (TOV) limiting mass for a cold, degenerate gas must be larger
than $1.97\ M_\odot$ \cite{NS2Msun} and may reach $2.2\ M_\odot$
\cite{2000ApJ...528L..29B,2004ApJ...610..941M,CooST94, APR,
1993PhRvL..70..379L,1988PhRvC..38.1010W,1975BAAS....7..240P,
BetheJohnson,Pandha1971}, depending on the equation of state (EOS).
One of the main goals of this work is to determine whether a WDNS
merger remnant will undergo prompt collapse to a black hole.

Population synthesis calculations \cite{Nelemans01,Cooray2004} show
that there are about $2.2\times 10^{6}$ WDNS binaries in our Galaxy, 
and that they have a merger rate between $10^{-6} \rm {\rm \ yr}^{-1}$ and $1.4\times 10^{-4} \rm {\rm \ yr}^{-1}$. 
Furthermore, these studies find that after a year of integration, 
LISA-like interferometers should be able to detect $1-40$ WDNS pre-merger binaries.
Recent work by Thompson et al. \cite{TThompson09} suggests that the lower limit on 
the merger rate of binary WDNSs in the Milky Way, at 95\% confidence, is 
$2.5\times 10^{-5}\rm {\rm \ yr}^{-1}$. Thompson et al. also suggest that the merger rate
in the local universe is $\sim 0.5 -  1 \times 10^4 \rm Gpc^{-3}
{\rm \ yr}^{-1}$.  Therefore, ignoring some uncertainties, all recent
population synthesis calculations suggest that LISA-like projects should
be able to detect a few WDNS pre-mergers per year.

\subsection{Previous WDNS work}

In \cite{WDNS_PAPERI} we focused on possible evolutionary scenarios
for circular WDNS binaries that have inspiraled sufficiently close
that they reach the termination point for equilibrium configurations. 
This is the Roche limit for WDNSs, at which point the WD fills its Roche lobe and 
may experience one of at least two possible fates: i) stable mass transfer (SMT) 
from the WD across the inner Lagrange point onto the 
NS, or ii) tidal disruption of the WD by the NS via unstable mass transfer (UMT). 

Note that once an UMT binary reaches the critical Roche separation, 
further inspiral and merger is governed by tidal effects and hydrodynamical
interactions and not GW emission.

We also studied key parameters that determine whether a system
will undergo SMT or UMT and found that, for a given NS mass, there
exists a critical mass ratio $q_{\rm crit} \approx 0.5$ that separates the UMT
and SMT regimes.  If the mass ratio $q=M_{\rm WD}/M_{\rm NS}$ of a
WDNS system is such that $q>q_{\rm crit}$, the WD quickly overfills its
Roche lobe, and the binary will ultimately undergo UMT. In the opposite
case, $q<q_{\rm crit}$, the system will undergo SMT. 
We showed that a
quasistationary treatment is adequate to follow the evolution of an SMT
binary during this secular phase and calculated the gravitational
waveforms. We also pointed out that WDNS observations suggest that there are known candidates
residing in both regimes.

In the case of tidal disruption (UMT), by contrast, 
the system will evolve on a hydrodynamical (orbital) time scale. 
In this scenario the NS may plunge into the WD and spiral 
into the center of the star, forming a quasiequilibrium
 configuration that resembles a
 Thorne-Zytkow object (TZO) \cite{ThorneZytkow77};
alternatively, the NS may be the receptacle of massive debris from the
 disrupted WD. WDNS mergers may also give rise to gamma ray bursts
\cite{2009A&A...498..501C,2011A&A...529A.130D}.

The ultimate fate of the merged WDNS 
depends on (1) the initial mass of the cold progenitor stars, (2) the 
degree of mass and angular momentum 
loss during the WD disruption and binary merger phases, (3) the angular momentum 
profile of the WDNS remnant, and (4) the
extent to which disrupted debris is heated by shocks and/or nuclear reactions as 
it settles onto the NS and forms an extended, massive mantle.
These are issues that require a hydrodynamic simulation to 
resolve. Note that Newtonian work on binaries with a WD component
has been performed analytically in \cite{Rappaport82,Rappaport83,Verbunt88,Podsiadlowski92,Marsh,WDNS_PAPERI} 
and via Newtonian hydrodynamic simulations in \cite{Benz1990, RasioShapiro95, Segretain1997, 
Guerrero2004, Yoon2007, Dan08}.
However, ascertaining whether or not the neutron star ultimately
undergoes catastrophic collapse (either prompt or delayed) to a black hole 
requires that such a simulation be performed in full general
relativity.  In fact, even the 
final fate of the NS in the alternative scenario in which there is 
a long epoch of SMT may also lead to catastrophic collapse, 
if the neutron star mass is close to the neutron star maximum mass.
This scenario too will require a general relativistic hydrodynamic simulation to track.

In \cite{WDNS_PAPERII} we employed the Illinois adaptive mesh refinement (AMR)
relativistic hydrodynamics code \cite{Etienne08,Illinois_new_mhd} to
perform the first simulations of these systems in full GR.
In particular, we studied the head-on collision from rest at large
separation of a massive WD and a NS. We focused on compact objects
whose total mass exceeds the maximum mass supportable by a cold EOS in
order to determine the outcome of such collisions.

The vast range of time and length scales involved 
in the WDNS problem make fully general relativistic simulations extremely challenging. 
In \cite{WDNS_PAPERII} we demonstrated that the length scales span four orders of magnitude,
as measured in neutron star radii, and that the associated time
scales span six orders of magnitude in $M$, the total system mass.
Current numerical relativity techniques and available computational 
resources make such simulations prohibitive. For this reason, we tackled
this problem using a different strategy. 

In particular, we constructed a six-parameter piecewise polytropic 
EOS which mimics realistic NS EOSs while, at the same time,
 scales down the size of the WD. We call these scaled-down WDs ``pseudo-WDs (pWDs)''. 
We chose all of the piecewise EOSs in such a way that the maximum NS mass is $1.8\ M_\odot$~\footnote{The
  $1.97\ M_\odot$ NS  \cite{NS2Msun}  was discovered after we started
  our pWDNS calculations. This is why the maximum mass of our EOS is
  smaller than $1.97\ M_\odot$.},
and the maximum WD mass is $1.43\ M_\odot$, i.e., the Chandrasekhar mass. Furthermore, we made sure these EOSs
preserve the qualitative shape of the central density--mass curves as well as the number
of stable and unstable NS and WD branches (see Figs. 1 and 2 in \cite{WDNS_PAPERII}).
Moreover, the scaling is performed so that all the length-scale and
time-scale inequalities of the realistic problem are left unchanged.
For a given set of EOS parameters, a realistic WD has a counterpart pWD
which has the same mass but is smaller in size. As a result, for every realistic WDNS system,
we can construct a pWDNS counterpart which involves the same (realistic) NS and the pWD counterpart of the WD.

Using pWDs we performed a sequence of head-on simulations in which the
EOS is changed so that the pWDs have the same mass ($0.98\ M_\odot$) but
decreasing compactions, while the compaction and mass of the NS
involved remains practically unchanged.  More precisely, while
keeping the masses of the binary components and the NS radius fixed,
the pWD compaction was modified so that the pWD:NS radius ratio varied
between 5:1 and 20:1.  We then scaled the results of our
simulations to predict the outcome in the realistic case: 500:1.

In addition to studying the effects of the pWD compaction, we also studied
the effects of NS mass. We considered NSs with masses $1.4\ M_\odot$,
$1.5\ M_\odot$, and $1.6\ M_\odot$.

All head-on collision simulations that we performed showed that after the collision,  
14\%-18\% of the initial total rest mass escapes to infinity.
In all cases, the remnant rest mass exceeded the maximum rest mass that
our {\it cold} EOS can support ($1.92\ M_\odot$), and no case led to
prompt collapse to a black hole.  This outcome arises because the 
final configurations become {\it hot}, due to shock heating. All our cases settle into a
spherical quasiequilibrium configuration consisting of a cold NS core
surrounded by a hot mantle. Hence, all remnants are Thorne-Zytkow-like Objects (TZlOs). 
Scaling our results to realistic WD compactions, we predict
that a realistic head-on collision will form a quasiequilibrium TZlO. 

Although the head-on collision simulations appear to lead to a
consistent result (the formation of a TZlO), these results cannot be
used to predict the final fate of WDNS systems in circular orbit. On
the one hand, one might expect that the remnant in the inspiraling case
will collapse to a black hole, because shock heating is not 
likely to be as intense as in the head-on case. On the other hand, the
large amount of angular momentum in the inspiraling binary case may work to 
prevent prompt collapse. Therefore, to predict
whether the merged WDNS remnant will collapse, promptly or following cooling,
we need to perform fully general relativistic
simulations of WDNS binaries through inspiral and merger.

\subsection{Goals and objectives}

The purpose of the current work is threefold: \\
a) We simulate the late inspiral and merger of a WDNS 
system consisting of a 1.4$M_\odot$ NS and a 0.98$M_\odot$ WD
initially in circular orbit and at the Roche limit. 
As in our head-on collision studies, we employ the pWD approximation
to make the computations feasible~\footnote{Even though the range of
  time and length scales is reduced when using pWDs, the
  resolution requirements to accurately follow the system and preserve
  angular momentum, rendered our calculations very expensive. In
  particular, it required 9 months of wallclock time for our simulations to finish.}. 
The pWD approximation is useful for predicting the ultimate fate of
a realistic WDNS merger using scaling. 
In particular, the collision velocity ($v_c$) and the pre-shocked 
WD sound speed $c_s$ both scale as $\sim (M/R_{\rm WD})^{1/2}$. 
This implies that the Mach number (${\cal M} = v_c/c_s$) is 
invariant under scaling of $R_{\rm WD}$ and so is the degree of shock heating. 
The thermal energy, as well as the rotational kinetic energy ($T$) 
and the gravitational potential energy ($W$) all scale 
as $\sim M^2/R_{\rm WD}$, when the binary merges. Thus,
$T/|W|$ is also invariant under scaling of $R_{\rm WD}$. 
These considerations simply mean that with respect to gravity
the relative importance of thermal and rotational support in a WDNS merger remnant 
is approximately invariant, when the masses of the binary components are fixed and the only 
quantity that changes is the WD radius. As a consequence, the results obtained
when adopting pWDNS systems can be scaled up to realistic WDNS systems.
Note that our compaction study in \cite{WDNS_PAPERII} confirms the 
above scaling with the Mach number.

b) We introduce a radiative cooling prescription and modify 
our adiabatic simulations by 
allowing for cooling to determine whether the
merger remnant will collapse without thermal support, if it fails to collapse promptly.  
Otherwise, angular momentum provides sufficient support to prevent collapse. 

c) We allow  cooling to occur in the TZlOs
formed in our WDNS head-on collision simulations \cite{WDNS_PAPERII}
to confirm that these remnants collapse to a black hole when the
excess thermal energy is radiated away.  In other words, we
demonstrate that it is thermal pressure alone that prevents these
objects from undergoing prompt collapse, since angular momentum
support is completely absent in head-on collisions.
Delayed collapse occurs on a cooling time scale in all cases, providing a consistency
check on our cooling implementation.

Our pWDNS merger calculations show that the inspiral remnant
is a spinning TZlO which is surrounded by a massive, extended, hot 
disk. In contrast to our head-on collisions, 
we do not find any outflows in the inspiraling case. Therefore, the final total mass
is greater than the maximum mass supportable by our cold EOS and many nuclear EOSs. 
However, the remnant does not collapse promptly to a BH. We find
that the remnant is both thermally and centrifugally supported. 
To determine whether centrifugal forces alone can support the remnant
we incorporate cooling and find that the
object does not collapse to a black hole. Therefore, the extra support
provided by rotation is sufficient for preventing the collapse.

Even though the TZlO does not collapse after cooling, we expect delayed collapse 
ultimately because the final total rest mass ($\sim 2.5\ M_\odot$) is larger than the maximum possible mass
supportable by our cold EOS (and many nuclear EOSs), even allowing for {\it uniform}
rotation. (The maximum gravitational mass of a uniformly rotating star with our adopted EOS is
$\simeq 2.1\ M_\odot$). We expect that collapse to a BH will take place after 
viscosity or magnetic fields redistribute the angular momentum, as in the case 
of a hypermassive neutron star \cite{SBS2000,2006PhRvL..96c1101D,2004PhRvD..69j4030D}. 
his conclusion will be true in the case of realistic WDNS mergers, unless the true nuclear EOS supports 
a uniformly rotating star with a rest mass exceeding the remnant mass. Many viable EOSs do not support 
rest masses as large as $2.5M_\odot$ \cite{2004ApJ...610..941M}, the remnant rest mass in our simulations.



This paper is organized as follows. In Sec.~\ref{sec:pWD}
the pWD approximation and the EOS adopted in our simulations are briefly reviewed.
Section ~\ref{sec:Init_data} outlines the initial data generation
technique. Sec.~\ref{sec:evolutions} summarizes the methods used
for evolving  the gravitational and matter fields. Sec.~\ref{sec:cooling} introduces
our radiative cooling formalism, which is then applied to the TZlO
remnants from our pWDNS head-on collision simulations in
Sec.~\ref{cool_TZlOs}. We present the results of our fully
relativistic hydrodynamic simulations of the binary pWDNS late inspiral and merger in
Sec.~\ref{sec:results}, and turn on cooling  in
Sec.~\ref{sec:mergerTZlOcooling}. In Sec.~\ref{sec:discussion} we
discuss possible effects of nuclear reactions in realistic WDNS mergers and give estimates
of realistic cooling and angular momentum redistribution time scales.
Sec.~\ref{sec:summary} concludes with a summary of the main findings. 
Throughout this work, geometrized
units are adopted, where $G=c=1$, unless otherwise specified.

\section{Equation of State}
\label{sec:pWD}

We employ the following 6-parameter piecewise polytropic cold EOS:
\labeq{EOS}{
\frac{P}{\rho_0} = 
\left\{
\begin{array}{ll}
\kappa_1 \rho_0^{1/n_1},  &  \rho_0 \leqslant  \rho_1  \\
& \\
\kappa_2 \rho_0^{1/n_2},  &  \rho_1<\rho_0 \leqslant  \rho_2 \ ,\\
& \\
\kappa_3 \rho_0^{1/n_3},  &  \rho_0 > \rho_2  
\end{array}
\right.
}
where $P$ is the pressure, $\rho_0$ is the rest-mass density
 and $\kappa_1, \kappa_2, \kappa_3, n_1 , n_2, n_3, \rho_1, \rho_2$ are the 
parameters of the EOS. The parameters in Eq.~\eqref{EOS} are 8 in number, but
continuity requires that the following conditions be true
\labeq{EOScond}{
\kappa_1=\kappa_2 \rho_1^{1/n_2-1/n_1}, \ \ 
\kappa_2=\kappa_3 \rho_2^{1/n_3-1/n_2}.
}
As a result, the adopted EOS 
has 6 free parameters $\kappa_3, n_1 , n_2, n_3, \rho_1$, and $\rho_2$.

Because of its multiple parameters, this EOS gives us the freedom to
capture the same characteristic curves and turning points on a TOV
mass-central density plot as for a realistic cold-degenerate EOS (see
\cite{Shapiro}), as shown in Fig. 1 in \cite{WDNS_PAPERII}. The EOS 
exhibits both stable ($dM/d\rho_{0,c}>0$) and unstable
($dM/d\rho_{0,c}<0$) branches for both WDs and NSs, as in the
realistic case.

Furthermore, this EOS 
allows us to adjust the size of a pWD of any given mass, thereby
shifting the pWD branch to smaller radii (see Fig.~2 in
\cite{WDNS_PAPERII}), while keeping the NS branch
approximately unchanged.  For more details about our EOS 
and pWDs we refer the interested reader to \cite{WDNS_PAPERII}.

In this work the EOS parameters correspond to the 10:1 EOS
we considered in \cite{WDNS_PAPERII}: 
$\kappa_3 =  4993$, $\Gamma_1= 1.515$, $\Gamma_2= 2.969$,
$\Gamma_3 = 0.714$, $\log(\rho_1/\rho_{\rm nuc})= -2.268$,
$\log(\rho_2/\rho_{\rm nuc})= 0.208$, where all values are in geometrized units
and $\Gamma_i = 1+1/n_i$, $\rho_{\rm nuc} = 1.485\times 10^{-4} {\rm km}^{-2}$.
These parameters are chosen such that the ratio of
the isotropic radius of a TOV $0.98\ M_\odot$ pWD to that of a TOV $1.5\ M_\odot$ NS is 10:1. 
In addition, the EOS has been constructed
so that the maximum gravitational mass of a NS is $1.8\ M_\odot$, i.e., the same as that
 for the AP2 version of the Akmal-Pandharipande-Ravenhall 
(APR) EOS \cite{APR,ReadLackey2009}, and the maximum gravitational mass
 of a pWD is $1.43\ M_\odot$, i.e., the maximum mass of a
 TOV WD obeying the Chandrasekhar EOS for mean molecular weight $\mu_e = 2$.

%
%

\section{Initial Data}
\label{sec:Init_data}

This section introduces the formalism adopted for generating valid
general relativistic initial data for binary pWDNS systems in circular
orbit.

\subsection{Gravitational Field Equations} \label{formalism_cts}

The spacetime metric in the standard 3+1 form \cite{ADM3plus1} is
written as
\begin{equation}\label{ADMmetric}
ds^2 = - \alpha^2 dt^2 + \gamma_{ij}(dx^i +\beta^i dt)
	(dx^j +\beta^j dt),
\end{equation}
where $\alpha$ is the lapse function, $\beta^i$ the 
shift vector and $\gamma_{ij}$ the three-metric on
 spacelike hypersurfaces of constant time $t$. 
Throughout the paper Latin indices run from 1 to 3, and Greek
indices run from 0 to 3. 

The three-metric $\gamma_{ij}$ is then conformally decomposed as
\begin{equation} \label{conflatmet}
\gamma_{ij} \equiv \Psi^4 f_{ij},
\end{equation}
where $\Psi$ is the conformal factor and $f_{ij}$ the conformal metric. 
We adopt the standard approximation of a conformally flat spacetime, so that 
$f_{ij} = \delta_{ij}$ in Cartesian coordinates.

We split the extrinsic curvature ($K^{ij}$) into trace ($K$)
and tracefree parts ($A^{ij}$)
\labeq{}{
K^{ij} = A^{ij}+\frac{1}{3}\gamma^{ij}K,
}
take the initial slice to be maximal
\begin{equation} \label{maxslice}
K = 0,
\end{equation}
and introduce a ``conformal'', 
traceless extrinsic curvature as
\labeq{confA}{
\bar A^{ij}\equiv \Psi^{10} A^{ij}.
}

Using Eqs.~\eqref{conflatmet}--\eqref{confA} and assuming the existence of 
an approximate helical Killing vector, the Hamiltonian
and momentum constraint equations assume the form of 
the conformal-thin-sandwich (CTS) equations \cite{BSBook}. The Hamiltonian constraint 
becomes
\begin{equation} \label{ham2}
\bar \nabla^2 \Psi = - \frac{1}{8} \Psi^{-7} \bar A_{ij} \bar A^{ij}
	- 2\pi\Psi^5 \rho.
\end{equation}
where $\bar\nabla^2$ is the flat Laplacian operator associated with $f_{ij}$.
Here the source term $\rho$ is defined as
\begin{equation} \label{rho}
\rho \equiv n^\alpha n^\beta T_{\alpha\beta},
\end{equation}
where $n^{\alpha}$ is the normal vector to a $t = {\rm constant}$ slice, and
$T_{\alpha\beta}$ is the stress-energy tensor of the matter.  
%

The momentum constraint yields
\begin{equation} \label{mom2}
\bar\nabla^2 \beta^i + \frac{1}{3} \bar\nabla^i (\bar\nabla_j \beta^j) 
= 2\bar A^{ij} \bar\nabla_j ( \alpha \Psi^{-6} )  + 16 \pi \alpha \Psi^4 j^i,
\end{equation}
where the source  term $j^i$ is given by
\begin{equation} \label{j}
j^{\alpha} \equiv - \gamma^{\alpha}_{~\beta}n_{\gamma} T^{\beta\gamma}.
\end{equation}

Taking the trace of the evolution equation for $K_{ij}$ (see Eq.~(2.106) in \cite{BSBook}), 
imposing the maximal slicing condition Eq.~\eqref{maxslice},
and combining the result  with Eq.~(\ref{ham2}), we obtain an equation for the lapse \cite{BSBook}
\begin{equation} \label{lap2}
\bar\nabla^2(\alpha \Psi) = \alpha \Psi \left( \frac{7}{8} \Psi^{-8} 
	\bar A_{ij} \bar A^{ij}  + 2 \pi \Psi^4 (\rho + 2S) \right).
\end{equation}
Here the source term $S$ is defined as
\begin{equation} \label{S}
S \equiv \gamma^{ij} T_{ij}.
\end{equation}
In all equations above 
\begin{equation} \label{k2}
\bar A^{ij} =  \frac{\Psi^6}{2\alpha} 
	\left( \bar\nabla^i \beta^j +\bar \nabla^j \beta^i
	- \frac{2}{3} f^{ij} \bar\nabla_k \beta^k \right),
\end{equation}
and $\bar A_{ij}=f_{im}f_{jn}\bar A^{mn}$.

Instead of solving Eq. \eqref{mom2} for the shift vector directly, it is convenient 
to decompose $\beta^i$ as a sum of a vector and a gradient (cf. \cite{BowenYork80})
\begin{equation} \label{shift}
\beta^i \equiv G^i - \frac{1}{4} \bar\nabla^i B.
\end{equation}
Eq.~(\ref{mom2}) can then be replaced by the two equivalent equations
\begin{equation} \label{g1}
\bar\nabla^2 G^i   
= 2  \bar A^{ij} \bar \nabla_j ( \alpha \Psi^{-6} ) +16 \pi \alpha \Psi^4 j^i
\end{equation}
and 
\begin{equation} \label{b1}
\bar\nabla^2 B = \bar\nabla_i G^i.
\end{equation}

Equations~\eqref{ham2}, \eqref{lap2}, \eqref{g1} and \eqref{b1} form a system of 6 coupled,
nonlinear elliptic equations for the 6 unknowns $\Psi$, $\alpha \Psi$, $G^i$ and $B$, which
must be solved iteratively. These equations are
elliptic and hence require outer boundary conditions to be specified. We impose
the same fall-off boundary conditions as in \cite{Baum98_NSNS}, except that here we choose the
binary components to be initially lined up on the $x$-axis and the
binary rotation axis parallel to the $z$-axis.
Table~\ref{tab:bconditions} lists the full set of outer boundary
conditions imposed in our initial data.


\begin{center}
\begin{table}
\caption{Outer boundary conditions imposed on the CTS variables when
  generating WDNS initial data.}
\begin{tabular}{cc}\hline\hline
\multicolumn{1}{p{3.0cm}}{\hspace{0.75 cm} Variable} & 
\multicolumn{1}{p{4.0cm}}{\hspace{0.75 cm} Fall-off condition}  \\ \hline
$\Psi-1$    	       &    $\sim 1/r$  \\  \hline
$\alpha-1$    	       &    $\sim 1/r$  \\  \hline
$G^x$    	       &    $\sim y/r^3$  \\  \hline
$G^y$    	       &    $\sim x/r^3$  \\  \hline
$G^z$    	       &    $\sim xyz/r^7$  \\  \hline
$B$     	       &    $\sim xy/r^3$  \\  \hline\hline
\end{tabular}
\label{tab:bconditions}
\end{table}
\end{center}

\subsection{Matter fields}

As we argued in \cite{WDNS_PAPERI} the WD in a WDNS binary with close separation likely will be tidally locked. 
For this reason we focus on corotating WDNS systems only. 

We assume that the matter is described by a perfect fluid stress-energy tensor:
\begin{equation}
T^{\alpha\beta} = (\rho_0 + \rho_i + P) u^\alpha u^\beta + P g^{\alpha\beta},
\end{equation}
where $g^{\alpha\beta}$ is the inverse of
the four-metric and $\rho_0, \rho_i, P, u^\alpha$ are the rest-mass density, 
internal energy density, pressure, and four-velocity of the fluid respectively. 
For all initial configurations, the pressure is given by the cold EOS
as specified in Eq.~\eqref{EOS}.
The internal energy density can be derived by integrating
\labeq{}{
d\bigg(\frac{\rho_i}{\rho_0}\bigg) = - P d\bigg(\frac{1}{\rho_0}\bigg),
}
and for Eq.~\eqref{EOS} the integration yields
\labeq{EOS_int}{
\frac{\rho_i}{\rho_0} = 
\left\{
\begin{array}{cl}
n_1\kappa_1 \rho_0^{1/n_1},  &  \rho_0 \leqslant  \rho_1  \\
& \\
n_2\kappa_2 \rho_0^{1/n_2} + c_2,  &  \rho_1<\rho_0 \leqslant  \rho_2  \\
& \\
n\kappa \rho_0^{1/n_3} + c_3,  &  \rho_0 > \rho_2
\end{array}
\right.
}
where 
\labeq{}{
c_2 = (n_1-n_2)\kappa_1\rho_1^{1/n_1}, 
\ \ \ c_3 = c_2 + (n_2-n_3)\kappa_2\rho_2^{1/n_2}, 
}

In Cartesian coordinates we choose the orbital plane of the binary to be the
$z=0$ plane, so that the fluid four-velocity takes the form \cite{BSBook}
\begin{equation}
u^\alpha = u^t(1,-\Omega y, \Omega( x-x_{\rm rot}),0),
\end{equation}
where $\Omega$ is the constant orbital angular velocity and $x_{\rm rot}$
is the $x$ coordinate of the axis of rotation. Following
\cite{Baum98_NSNS}, we introduce a vector
\begin{equation} \label{xi}
\xi^\alpha = (0,-y,( x-x_{\rm rot})),
\end{equation}
and rewrite the four-velocity as
\begin{equation}
u^\alpha = u^t (\alpha n^\alpha + \Omega\xi^\alpha + \beta^\alpha).
\end{equation}

The source term $\rho$ in Eq.~(\ref{rho}) can then be written
\begin{equation} \label{rho2}
\rho 	= \frac{\rho_0 + \rho_i + P}{1 - v^2} - P,
\end{equation}
where $v$ is the magnitude of the three-velocity of the fluid.
Using $u^\alpha u_\alpha \equiv -1$, it can be shown that $v^2$ is given by
\begin{equation}\label{vsqr}
v^2 = \frac{\Psi^4}{\alpha^2} \left[ (\Omega y - \beta^x)^2 +
	\big(\Omega(x-x_{\rm rot})+\beta^y\big)^2 + (\beta^z)^2 \right].
\end{equation}
The momentum source $j^i$ in Eq.~(\ref{j}) becomes
\begin{equation} \label{j2}
j^i 	= \frac{(\rho_0 + \rho_i + P)}{\alpha} 
	\frac{(\Omega \xi^i + \beta^i)}{1-v^2},
\end{equation}
and $S$ in Eq.~(\ref{S}) is given by
\begin{equation} \label{S2}
S 	= (\rho_0 + \rho_i + P)\frac{v^2}{1 - v^2} + 3P.
\end{equation}
%





\subsection{Computational methods}

We solve the nonlinear elliptic equations~\eqref{ham2}, 
\eqref{lap2}, \eqref{g1} and \eqref{b1} using a fixed-mesh-refinement
(FMR) finite difference code we developed, which is
based on the Portable, Extensible Toolkit for Scientific Computation
(PETSc) library \cite{petsc-web-page,petsc-user-ref,petsc-efficient}.
A full description of our code may be found in
\cite{WDNS_PAPERII}. Here we summarize the basic features.

The grid structure used in our FMR elliptic code is a multi-level set of properly nested, 
uniform grids. We use standard cell-centered, second-order accurate finite difference stencils
for the Laplacian operator and the derivatives of the variables, using first-order interpolation 
across the refinement level boundaries when necessary. We calculate the solution across the entire 
grid, and only on leaf cells (i.e. cells within which there exist no higher resolution cells). 
In \cite{WDNS_PAPERII} we performed a series of tests involving single NSs, and we 
demonstrated that the code converges to the expected solutions at
second order.


Given the matter distribution, $\Omega$ and $x_{\rm rot}$ we solve the CTS equations iteratively, addressing the
non-linearity of Eq.~\eqref{ham2} by performing Newton-Raphson
iterations, until the residuals of all six equations become smaller
than some set tolerance (usually set to $10^{-15}$).

We obtain the WD rest-mass density distribution, $\Omega$ and $x_{\rm rot}$ at the Roche limit for 
equilibrium, corotating binary WDNSs in circular orbit obeying our cold EOS
using the unigrid Newtonian code we developed and tested in \cite{WDNS_PAPERI}. 
At the Roche limit, the binary separation is large enough so that
the tidal effects on the NS are negligible, and hence the NS
will be spherical to a high degree and point-like from the point of
view of the WD.  Thus, in the Newtonian code we model the NS as a
point mass and we self-consistently solve for the WD rest-mass density distribution
via the integrated Euler equation. We use the Newtonian equations for this step,
because it is computationally simple and fast. Also, the large separation at the Roche limit
ensures that the WD and NS interaction lies in the Newtonian regime,
so that our initial configuration is nearly in equilibrium.


After the WD rest-mass density distribution has been calculated, the point-mass NS 
is replaced by a TOV NS with gravitational mass equal to
that of the point-mass NS, centered at the position of the
point mass. For simplicity, we model the NS as corotational because there is no 
essential difference between an irrotational and a corotational NS at such large separations. 
The spin of a corotating NS is very small. To understand this, consider the ratio of 
the angular velocity of the corotating
NS ($\Omega_{\rm cor}$) to that at the mass-shedding ($\Omega_{\rm ms}$) limit: 
\labeq{}{
\frac{\Omega_{\rm cor}}{\Omega_{\rm ms}}= 
\sqrt{\frac{M_{\rm total}}{M_{\rm NS}}} \bigg(\frac{R_{\rm NS}}{A_{\rm R}}\bigg)^{3/2} 
\approx 1.3\bigg(\frac{R_{\rm NS}}{A_{\rm R}}\bigg)^{3/2},
}
where $A_{\rm R}$ is the Roche limit separation. For the typical system we
consider $R_{\rm NS}/A_{\rm R} \approx R_{\rm NS}/3 R_{\rm WD}$. For realistic
massive WDs $R_{\rm WD}/R_{\rm NS} \approx 500$, and for pWDs $R_{\rm pWD}/R_{\rm NS} \approx 10$. 
Thus,  $\Omega_{\rm cor}/\Omega_{\rm ms} \approx 10^{-5}$ 
for realistic WDNSs and $\Omega_{\rm cor}/\Omega_{\rm ms} \approx 10^{-2}$ for pWDNSs.
Therefore, the corotation spin the NS acquires is very small and has no physical significance.

Having prescribed the NS and pWD rest-mass density, using second-order polynomial interpolation, we
interpolate the NS and pWD matter distribution on the grid of our FMR elliptic
code and solve the CTS equations.



\section{Evolution of WDNS systems}
\label{sec:evolutions}

\subsection{Basic Equations}
\label{sec:basic_eqns}

The formulation and numerical scheme for our simulations are
the same as those reported
in~\cite{PhysRevD.72.024028,Etienne08a,Illinois_new_mhd,WDNS_PAPERII},
to which the reader may refer for details.  Here we introduce our
notation and summarize our method.

We use the 3+1 formulation of general relativity, in which
the metric is decomposed as in Eq.~\eqref{ADMmetric}.  In this
formalism, the fundamental dynamical variables for the metric
evolution are the spatial three-metric $\gamma_{ij}$ and extrinsic
curvature $K_{ij}$.  The Baumgarte-Shapiro-Shibata-Nakamura (BSSN) 
formalism~\cite{ShibNakamBSSN,BaumShapirBSSN,BSBook} is adopted.  The
BSSN evolution variables are the conformal exponent $\phi
\equiv \ln (\gamma)/12$, the conformal 3-metric $\tilde
\gamma_{ij}=e^{-4\phi}\gamma_{ij}$, three auxiliary functions
$\tilde{\Gamma}^i \equiv -\tilde \gamma^{ij}{}_{,j}$, the trace of
the extrinsic curvature $K$, and the trace-free part of the conformal extrinsic
curvature $\tilde A_{ij} \equiv e^{-4\phi}(K_{ij}-\gamma_{ij} K/3)$.
Here $\gamma={\rm det}(\gamma_{ij})$. The full spacetime metric $g_{\mu \nu}$
is related to the three-metric $\gamma_{\mu \nu}$ by $\gamma_{\mu \nu}
= g_{\mu \nu} + n_{\mu} n_{\nu}$, where the future-directed, timelike
unit vector $n^{\mu}$ normal to the time slice can be written in terms
of the lapse $\alpha$ and shift $\beta^i$ as $n^{\mu} = \alpha^{-1}
(1,-\beta^i)$. The evolution equations of these BSSN variables are 
given by Eqs.~(9)--(13) in~\cite{Etienne08a}.

We adopt standard puncture gauge conditions: an advective
``1+log'' slicing condition for the lapse and a 
``Gamma-freezing'' condition for the shift~\cite{PunctureGauge}. 
Thus, we have 
\beqn
  \partial_0 \alpha &=& -2\alpha K  \ , \label{eq:1+log} \\ 
  \partial_0 \beta^i &=& (3/4) B^i \ , \\ 
  \partial_0 B^i &=& \partial_0 \tilde{\Gamma}^i - \eta B^i \ ,
\label{puncturegauge}
\eeqn
where $\partial_0 \equiv \partial_t - \beta^j \partial_j$. We
set the $\eta$ parameter to $0.01{\rm \ km}^{-1}$ 
for all simulations presented in this work.

The fundamental matter variables are the rest-mass density 
$\rho_0$, specific internal energy $\epsilon$, pressure $P$, and 
four-velocity $u^{\mu}$. We write the stress-energy tensor as
\beq
  T_{\mu \nu} = \rho_0 h u_\mu u_\nu + P g_{\mu \nu} \ ,
\eeq
where $h=1+\epsilon+P/\rho_0$ is the specific enthalpy
and $\epsilon$ is the specific internal energy. 
In our numerical implementation of the hydrodynamics 
equations, we evolve the following ``conservative'' variables:
\beqn
&&\rho_* \equiv - \sqrt{\gamma}\, \rho_0 n_{\mu} u^{\mu} \ ,
\label{eq:rhos} \\
&& \tilde{S}_i \equiv -  \sqrt{\gamma}\, T_{\mu \nu}n^{\mu} \gamma^{\nu}_{~i}
\ , \\
&& \tilde{\tau} \equiv  \sqrt{\gamma}\, T_{\mu \nu}n^{\mu} n^{\nu} - \rho_* \ .
\label{eq:S0} 
\eeqn
The evolution equations for these variables are given by Eqs.~(27)--(29) 
in~\cite{Illinois_new_mhd}.

The EOS we adopt for the evolution has both a thermal and cold
contribution, and can therefore be written
\labeq{Ptot}{
P = P_{\rm th} + P_{\rm cold},
}
where $P_{\rm cold}$ is given by Eq.~\eqref{EOS} and
 the thermal pressure is given by
\labeq{Pthermal}{
P_{\rm th} = (\Gamma_{\rm th} - 1)\rho_0(\epsilon-\epsilon_{\rm cold}),
}
where  
\labeq{}{
\epsilon_{\rm cold} = - \int P_{\rm cold}d(1/\rho_0).
}
We set $\Gamma_{\rm th} = 1.66$ ($\simeq 5/3$) in all our simulations.
That is, we set $\Gamma_{\rm th}$ to the $\Gamma_1$ exponent of the
10:1 EOS, appropriate either for nonrelativistic cold, degenerate
electrons or (shock) heated, ideal nondegenerate baryons.
Equation~\eqref{Ptot} reduces to our piecewise polytropic law 
Eq.~\eqref{EOS} for the initial (cold) NS and pWD matter. 

\subsection{Evolution of the metric and hydrodynamics}
\label{sec:num_metric_hydro}

We evolve the BSSN equations using
fourth-order accurate, cell-centered finite-differencing stencils,
except on shift advection terms, where fourth-order accurate
upwind stencils are applied.  We apply Sommerfeld outgoing wave boundary
conditions on all BSSN fields, as in~\cite{Etienne08a}.  Our code is embedded in
the Cactus parallelization framework~\cite{Cactus}, and our
fourth-order Runge-Kutta timestepping is managed by the {\tt MoL}
(Method of Lines) thorn, with the Courant-Friedrichs-Lewy number
set to 0.45 in all pWDNS simulations.  We use the
Carpet~\cite{Carpet} infrastructure to implement the moving-box
adaptive mesh refinement. In all AMR simulations presented here, we
use second-order temporal prolongation, coupled with fifth-order
spatial prolongation, and impose equatorial symmetry to reduce the computational 
cost.

We write the general relativistic hydrodynamics equations in
conservative form. They are evolved via a high-resolution
shock-capturing (HRSC) technique~\cite{PhysRevD.72.024028,Illinois_new_mhd}
 that employs the
piecewise parabolic (PPM) reconstruction scheme~\cite{PPM}, coupled to
the Harten, Lax, and van Leer (HLL) approximate Riemman solver~\cite{HLL}. 
The adopted hydrodynamic scheme is second-order accurate. 
To stabilize our hydrodynamic scheme in regions where there is no
matter, a tenuous atmosphere is maintained on our grid, with a density
floor $\rho_{\rm atm}$ set to $10^{-10}$ times the initial
maximum density on our grid. The average density of the pWD is $10^{11} \rm gr/cm^3$, 
and at least six orders of magnitude larger than that of the artificial atmosphere. Thus,
the atmosphere poses no problem in evolving the pWD.
The initial atmospheric pressure
$P_{\rm atm}$ is set by using the cold EOS~\eqref{EOS}.
Throughout the evolution, we impose limits on the pressure to prevent
spurious heating and negative values of the internal energy
$\epsilon$. Specifically, we require $P_{\rm min}\leq P \leq P_{\rm max}$, 
where $P_{\rm max}=10 P_{\rm cold}$ and $P_{\rm min}=0.8P_{\rm cold}$,
 where $P_{\rm cold}$ is 
the pressure calculated using the cold EOS~\eqref{EOS}.
Whenever $P$ exceeds $P_{\rm max}$ or drops below $P_{\rm min}$, we 
reset $P$ to $P_{\rm max}$ or $P_{\rm min}$, respectively.  
Following~\cite{Etienne08} we impose the upper pressure limits only 
in regions where the rest-mass density remains very low ($\rho_0 < 100
\rho_{\rm atm}$), but we impose the lower limit everywhere on our grid.
We impose the pressure floor everywhere, because numerical error 
sometimes leads $\epsilon - \epsilon_{\rm cold}$ slightly below zero, resulting in negative thermal pressure. 
We have found experimentally that if this situation arises, it can be avoided in 
the subsequent timesteps by imposing the pressure floor.

At each timestep, 
the ``primitive variables'' 
$\rho_0$, $P$, and $v^i$ must be recovered from the ``conservative'' variables 
$\rho_*$, $\tilde{\tau}$, and $\tilde{S}_i$. We perform the 
inversion numerically as specified in~\cite{Illinois_new_mhd}.  We use 
the same technique as
in ~\cite{FontFix,SSL2008} to ensure that the values of $\tilde{S}_i$ and
$\tilde{\tau}$ yield physically valid primitive variables.



\begin{center}
\begin{table*}[t]
\caption{Summary of initial configurations. $M_{\rm NS}$ ($M_{\rm WD}$)
is the ADM mass of an isolated NS (pWD)$^{(a)}$, $R_{\rm NS}$ ($R_{\rm WD}$) 
the isotropic radius of an isolated NS (pWD), $C_{\rm NS}$
the compaction of an isolated NS, where the compaction is
the ratio of the ADM mass of the isolated star to its areal radius. 
All pWDs considered here have $C_{\rm WD}=0.01$.
$M_{\rm ADM}$ is the ADM mass of the system and $A$ the initial binary separation 
in isotropic coordinates. Cases A1 and A3 are the same as the head-on collision cases we studied
in \cite{WDNS_PAPERII}. The initial coordinate separation for these cases was set to $587{\rm \ km}$.  
Case A corresponds to our simulation of a binary pWDNS in circular orbit starting at the Roche limit.
For this case $\Omega M_{\rm ADM}= 6.95\times 10^{-4}$.
All cases have been produced with the 10:1 EOS of \cite{WDNS_PAPERII}.}
\begin{tabular}{ccccccccc}\hline\hline
\multicolumn{1}{p{1.8cm}}{\hspace{0.45 cm} Case } & 
\multicolumn{1}{p{1.8cm}}{\hspace{0.15 cm} $M_{\rm NS}/M_\odot$ \quad}  &
\multicolumn{1}{p{1.8cm}}{\hspace{0.1 cm} $M_{\rm WD}/M_\odot$ \quad}  &
\multicolumn{1}{p{1.8cm}}{\hspace{0.4 cm} $C_{\rm NS}$ \quad}  &
\multicolumn{1}{p{1.8cm}}{\hspace{0.1 cm} $J/M_{\rm ADM}^2$ \quad}  &
\multicolumn{1}{p{1.8cm}}{\hspace{0.05 cm} $R_{\rm WD}/R_{\rm NS}$ \quad}  &
\multicolumn{1}{p{2.1cm}}{\hspace{0.1 cm} $R_{\rm WD}/M_{\rm ADM}$}  &
\multicolumn{1}{p{1.95cm}}{\hspace{0.05 cm} $M_{\rm ADM}/M_\odot$ }  &
\multicolumn{1}{p{1.6cm}}{\hspace{0.2 cm} $A/R_{\rm WD}$ \quad}  \\  \hline 
A1    	               &    1.4    &  0.98  & 0.11    & 0.      &  8.88    & 41.18    & 2.41  &   4.00        \\  \hline
A3    	               &    1.6    &  0.98  & 0.15    & 0.      &  11.15   & 37.46    & 2.65  &   4.00        \\  \hline
A    	               &    1.4    &  0.98  & 0.11    & 2.88      &  8.88    & 40.05    & 2.48  &   3.14       \\
\hline\hline 
\end{tabular}
\begin{flushleft}
$^{(a)}$ Here we list the ADM masses, isotropic radii and compactions of the isolated NS stars, whose
rest-mass density profiles were used to generate initial data. The same holds for the pWDs in 
cases A1 and A3. In case A the pWD rest-mass density profile, the Roche limit separation and $\Omega$ 
were generated by a Newtonian binary WDNS code and then used in our CTS solver.
\end{flushleft}
\label{tab:cases}
\end{table*}
\end{center}

\subsection{Diagnostics}
\label{sec:diagnostics}

During the evolution, we monitor the normalized Hamiltonian and momentum
constraints as defined in Eqs.~(40)--(43) of ~\cite{Etienne08a}. We also monitor the 
ADM mass and angular momentum of the system.
The equations used to
calculate the ADM mass and angular momentum with minimal numerical noise
are as follows \cite{BSBook}
\beqn
  M&=& \int d^3x \left(\psi^5\rho + {1\over16\pi}\psi^5 \tilde{A}_{ij}
\tilde{A}^{ij} - {1\over16\pi}\tilde{\Gamma}^{ijk}\tilde{\Gamma}_{jik} \right.
\ \ \label{eq:M_sur_vol} \nonumber \\
&& \left. + {1-\psi\over16\pi}\tilde{R} - {1\over24\pi}\psi^5K^2\right), 
\eeqn

\beqn
  J_i &=& {1\over8\pi} \epsilon_{ij}{}^n\int d^3x
           \bigl[\psi^6(\tilde{A}^j{}_n + {2\over3}x^j\partial_nK 
\label{eq:J_sur_vol}  \nonumber \\
         && - {1\over2} x^j\tilde{A}_{\ km}\partial_n\tilde{\gamma}^{\ km}) 
         + 8\pi x^j S_n\bigr].  
\eeqn
Here
$\psi = e^\phi$, 
$\rho=n_\mu n_\nu T^{\mu \nu}$, $S_i = -n_\mu \gamma_{i \nu} T^{\mu \nu}$, 
$\tilde{R}$ is the Ricci scalar associated with $\tilde{\gamma}_{ij}$, 
and $\tilde{\Gamma}_{ijk}$ are Christoffel symbols associated with 
$\tilde{\gamma}_{ij}$. 

When hydrodynamic matter is evolved on a fixed uniform grid, our
hydrodynamic scheme guarantees that the rest mass $M_0$ is conserved
to machine roundoff error.  This strict conservation is no longer maintained
in an AMR grid, where spatial and temporal prolongation is performed
at the refinement boundaries.  Hence, we also monitor the
rest mass
\beq
  M_0 = \int \rho_* d^3x
\label{eq:m0}
\eeq
during the evolution. Rest-mass conservation is also violated whenever 
$\rho_0$ is reset to the atmosphere value. This usually happens only in the 
very low-density atmosphere.  The low-density regions do not affect rest-mass 
conservation significantly. 

In all simulations we present in this work 
the normalized Hamiltonian constraint violations remain smaller
than $0.9\%$ and the normalized momentum constraint violations
smaller than $2.1\%$. Rest mass is conserved to within 
$4\%$ and angular momentum to within $10\%$. 

Shocks occur when the stars collide. We measure the
 entropy generated by shocks via the quantity
$K\equiv P/P_{\rm cold}\geq 1$, where $P_{\rm cold}$ is
the pressure associated with the cold EOS that characterizes the
initial matter (see Eq.~\eqref{EOS}).

\section{Radiative Cooling}
\label{sec:cooling}

Our binary WDNS head-on collision studies in \cite{WDNS_PAPERII} demonstrate that
the hot, quasiequilibrium TZlO remnants do not collapse promptly to a black hole, 
even though the final total mass is larger than the maximum mass supportable by 
the cold EOS. This outcome might also arise in the case of WDNS mergers in circular orbit, 
and can be due to additional support provided by thermal pressure 
and/or rapid rotation. In order to determine whether thermal support is dominant, 
we add radiative cooling to the GR hydrodynamic equations. 
We now describe our formalism for implementing this.

\begin{table*}[t]
\caption{Grid configurations used in our simulations. Here $M$ is the
  sum of the ADM masses of the isolated stars, $\Delta x$ is the grid spacing in the innermost refinement 
box surrounding the NS, $N_{\rm NS}$ denotes the number of
grid points covering the diameter of  the NS initially, and $N_{\rm WD}$
denotes the number of grid points covering the (smallest) diameter
of the pWD initially. The outer boundary distance to the center of mass is approximately
$1020M$ in cases A1 and A3, and $540M$ in case A.}
\begin{tabular}{cccccc}
  \hline \hline
  Case & $M/M_{\odot}$ & Grid Hierarchy (in units of $M$)$^{(a)}$
  & $\Delta x$ & $N_{\rm NS}$ & $N_{\rm WD}$ \\
  \hline \hline
  A1  &  2.38   & (534.33, 267.16, 133.58, 66.79, 35.78, 19.08, 10.44,
  7.156) & $M/6.71$ & 63 & 35 \\
  A3  &  2.58 & (467.27, 233.64,116.82, 58.41, 29.20 , 15.58, 8.518,
  5.841) & $M/8.22$ & 56 & 38 \\
  A   &  2.38   & (270.87, 135.44, 67.72, 36.28[N/A], 19.35[N/A], 9.674[N/A], 5.744[N/A]) & $M/13.2$ & 124 & 73 \\
  \hline \hline
\end{tabular}

\begin{flushleft}
$^{(a)}$ There are two sets of nested refinement boxes: one centered
  on the NS and one on the pWD.  This column specifies the half-length
of the sides of the refinement boxes centered on both the NS and pWD. 
  When there is no corresponding pWD
  refinement box (as the pWD is much larger than the NS), 
  we write [N/A] for that box.
\end{flushleft}
\label{table:GridStructure}
\end{table*}

The dynamics of radiation is governed by \cite{MihalasBook,CollapseShapiro1996,BFarris2008}
\labeq{rad_dyn}{
\nabla_\alpha R^{\alpha\beta}=-G^{\beta},
}
where $R^{\alpha\beta}$ is the radiation stress-energy tensor given by
\labeq{}{
R^{\alpha\beta}=\int d\nu d\Omega I_{\nu}N^{\alpha}N^{\beta},
}
and $G^{\alpha}$ is the radiation four-force density given by
\labeq{}{
G^{\alpha}=\int d\nu d\Omega (\chi_\nu I_{\nu}-j_\nu)N^{\alpha}.
}
In the equations above $d\Omega$ is the solid angle, $\nu$ and $I_{\nu}=I_{\nu}(x^\alpha,N^i,\nu)$ 
are the radiation frequency and
specific intensity of radiation at $x^\alpha$ moving in direction $N^\alpha=p^\alpha/h\nu$, respectively. 
All quantities are measured in the local Lorentz frame of a 
fiducial observer with four-velocity
$u^{\alpha}_{fid}$, i.e., 
\labeq{}{
h\nu = -p_{\alpha} u^{\alpha}_{fid},
}
where $p^{\alpha}$ is the photon four-momentum and $h$ denotes Planck's constant. The
energy-momentum conservation equation then becomes
\labeq{}{
\nabla_{\alpha} (T^{\alpha\beta}+R^{\alpha\beta}) =0
}
or after using Eq.~\eqref{rad_dyn}
\labeq{divT}{
\nabla_{\alpha} T^{\alpha\beta} =G^{\beta}.
}
After projecting this equation using the fluid four-velocity
$u^{\alpha}$, we obtain the modified energy equation:
\labeq{energy}{
u^\alpha\nabla_\alpha \varepsilon = -(\varepsilon+P)\nabla_{\alpha}u^{\alpha} - u^{\alpha}G_{\alpha},
}
where the perfect fluid stress-energy tensor has been written as
\labeq{}{
T^{\mu\nu} = (\varepsilon+P)u^{\mu}u^{\nu}+Pg^{\mu\nu},
}
and $\varepsilon$ is the total energy density.
Using the continuity equation
\labeq{cont}{\nabla_\alpha(\rho_0 u^{\alpha}) =0,}
Eq. \eqref{energy} becomes
\labeq{energy2}{
u^\alpha\nabla_\alpha \varepsilon = \frac{\varepsilon+P}{\rho_0}u^{\alpha}\nabla_{\alpha}\rho_0 - u^{\alpha}G_{\alpha}.
}
The total energy density is related to the specific thermal energy via the following equation
\labeq{rho_of_eth}{
\varepsilon=\rho_0(1+\epsilon_{\rm th}+\epsilon_{\rm cold}).
}
Using Eqs.~\eqref{energy2} and~\eqref{rho_of_eth} we find that the specific thermal energy evolves as
\labeq{thermal_def}{
u^\alpha\nabla_\alpha \epsilon_{\rm th} = \frac{P_{\rm th}}{\rho_0^2}u^\alpha\nabla_\alpha\rho_0-\frac{1}{\rho_0}u^\alpha G_{\alpha},
}
where we have used $d\epsilon_{\rm cold}/d\rho_0 = P_{\rm cold}/\rho_0^2$, and $P_{\rm th}= P-P_{\rm cold}$.

In the comoving reference frame $u^{\alpha}\nabla_{\alpha} = d/d\tau$, where $\tau$ is the proper time. 
Thus, in the comoving frame Eq. \eqref{thermal_def} becomes 
\labeq{thermal2}{
\frac{d}{d\tau}\epsilon_{\rm th} = \frac{P_{\rm th}}{\rho_0^2}\frac{d}{d\tau}\rho_0-\frac{1}{\rho_0}u^\alpha G_{\alpha}.
}
In order to achieve cooling, the radiation term in \eqref{thermal2} 
must be specified so that thermal energy can be removed. For this reason we choose the following cooling law that 
gives rise to exponential cooling 
\labeq{uG}{
u^\alpha G_{\alpha}=\epsilon_{\rm th}\rho_0/\tau_c,
}
where $\tau_c>0$ is the cooling time scale.
Substituting Eq.~\eqref{uG} in Eq.~\eqref{thermal2} we obtain
\labeq{thermal2a}{
\frac{d}{d\tau}\epsilon_{\rm th} = \bigg[\frac{(\Gamma_{\rm th}-1)}{\rho_0}\frac{d\rho_0}{d\tau}-\frac{1}{\tau_c}\bigg]\epsilon_{\rm th},
}
where we used Eq.~\eqref{Pthermal} to substitute for the thermal pressure.

The first term in brackets on the RHS of Eq.~\eqref{thermal2a} results from
adiabatic compression or expansion. 
The second term results from cooling and radiates away thermal energy exponentially. Thus, if initially we have a 
quasiequilibrium spherical object which is thermally supported, and we cool it quasistatically,
i.e., choose a cooling time scale much longer than the free-fall time scale of the star, 
it will radiate thermal energy away and contract. While the contraction generates extra heat,
radiation tends to remove any residual thermal energy. 
Thus, after cooling, TZlOs are expected either to collapse to a BH or 
be supported by the residual cold pressure and centrifugal force.

\begin{figure*}
\centering
%
\subfigure{\includegraphics[width=0.325\textwidth]{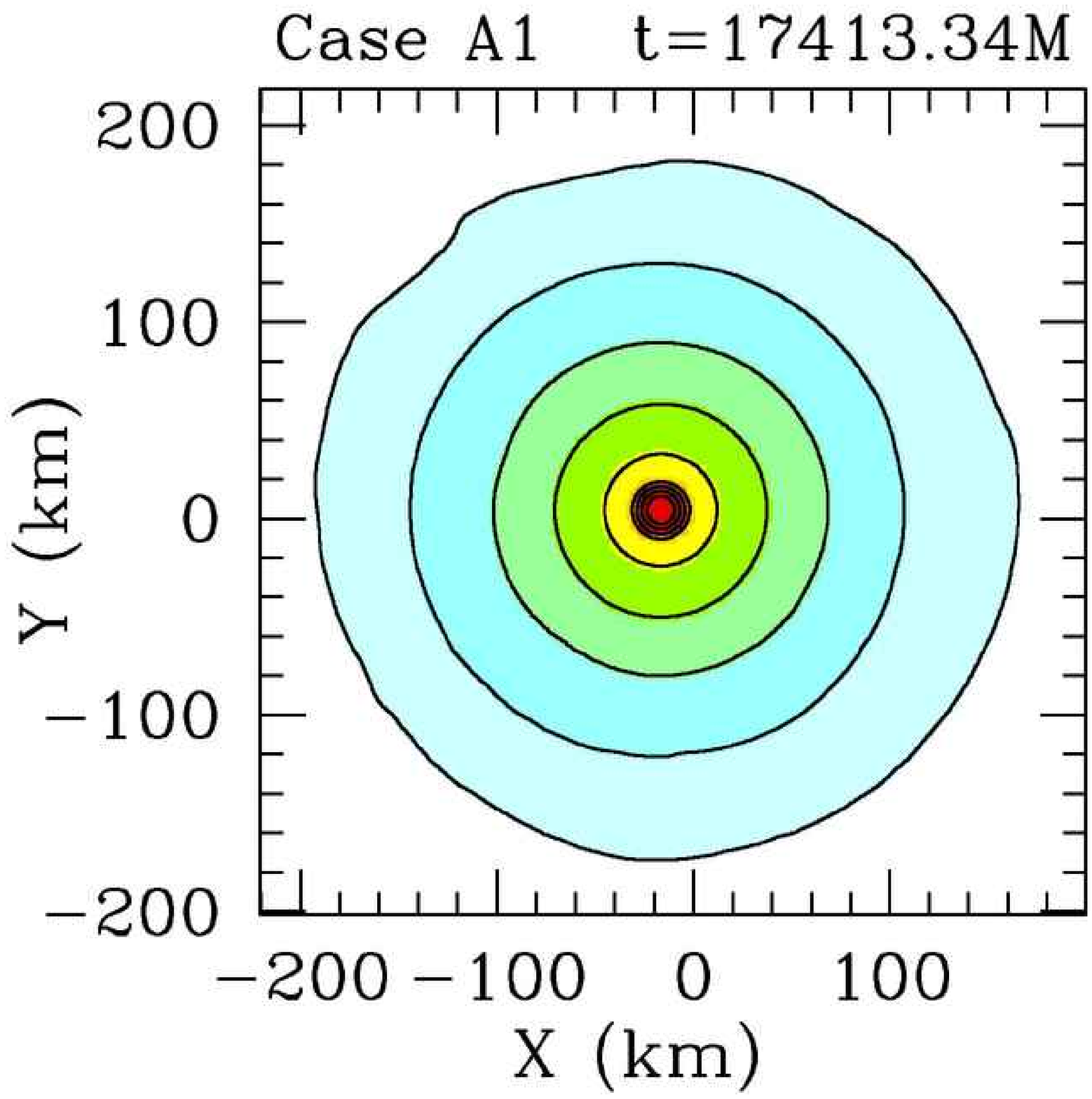}}
\subfigure{\includegraphics[width=0.325\textwidth]{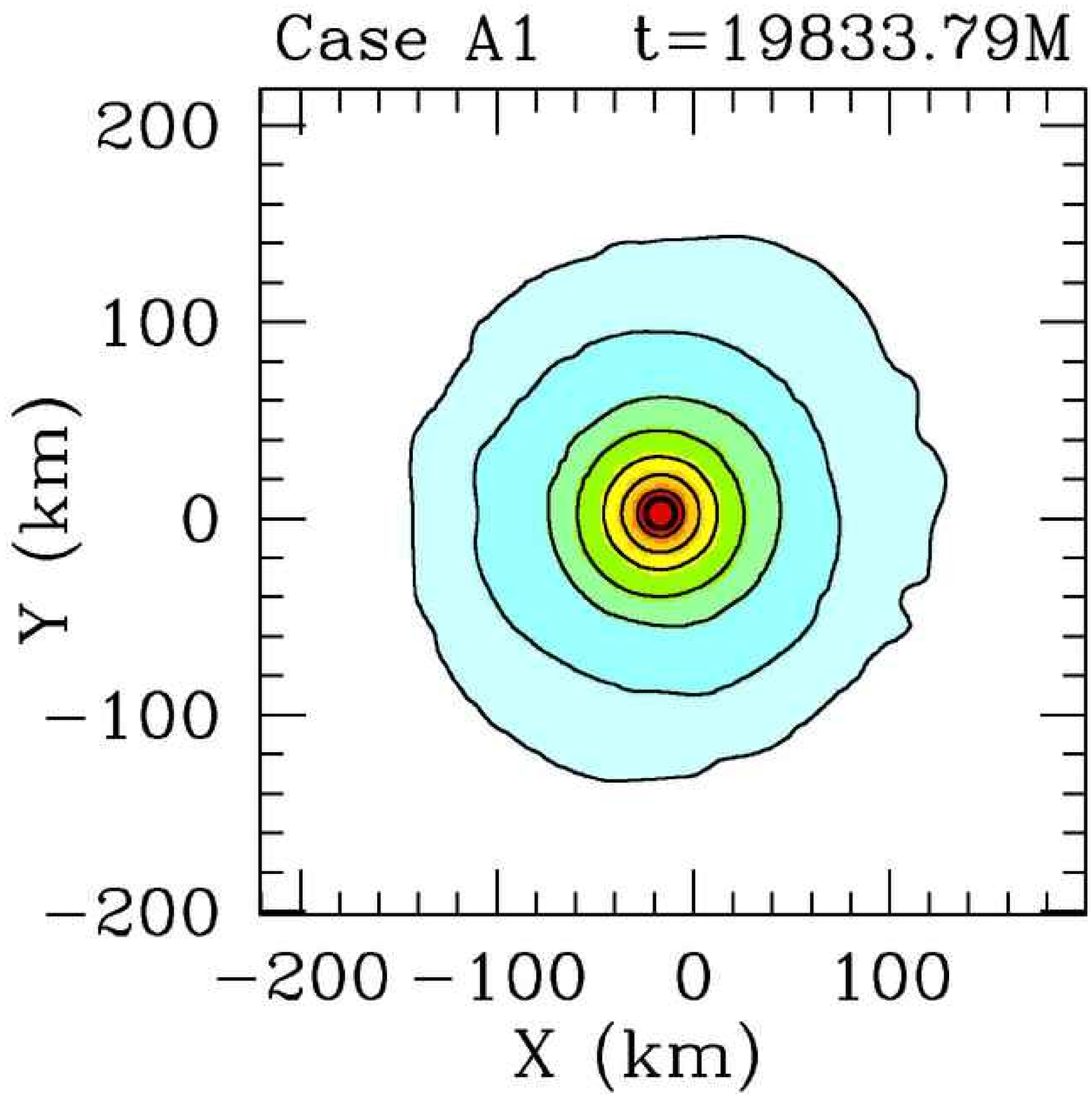}}
\subfigure{\includegraphics[width=0.325\textwidth]{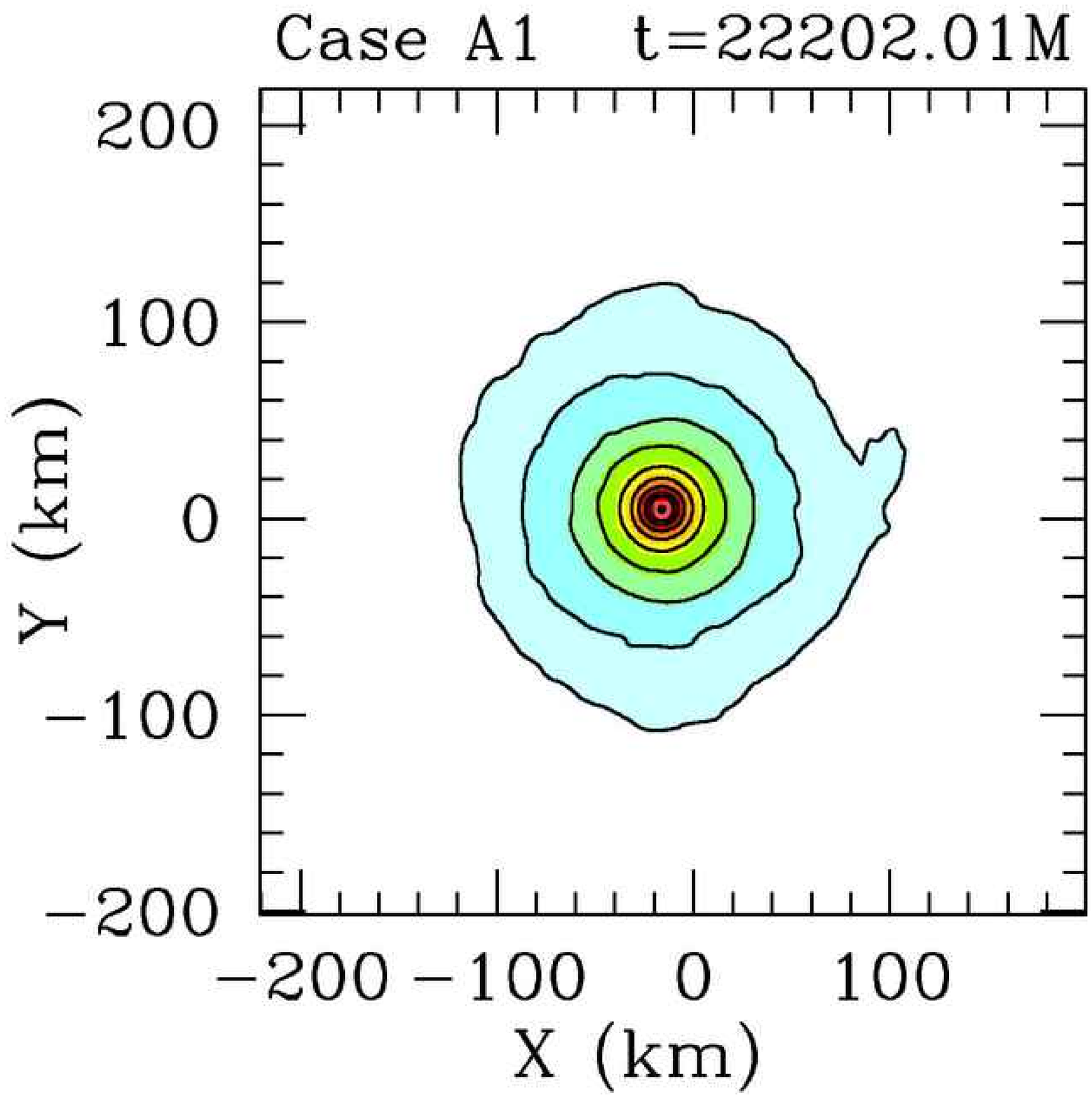}}
\subfigure{\includegraphics[width=0.325\textwidth]{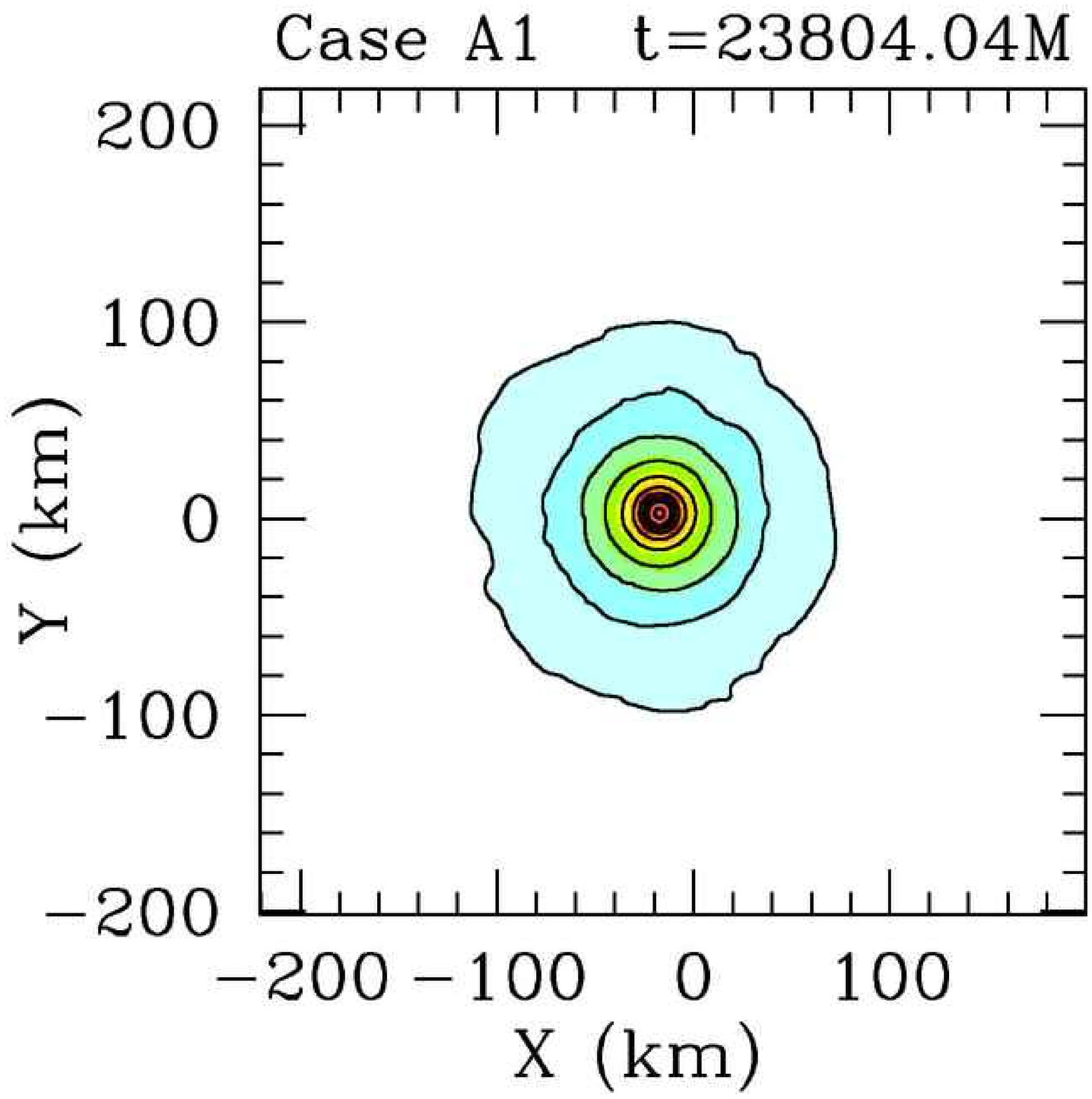}}
\subfigure{\includegraphics[width=0.325\textwidth]{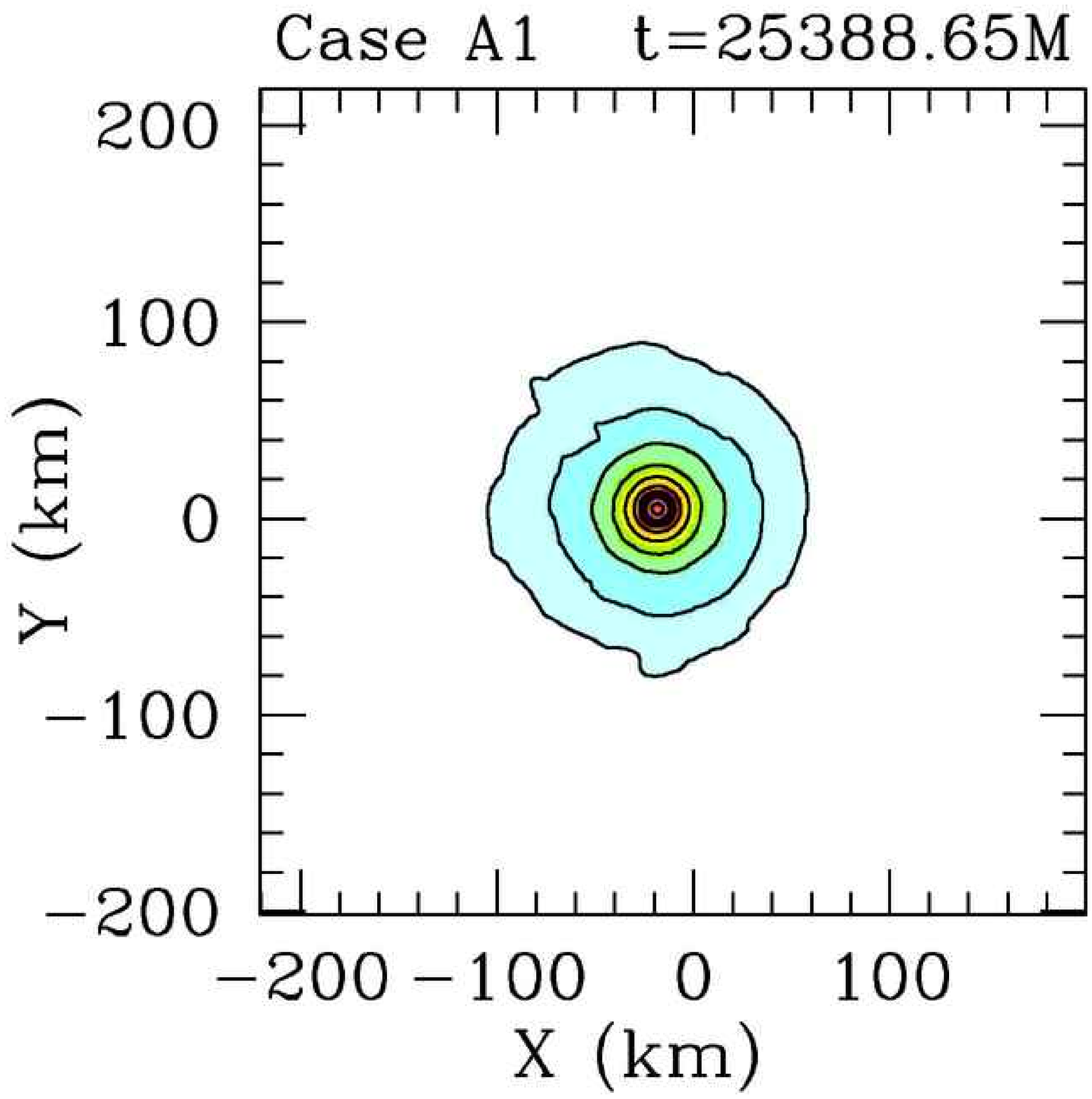}}
\subfigure{\includegraphics[width=0.325\textwidth]{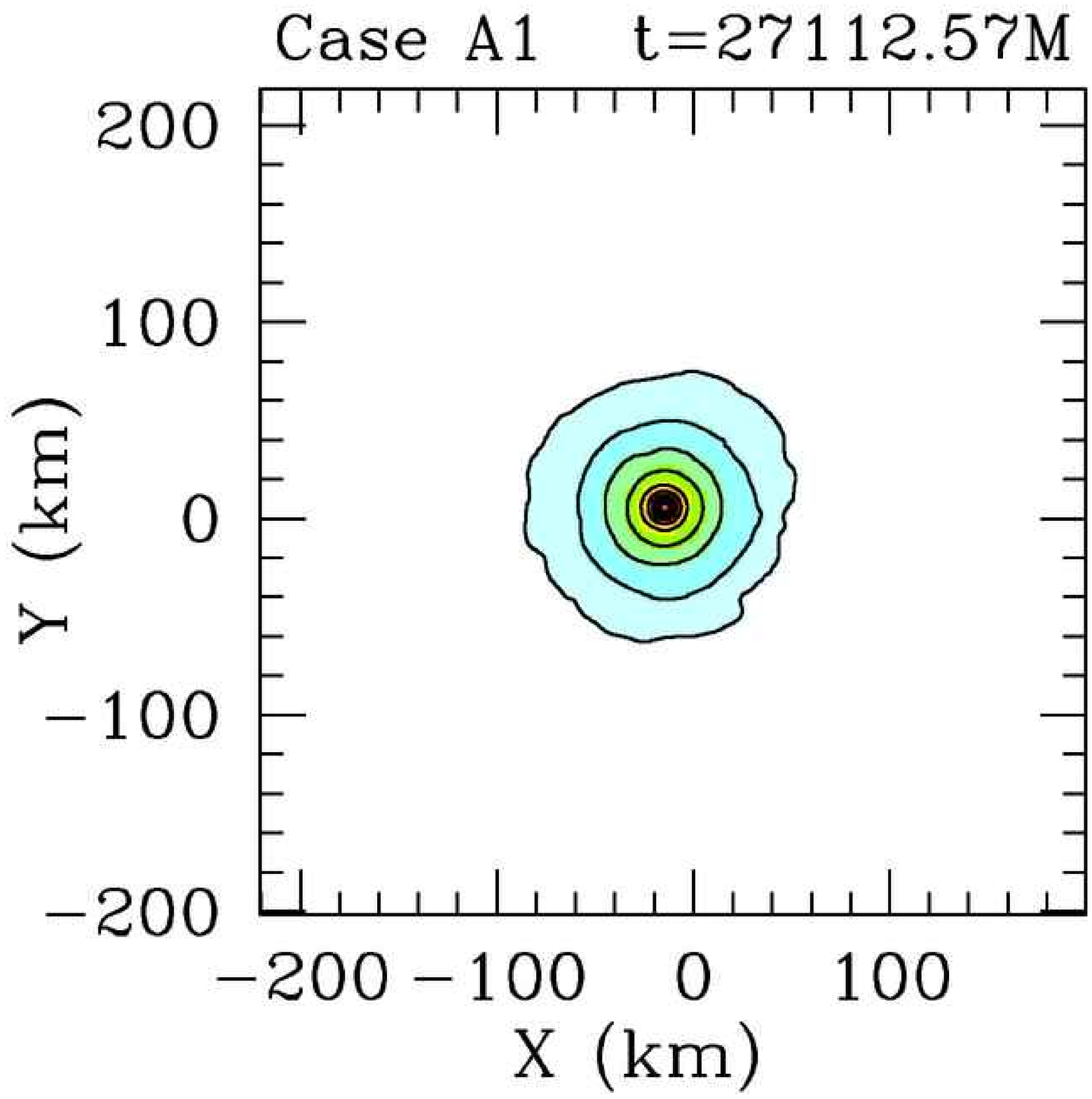}}
\caption{
Snapshots of rest-mass density profiles at selected times for head-on case A1 after cooling is turned on. 
The contours represent the rest-mass density in the orbital plane, plotted 
according to $\rho_0 = \rho_{0,\rm max} 10^{-0.68j-0.16}\ (j=0,1,\ldots, 7)$. The color sequence
dark red, red, orange, yellow, green, light green, blue and light blue 
implies a sequence from higher to lower values. This roughly corresponds to darker grey-scaling for higher values.
Here $\rho_{0,\rm max} = 4.645\rho_{\rm nuc}$, where $\rho_{\rm nuc}=2\times 10^{14}{\rm \ g/cm}^3$.
These snapshots clearly demonstrate that the entire TZlO remnant
collapses once cooling is turned on.
Here $M = 2.38\ M_\odot = 3.52{\rm \ km} = 1.17\times 10^{-5}\ s $
is the sum of the ADM masses of the isolated stars. 
\label{fig:A1xy}
}
\centering
\end{figure*}


In the optically thin regime, assuming the fluid radiates isotropically in its rest frame, and 
using $u^\alpha_{fid}=u^\alpha$, the source term $G^\alpha$ is generally expressed as \cite{CollapseShapiro1996,BFarris2008}
\labeq{Ga}{
G^\alpha=u^\alpha(\Gamma-\Lambda)+\int d\nu (\chi_\nu^a+\chi_\nu^s)F^\alpha_\nu, 
}
where $\Gamma$  and $\Lambda$ are the heating and cooling terms, respectively, given by
\labeq{}{
\begin{split}
\Gamma = & \int d\nu d\Omega\chi_\nu^a I_\nu, \\
\Lambda = & \int d\nu d\Omega j_\nu,
\end{split}
}
and where $\chi_\nu^a, \chi_\nu^s$ are the absorption and scattering coefficients, respectively, 
and $j_\nu$ is the emissivity. 
Finally, $F^\alpha_\nu$ is the total radiation flux four-vector
\labeq{}{
F^\alpha_\nu = P^\alpha{}_\beta \int d\Omega I_\nu N^\beta,
} 
where the projection operator $P^\alpha{}_\beta$ is defined as
\labeq{uproj}{
P^\alpha{}_\beta \equiv \delta^{\alpha}{}_{\beta}+u^\alpha u_\beta.
}

If we assume that there is no absorption and no scattering,
Eq.~\eqref{Ga} becomes
\labeq{isoG}{
G^\alpha=-u^\alpha\Lambda.
}
As a result,
\labeq{uG2}{
u_\alpha G^\alpha=\Lambda.
}
A straightforward comparison of Eqs.~\eqref{uG} and \eqref{uG2} shows 
that the integrated emissivity in our cooling model is given by
\labeq{emis}{
\Lambda = \frac{\rho_0}{\tau_c}\epsilon_{\rm th}.
}

\begin{figure*}
\centering
\subfigure{\includegraphics[width=0.325\textwidth]{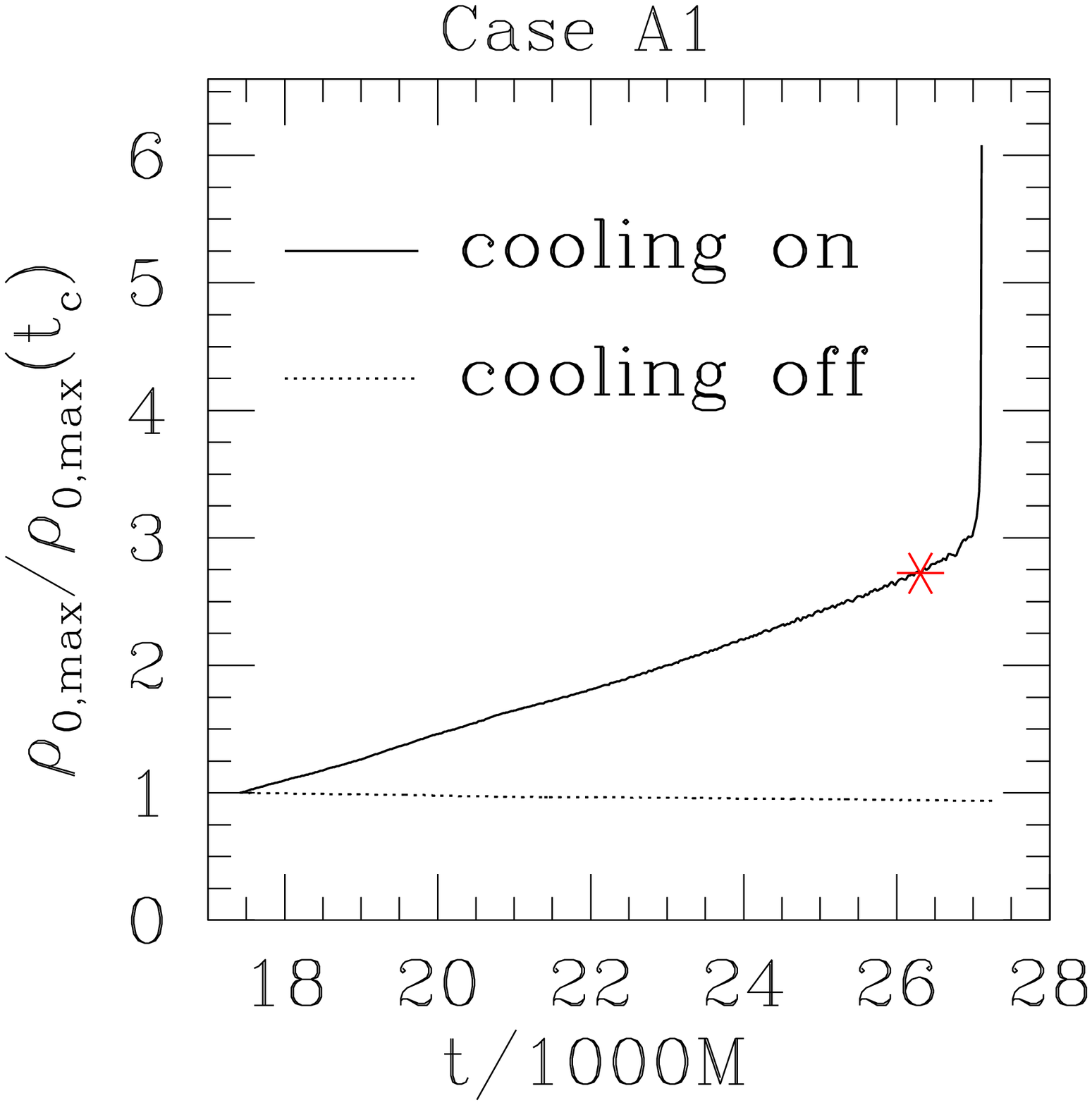}}
\subfigure{\includegraphics[width=0.325\textwidth]{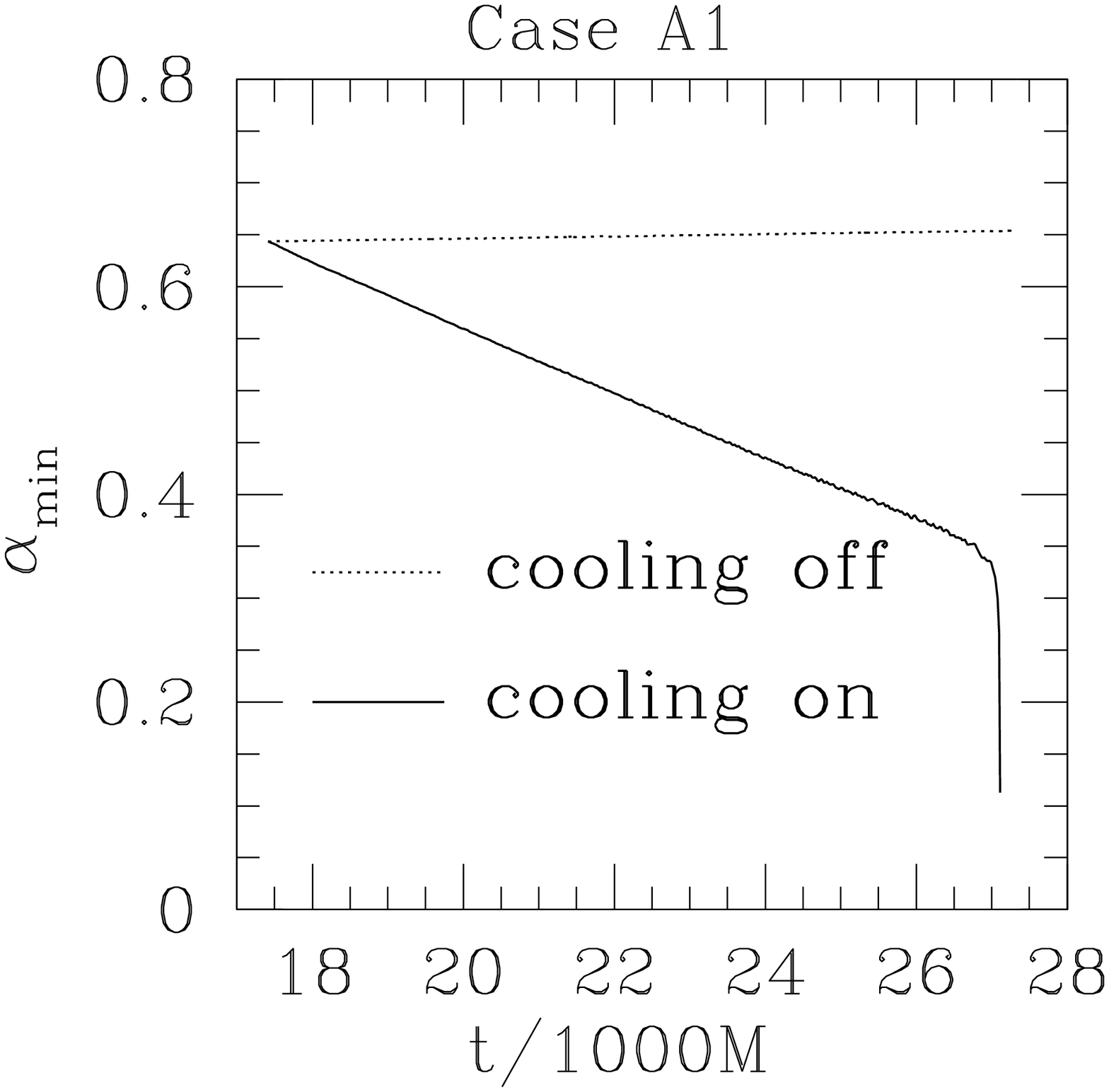}}
\subfigure{\includegraphics[width=0.325\textwidth]{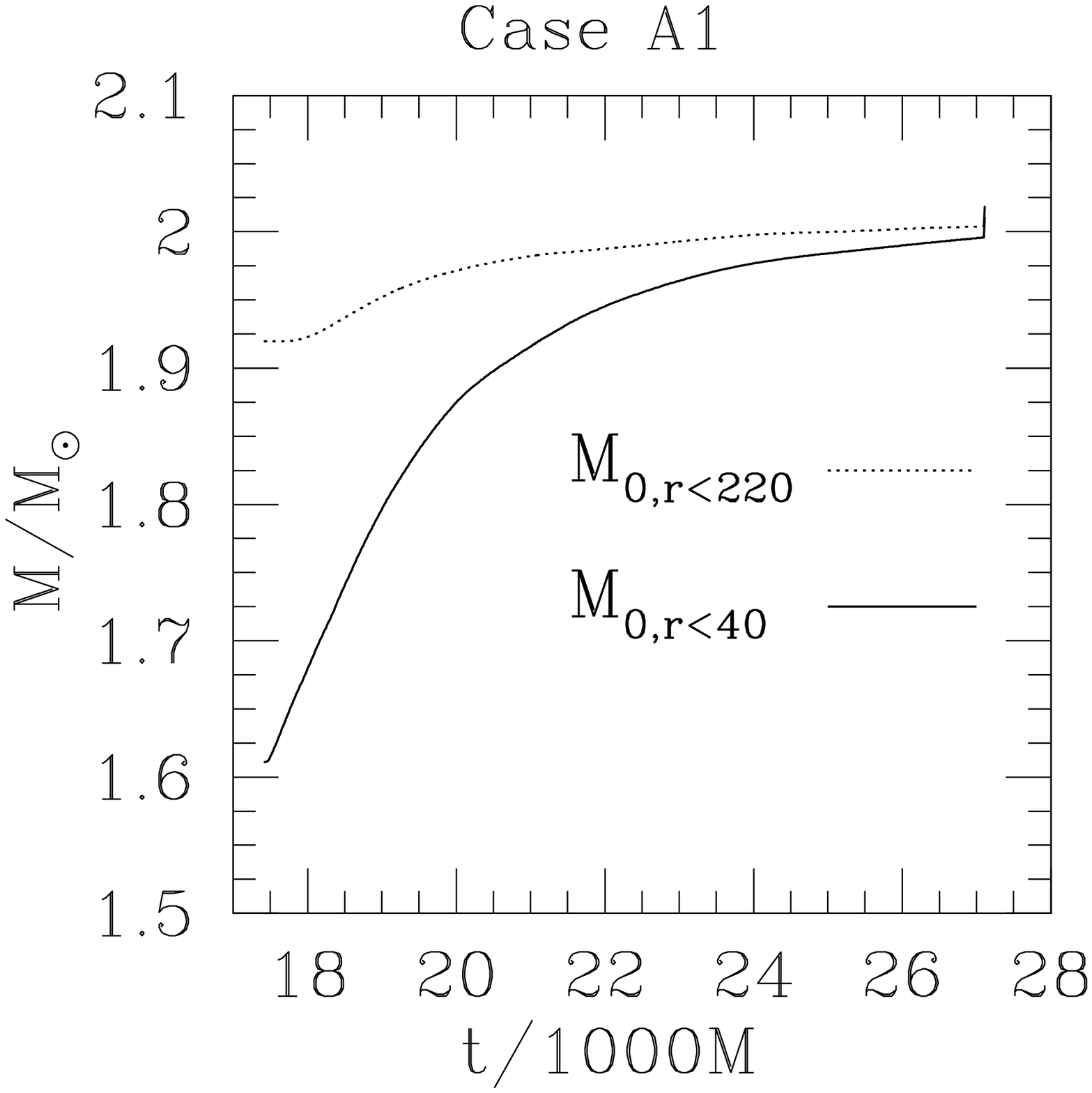}}
\caption{
Left panel: Evolution of maximum value of rest-mass density with cooling (solid curve) and without 
cooling (dotted curve) for the case shown in Fig.~\ref{fig:A1xy}. Here $\rho_{0,{\rm max}}$ is the maximum value of the rest-mass density, 
$\rho_{0,{\rm max}}(t_{c})=5.879\times 10^{-4} {\rm \ km}^{-2}=7.919\times 10^{14} {\rm \ g/cm}^3$ 
is the maximum value of rest-mass density at the time ($t_{c}=17413M$) when cooling is turned on. 
The asterisk on the curve denotes the value of the central density corresponding to the maximum-mass TOV NS 
($\rho_{\rm max, TOV}=2.16\times 10^{15}{\rm \ g/cm}^3$). Soon after 
the maximum density of the TZlO crosses $\rho_{\rm max, TOV}$, the remnant collapses to a BH.
Middle panel: Evolution of minimum value of the lapse function ($\alpha_{\rm min}$) with cooling (solid curve) and without cooling (dotted curve). 
Right panel: Evolution of rest-mass contained within different radii with cooling turned-on. 
Here $M_{0,\rm r < r_0}$  stands for the rest mass contained within a coordinate sphere of
radius $r_0$ in units of {\rm \ km}, centered on the remnant's center of mass. These plots demonstrate that the TZlO remnant collapses
as a whole. All plots correspond to case A1, and $M = 2.38\ M_\odot = 3.52{\rm \ km} = 1.17\times 10^{-5}\ s$.
\label{fig:A1mass_density}
}
\centering
\end{figure*}

If we project Eq.~\eqref{divT} using the timelike unit vector $n^\alpha$ 
normal to spacelike hypersurfaces and the projection operator 
$h^\alpha{}_\beta=\delta^\alpha{}_\beta + n^\alpha n_\beta$, we find that
the 3+1 GRHD equations become
\labeq{Stilde}{
\partial_t \tilde S_i +\partial_j(\alpha\sqrt{\gamma}T^j{}_{i})=
               \half\alpha\sqrt{\gamma} T^{\alpha\beta}g_{\alpha\beta,i}
	       -\alpha\sqrt{\gamma}u_i \Lambda,
}
and 
\labeq{tautilde}{
\partial_t \tilde \tau +\partial_i(\alpha^2\sqrt{\gamma}T^{0i}-\rho_*v^i)= s
	       -\alpha^2\sqrt{\gamma}u^0 \Lambda,
}
where we have used Eq.~\eqref{isoG} and $\Lambda$ is given by Eq. \eqref{emis}. Thus, 
cooling enters as a source term in the GRHD equations, which is precisely 
how cooling is implemented in our HRSC code. 

Note that the optically-thin approximation employed here is valid for neutrino cooling in the 
WDNS merger scenario (head-on or otherwise). According to our analysis in \cite{WDNS_PAPERII}, the 
temperatures and densities of the hot mantle 
of a TZlO are such that thermal neutrino emission likely will be
the dominant source of cooling. The diffuse TZlO mantle composed of the 
WD debris is optically thin to neutrinos, justifying the above approximation.

Finally, note that the self-gravity of the radiation is neglected, i.e., we assume that radiation 
does not affect the spacetime structure, so only 
the perfect fluid stress energy tensor contributes to the BSSN source terms. 
This is a good approximation as long as the radiation energy density is subdominant 
($n_\alpha n_\beta R^{\alpha\beta} \ll n_\alpha n_\beta T^{\alpha\beta}$). 
This is indeed the case in a WDNS scenario because the rest mass
dominates the mass-energy as can be inferred by the local constraint violations.
In addition, the NS (i.e. the most compact object in our scenario) remains almost 
unaffected and cold throughout the evolution, i.e., radiation has no effect on the spacetime structure
in the vicinity of the NS. In all simulations we present here 
for radii ($r$) such that $r \gtrsim R_{\rm core} \approx 20\rm km$ the local constraint violations 
remain smaller than 0.1\% throughout the evolution after cooling is turned on, 
justifying our neglecting of the radiation self-gravity.

\section{Cooling of TZ$\rm l$Os formed in head-on collisions}
\label{cool_TZlOs}

We found in \cite{WDNS_PAPERII} that all our binary pWDNS head-on collisions 
formed hot, quasiequilibrium TZlOs, which were more massive than the maximum mass
our cold EOS
can support. However, these remnants did not collapse promptly to a black hole. 
As there is no angular momentum involved in a head-on collision, the additional support that prevents
collapse arises from thermal pressure alone. Therefore, if one were to cool these objects, one would expect
that they would eventually collapse on a cooling time scale. We check this expectation here so that we may 
implement the same cooling mechanism in the inspiral case, where the outcome is not so certain.

To determine the dominant cooling mechanism, and hence the cooling time scale,
one needs to know the density and temperature of the matter.
We estimated the temperature of realistic TZlOs 
to be of order $10^9 \rm\ K$. Given that
typical WD densities are of order $10^6 \rm {\rm \ g/cm}^3$, it is likely that 
cooling will be dominated by thermal neutrino processes. Realistic neutrino cooling 
time scales are at best of order $1$ {\rm \ yr} (see discussion in Sec.~\ref{sec:discussion}), or equivalently 
$\sim 10^7$ TZlO dynamical time scales. This slow cooling rate ensures that 
the collapse of TZlOs will be quasistatic. However, realistic cooling time scales are so 
long that it would be impossible to follow this secular phase with hydrodynamic simulations
in full GR because of the prohibitive computational cost.

Nevertheless, to confirm that these TZlOs collapse to BHs
after they have cooled, we can simply scale up the cooling law, as long as we keep 
the cooling time scale longer than the hydrodynamical time scale. 

We can estimate the dynamical time scale of a TZlO as
\labeq{tTZlO}{
t_{\rm TZlO} = \sqrt{\frac{R_{\rm TZlO}^3}{M_{\rm TZlO}}},
} 
where $R_{\rm TZlO}$ and $M_{\rm TZlO}$ are the radius and mass of the
remnant, respectively. 
Here we define the radius of TZlOs by the radius of a
coordinate sphere that contains 90\% of the remnant's total mass. 
For case A1 we find 
$t_{\rm TZlO} \approx 2300 {{\rm \ km}} \approx 654.5 \rm M$. 
The TZlO dynamical time scale in case A3 is approximately the same as
 that of case A1.
We choose $\tau_c= 6000{{\rm \ km}}$ for both cases, so that initially 
$\tau_c \approx 2.6 t_{\rm TZlO}$. This way we reduce the simulation time, 
while maintaining the time-scale inequality. 

Though we choose $\tau_c$ to be only a few times $t_{\rm TZlO}$, it
is $\gtrsim 100$ times larger than the dynamical time scale ($t_{\rm
  core}$) of the innermost, cold, NS core, which satisfies $t_{\rm
  core}/t_{\rm TZlO} \sim 15^{3/2}\approx 60$. 
This property is important because the core of the TZlO collapses first.
Finally, note that as the entire remnant contracts, its dynamical time
scale decreases so that the required inequalities are better satisfied with 
increasing time.

Our expectation is that these TZlOs will collapse following cooling 
and that 
collapse should occur on a cooling timescale. Given that (a) the
 shock-induced thermal 
energy is comparable to the gravitational binding energy, and (b) the 
total remnant mass 
is close to the maximum mass of a cold configuration allowed by our adopted EOS,
we also expect that a large fraction of the thermal energy needs to be removed
to induce collapse.
Note that this expectation
applies solely to TZlOs formed in head-on collisions. When TZlOs form 
after the merger 
of WDNSs in an initially circular orbit, the remnant rotates. Hence,
we cannot tell a priori that collapse will take place because of 
the additional centrifugal support.

Table~\ref{tab:cases} summarizes the physical parameters of the 
head-on collision cases we study, 
and Table~\ref{table:GridStructure} presents the AMR grid structure 
used in each case.
All our simulations with cooling turned on show that the TZlOs 
remnants formed in the head-on collisions begin to contract, and within a few
cooling time scales collapse to a black hole. In contrast, we find that 
when cooling is turned off the remnant does not collapse and remains
 in quasiequilibrium.
All these results can be clearly seen in Figs.~\ref{fig:A1xy}
and~\ref{fig:A1mass_density}, which correspond to case A1. Case A3 is
qualitatively similar.

In Fig.~\ref{fig:A1xy} rest-mass density contours are plotted in 
the orbital plane at selected times. Figure~\ref{fig:A1mass_density}
shows the evolution of the maximum rest-mass density with and without cooling, 
the minimum value of the lapse function, and the 
rest-mass contained within different radii from the center of mass. 
If the cooling mechanism remains off, both the maximum value of 
the density and the minimum value of the lapse 
remain constant with time (see left and middle panels of
 Fig.~\ref{fig:A1mass_density}). We find that the same
holds true for the rest-mass contained within $40${\rm \ km} and 
$220${\rm \ km}.

\begin{figure*}
\centering
\subfigure{\includegraphics[width=0.45\textwidth]{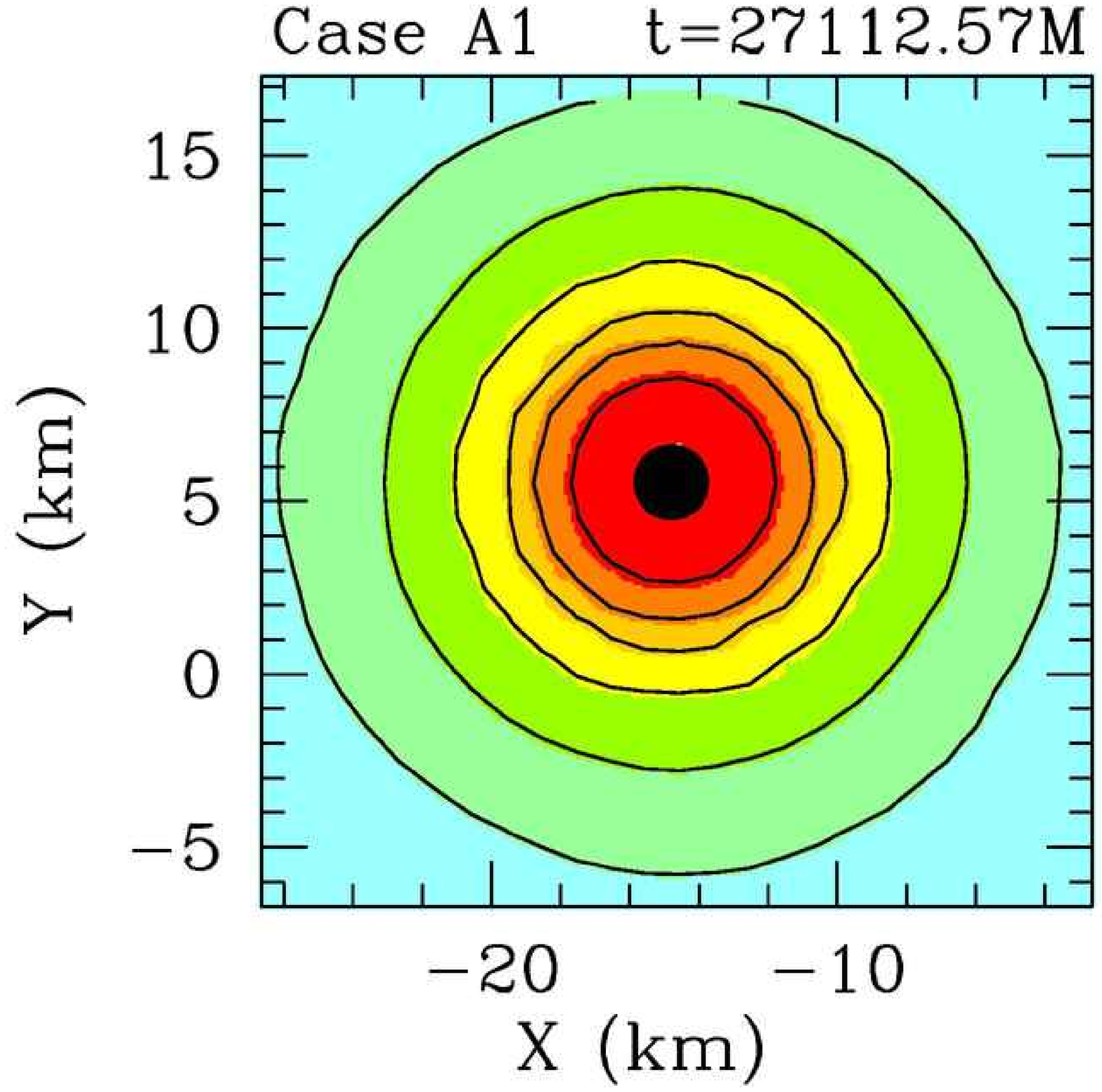}}
\subfigure{\includegraphics[width=0.45\textwidth]{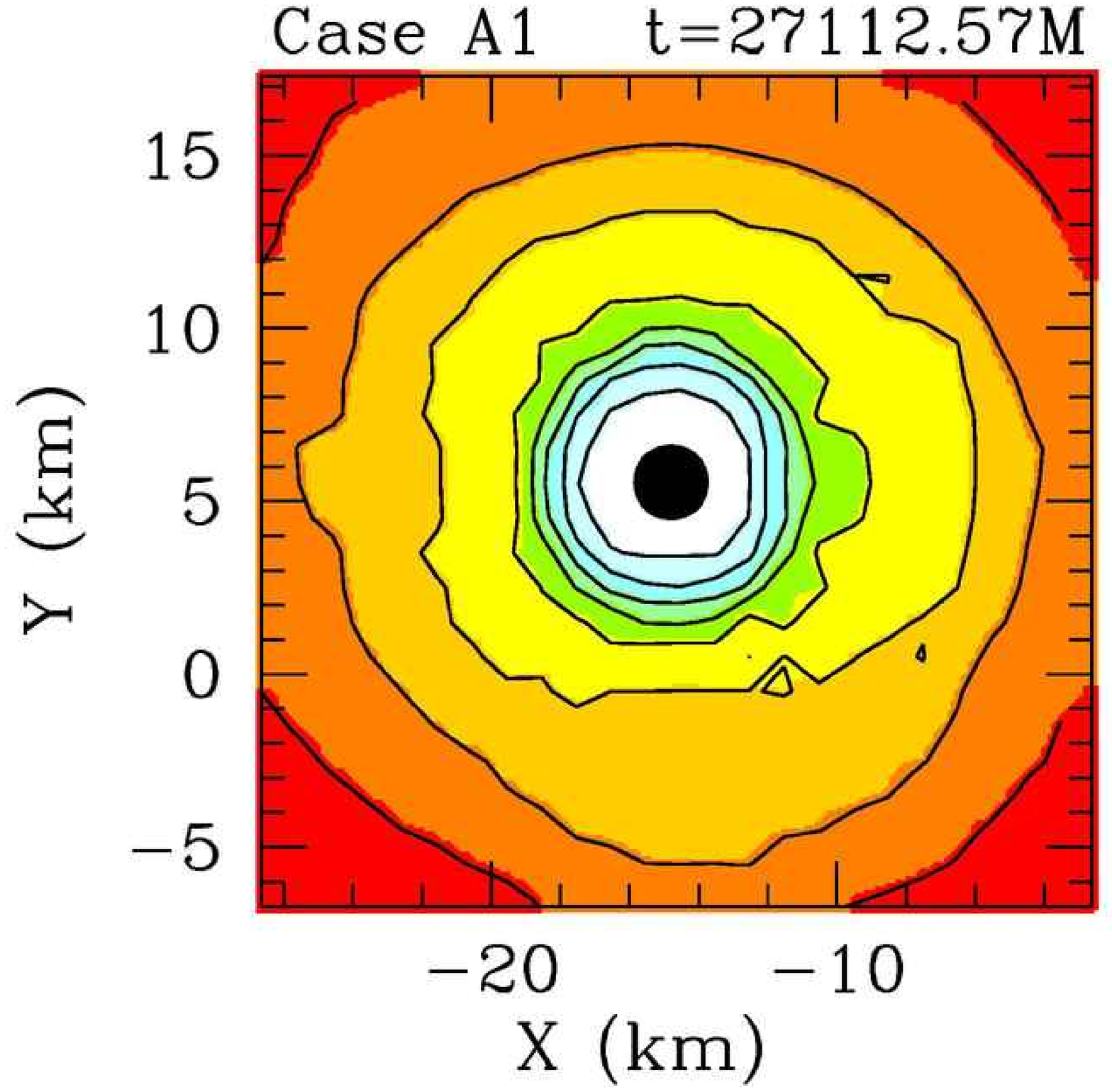}}
\caption{
Left: Snapshot of rest-mass density profile at the end of the A1 
simulation shown in Fig.~\ref{fig:A1xy}.
Contours represent the rest-mass density in the orbital plane, plotted 
according to $\rho_0 = \rho_{0,\rm max} 10^{-0.5j-0.088}\ (j=0,1,\ldots, 7)$. 
Here $\rho_{0,\rm max} = 4.645\rho_{\rm nuc}$, where $\rho_{\rm nuc}=2\times 10^{14}{\rm \ g/cm}^3$.
Right: Snapshot of $K=P/P_{\rm cold}$ profile at the end of the A1 simulation.  
The contours represent $K$ in the orbital plane, 
plotted according to $K = 10^{-0.1125j+0.9}\ (j=0,1,\ldots, 7)$. The plot demonstrates that 
the matter near the BH is cold ($K\approx 1$), as expected, and $K$ increases with the distance from the core.
The color coding is the same as that used in Fig.~\ref{fig:A1xy}, with white indicating
$K\approx 1$. The black disk in the center denotes the BH apparent horizon. 
The plots focus in the innermost $12{\rm \ km}$ from the TZlO center of mass, where the object 
is approximately spherical, and for this reason we do not show XZ and YZ meridional slices.
Here $M = 2.38\ M_\odot = 3.52{\rm \ km} = 1.17\times 10^{-5}\ s $
is the sum of the ADM masses of the isolated stars. 
\label{fig:A1KBHxy}
}
\centering
\end{figure*}

\begin{figure*}
\centering
%
\subfigure{\includegraphics[width=0.325\textwidth]{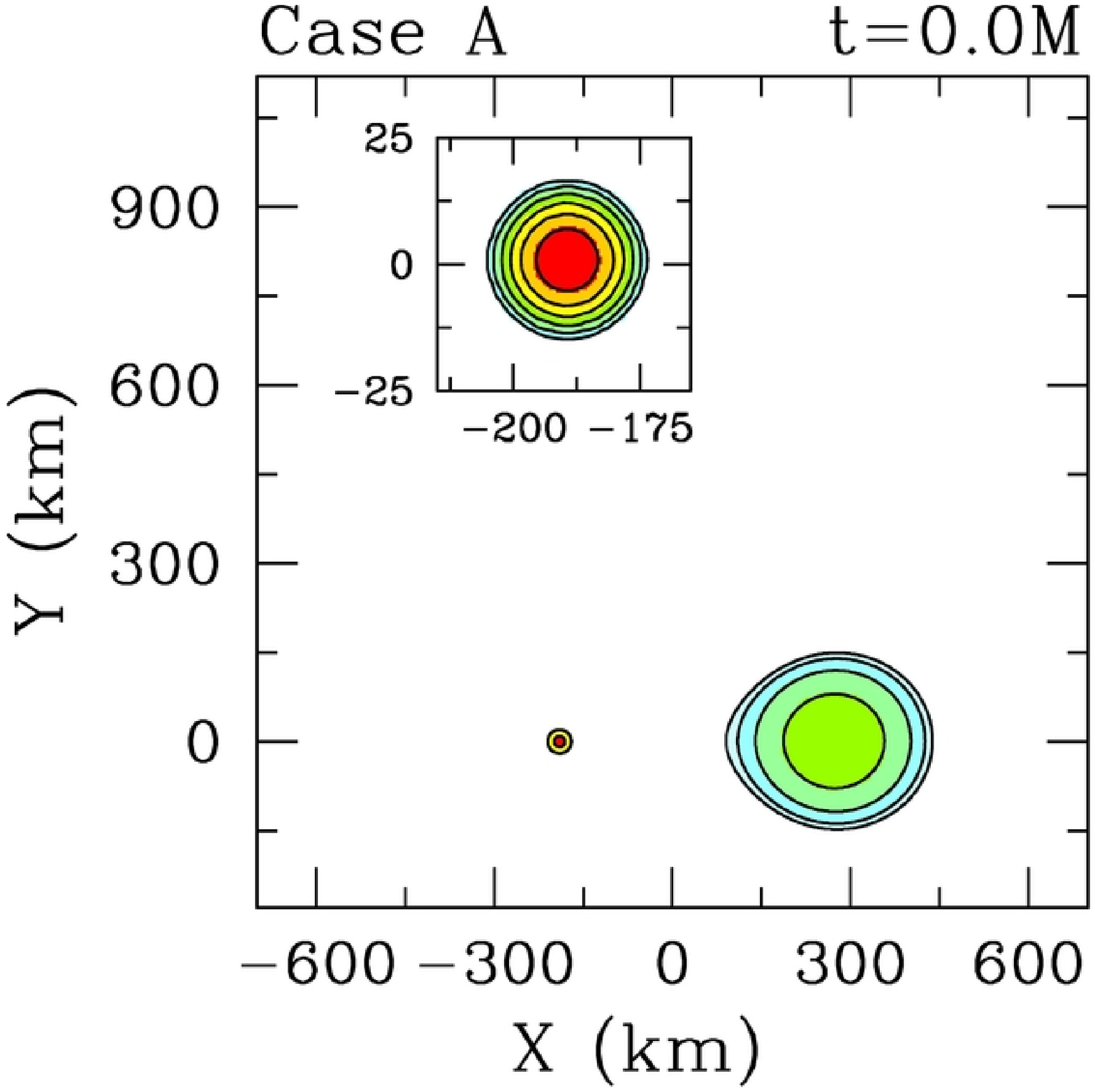}}
\subfigure{\includegraphics[width=0.325\textwidth]{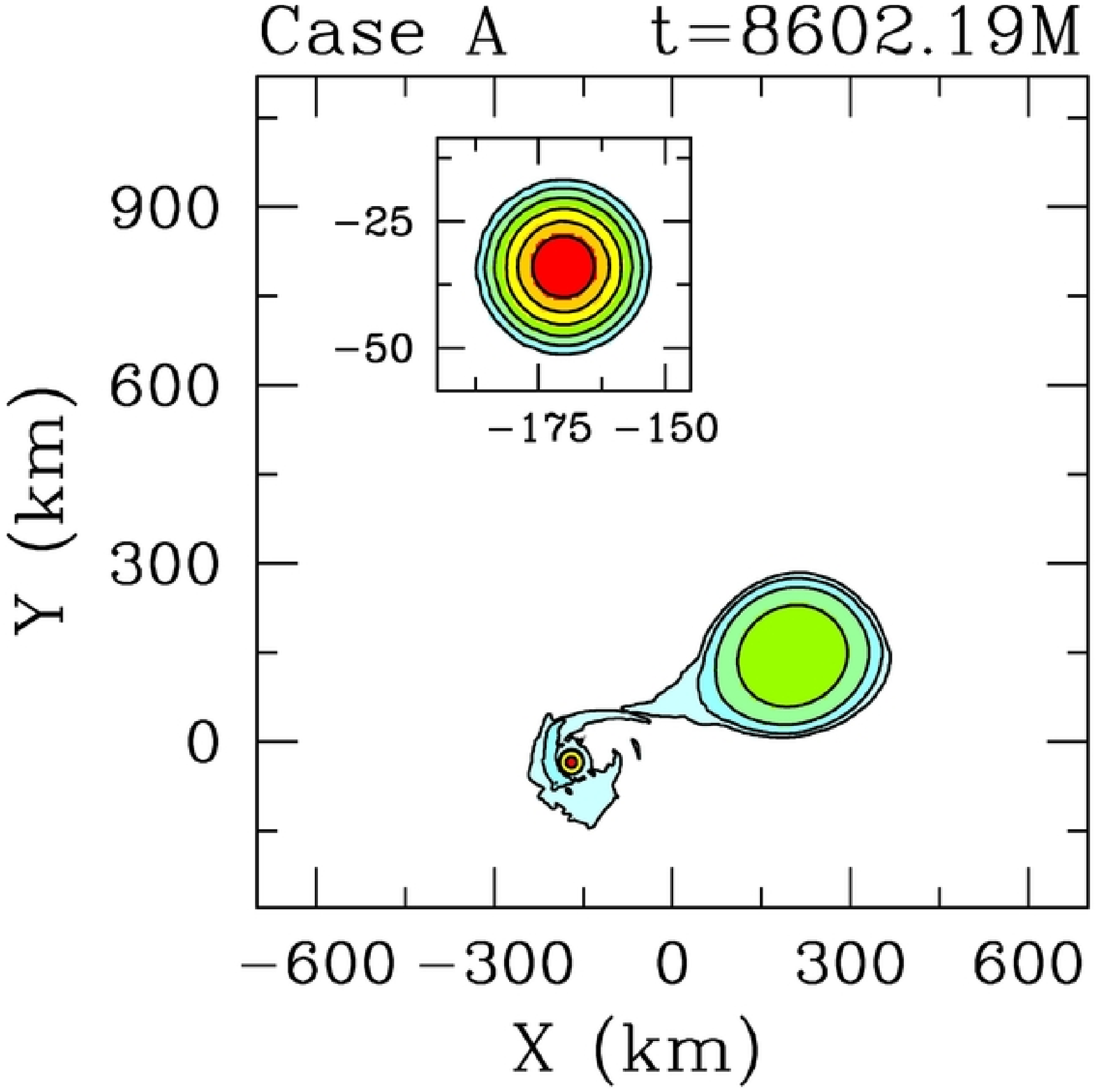}}
\subfigure{\includegraphics[width=0.325\textwidth]{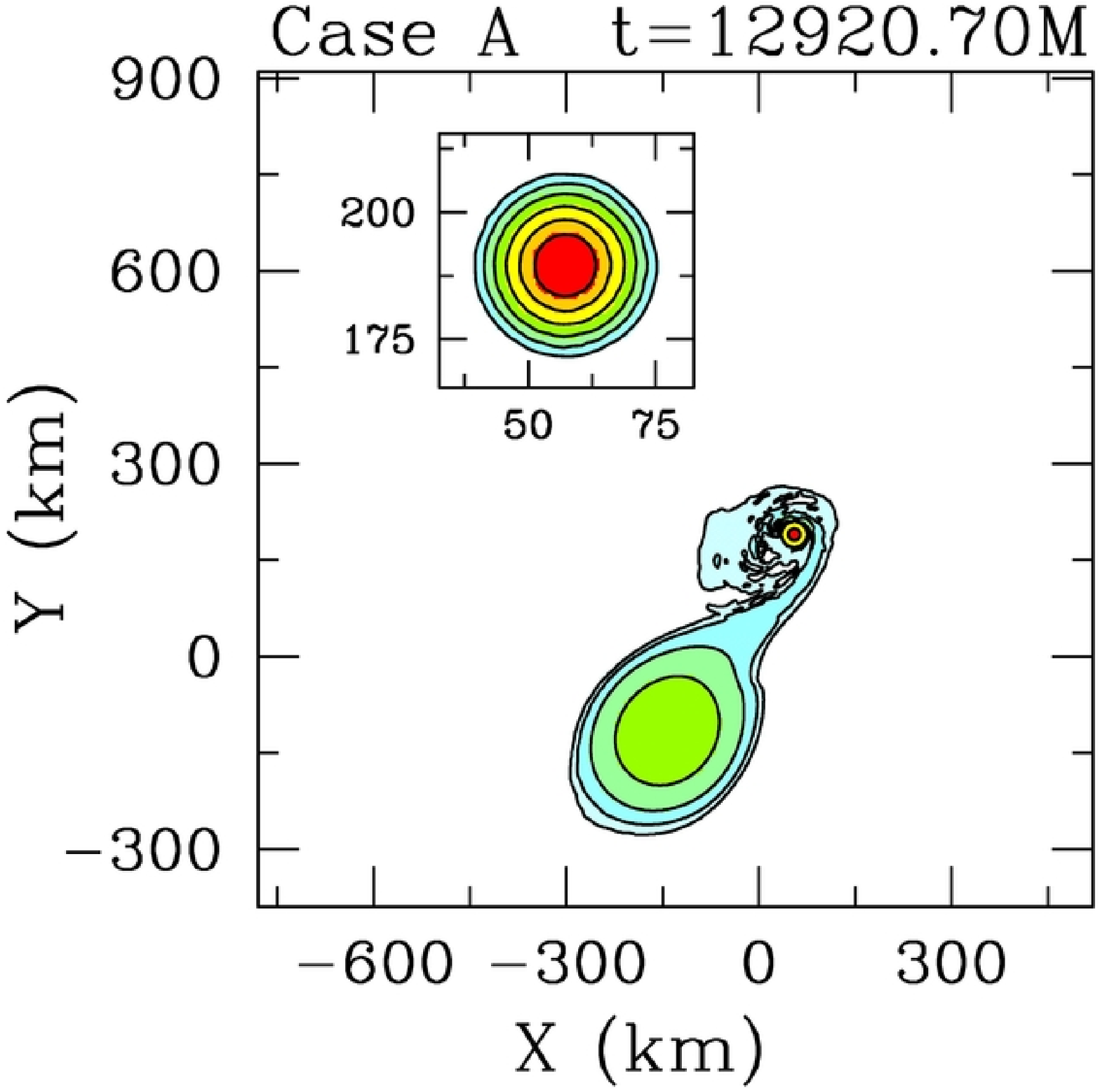}}
\subfigure{\includegraphics[width=0.325\textwidth]{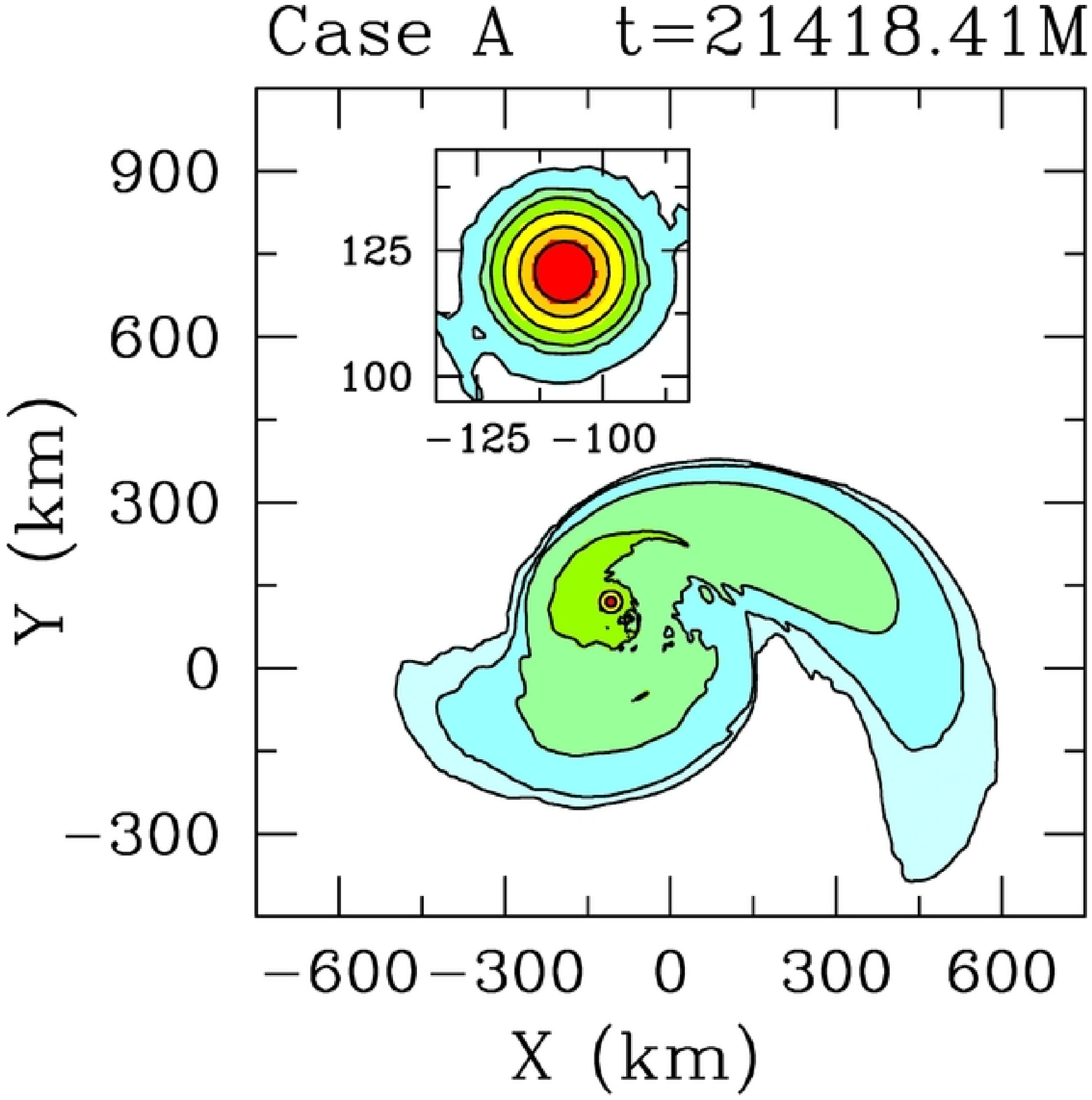}}
\subfigure{\includegraphics[width=0.325\textwidth]{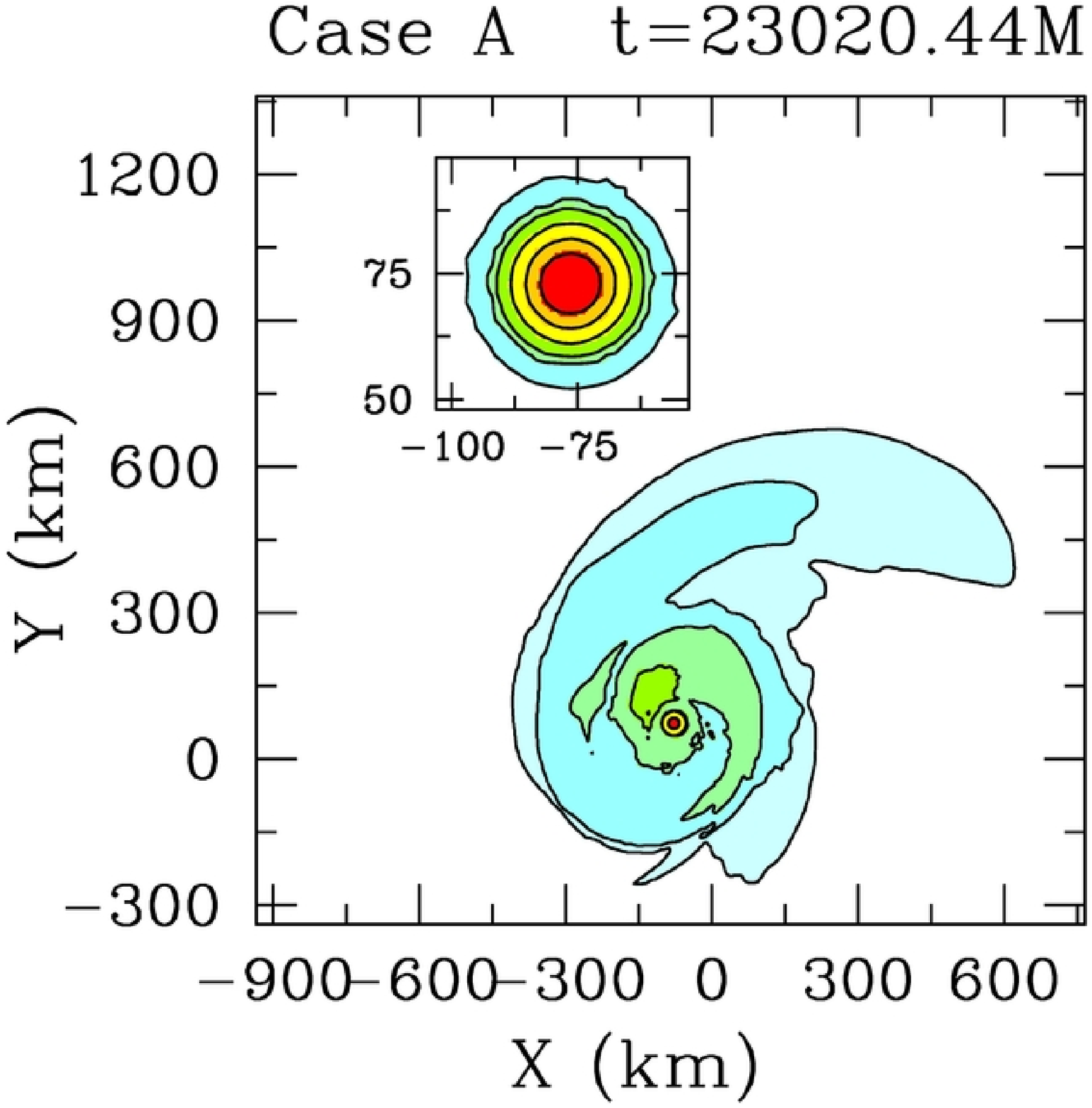}}
\subfigure{\includegraphics[width=0.325\textwidth]{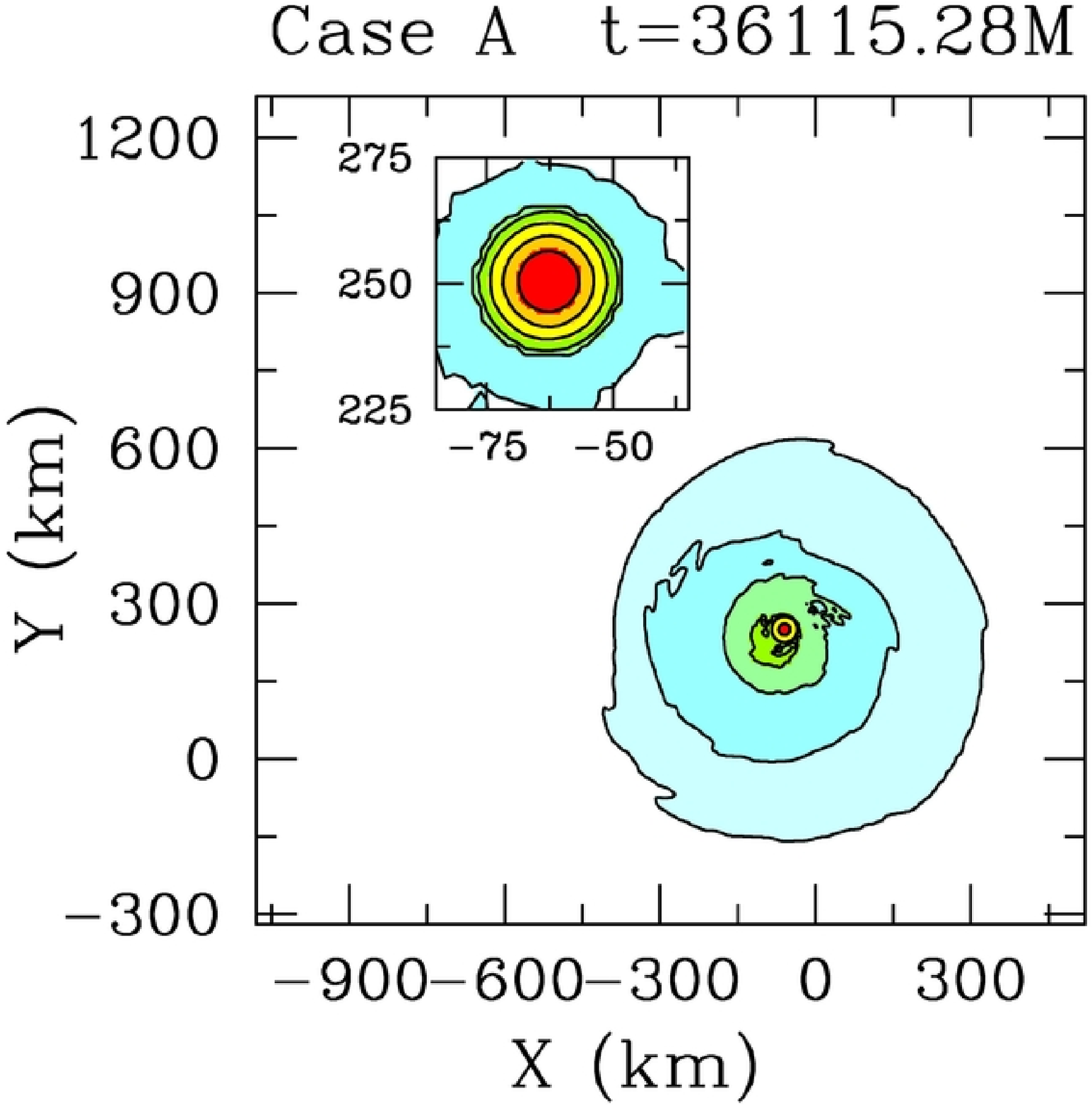}}
\caption{
Snapshots of rest-mass density profiles at selected times for case A. 
The contours represent the rest-mass density in the orbital plane, plotted 
according to $\rho_0 = \rho_{0,\rm max} 10^{-0.66j-0.16}\ (j=0,1,\ldots, 9)$. The insets
focus on the NS and demonstrate it is nearly unaffected by the merger. The inset density 
contours are plotted according to $\rho_0 = \rho_{0,\rm max} 10^{-0.525j-0.861}\ (j=0,1,\ldots, 7)$
In both cases $\rho_{0,\rm max} = 4.645\rho_{\rm nuc}$, where $\rho_{\rm nuc}=2\times 10^{14}{\rm \ g/cm}^3$.
The color coding is the same as that used in Fig.~\ref{fig:A1xy}, with
white indicating near vacuum.
Here $M = 2.38\ M_\odot = 3.52{\rm \ km} = 1.17\times 10^{-5}\ s $
is the sum of the ADM masses of the isolated stars. 
\label{fig:Axy}
}
\centering
\end{figure*}

By contrast, if cooling is turned on, the maximum density (minimum lapse) increases (decreases) with time.
Moreover, the rest mass contained within $40${\rm \ km} and $220${\rm \ km} increases
with time, indicating that the outer layers are also contracting as
the TZlO cools 
(see right panel in Fig.~\ref{fig:A1mass_density}). 
Initially, the maximum density (minimum lapse) increases (decreases) almost linearly with time, 
until $\rho_{0,\rm max}$ crosses the value of $2.16\times 10^{15}{\rm \ g/cm}^3$, which corresponds to that
of a maximum mass NS configuration built with our cold EOS.
Soon after this point, the remnant essentially
free-falls, the density blows up, the lapse function plummets and a BH
is eventually formed. 
The BH in case A1 can be seen in Fig.~\ref{fig:A1KBHxy}, where we plot rest-mass density contours 
and $K=P/P_{\rm cold}$ contours in the orbital plane in the innermost $12${\rm \ km} of the remnant, which contain 
about $85\%$ of the total mass at the time of BH formation. 
The $K$ contours show that the matter around the BH is cold, i.e., $K\approx 1$, as expected.

\begin{figure*}
\centering
\subfigure{\includegraphics[width=0.45\textwidth]{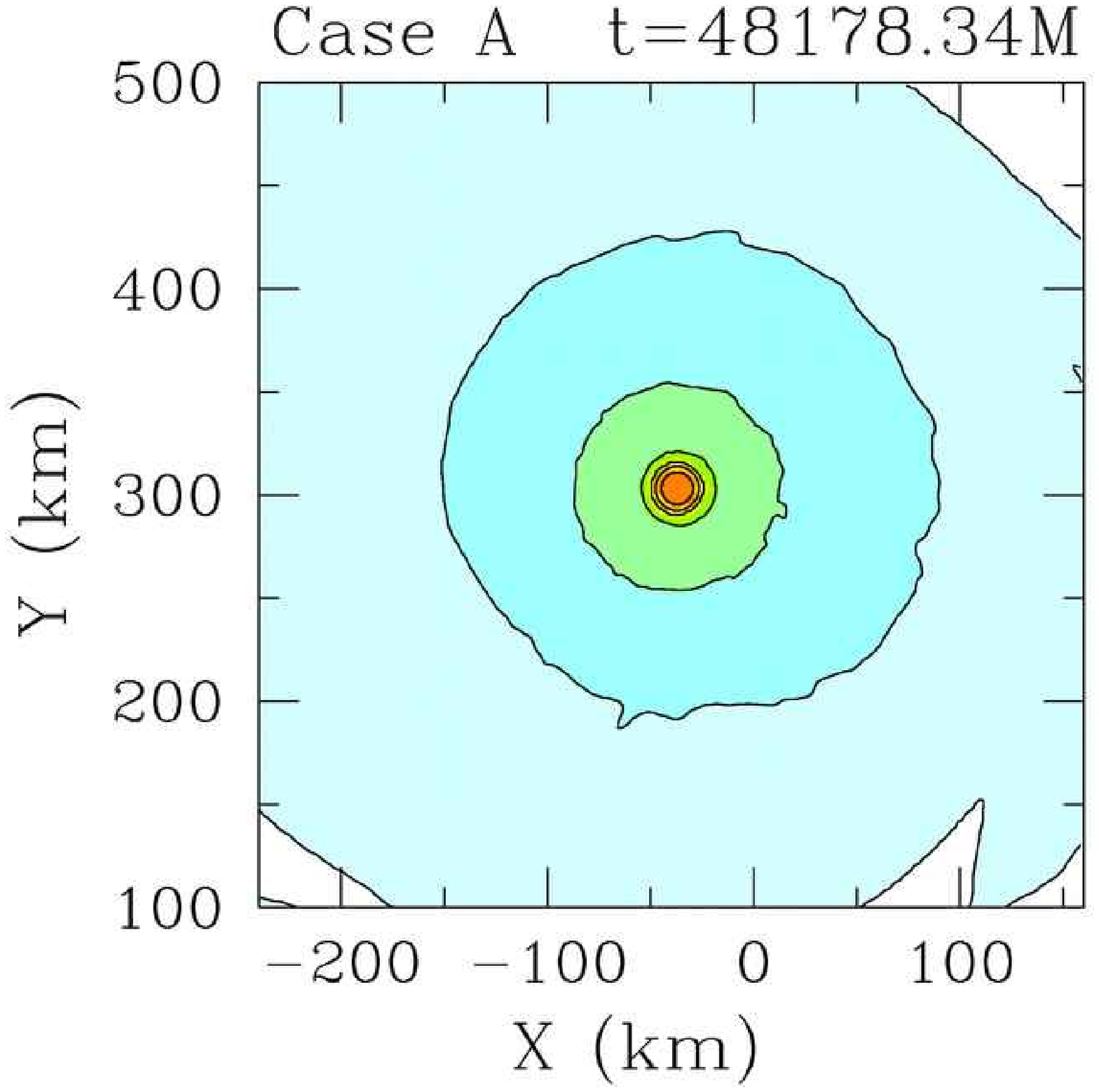}}
\subfigure{\includegraphics[width=0.45\textwidth]{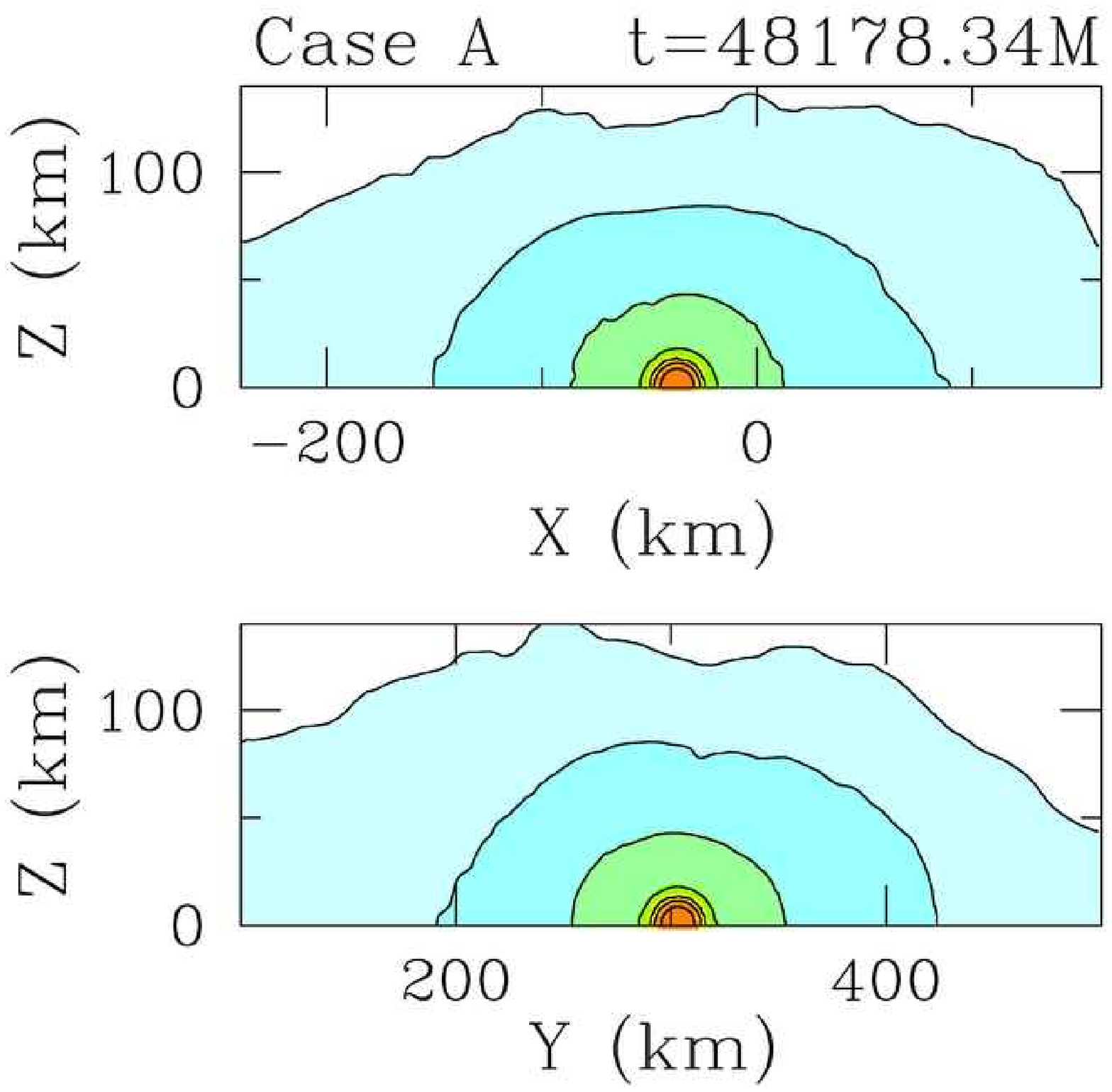}}
\subfigure{\includegraphics[width=0.45\textwidth]{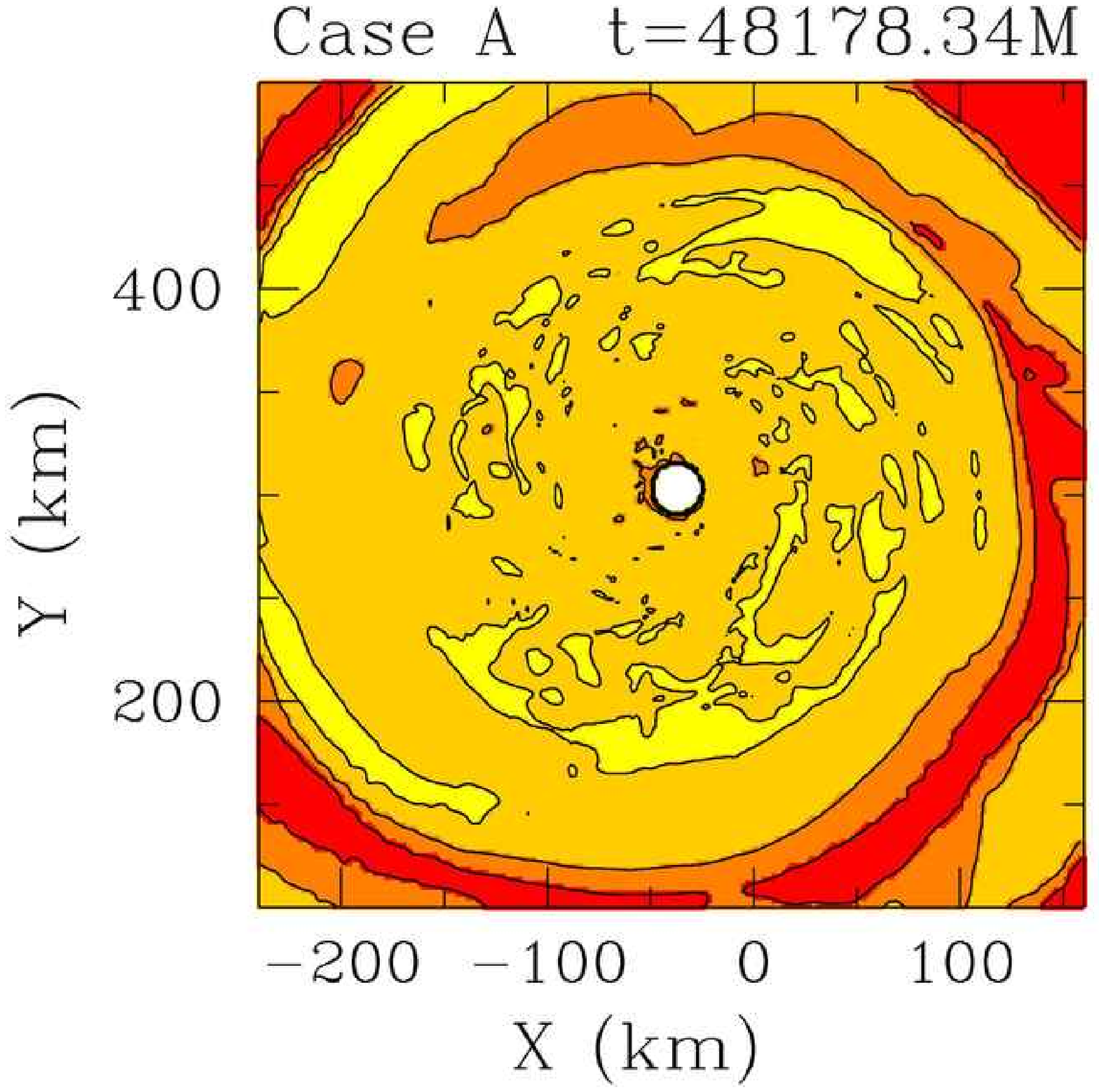}}
\subfigure{\includegraphics[width=0.45\textwidth]{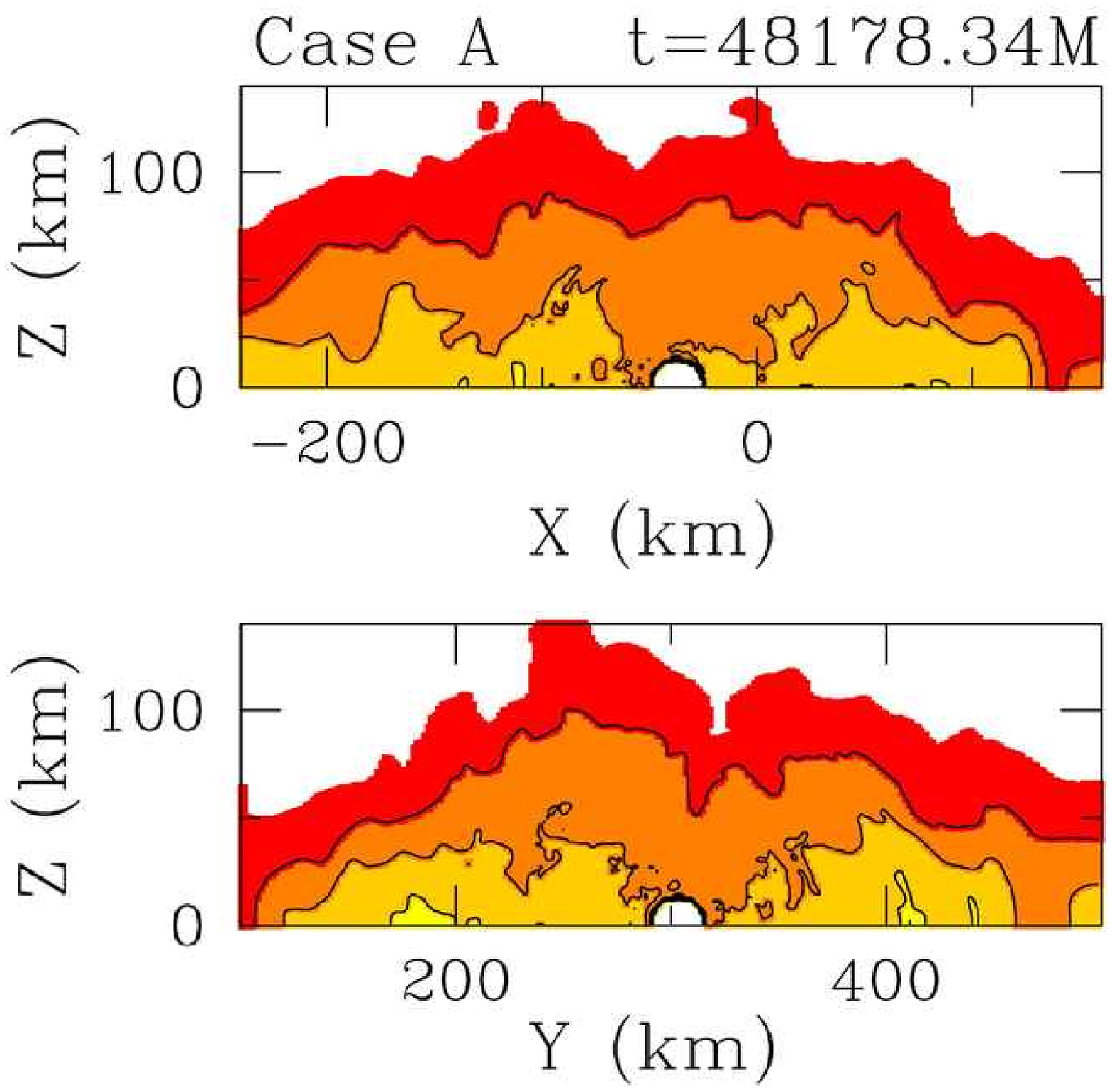}}
\caption{
First row: Snapshots of rest-mass density profiles at selected times for case A. 
The contours represent the rest-mass density in the orbital plane and the XZ and YZ meridional planes, 
plotted according to $\rho_0 = \rho_{0,\rm max} 10^{-0.69j-0.16}\ (j=0,1,\ldots, 7)$, where
$\rho_{0,\rm max} = 4.645\rho_{\rm nuc}$, and $\rho_{\rm nuc}=2\times 10^{14}{\rm \ g/cm}^3$.
Second row: Snapshots of $K=P/P_{\rm cold}$ profiles at selected times for case A. 
The contours represent $K$ in the orbital plane and the XZ and YZ meridional planes, 
plotted according to $K = 10^{-0.1125j+0.9}\ (j=0,1,\ldots, 7)$. The plots show that 
the remnant NS core is approximately spherical and cold ($K\approx 1$). Far from the core 
the remnant is hot. $K$ increases as we move away from the core and the orbital plane. 
All plots focus in the innermost $200{\rm \ km}$ from the TZlO center of mass,
The color code used is the same as that defined in Fig.~\ref{fig:A1xy}, with white color
in the second row indicating $K\approx 1$. Here $M = 2.38\ M_\odot = 3.52{\rm \ km} = 1.17\times 10^{-5}\ s $.
\label{fig:AKxy}
}
\centering
\end{figure*}

Cases A1 and A2 collapse to a black hole after about 5 and 3 cooling time scales, 
respectively, which is expected as the collapse proceeds without 
additional shock heating.
The mass of the black hole when an apparent 
horizon forms for the first time is $M_{\rm BH}\approx 1.4\ M_{\odot}$ and the coordinate radius
of the BH (in our adopted gauge) is $R_{\rm BH}\approx 1.08 {\rm \ km} \approx 0.53 M_{\rm BH}$. 
Thus we have demonstrated that our cooling mechanism yields results
which are consistent with our theoretical expectations.

\section{Binary WDNS inspiral}
\label{sec:results}

To predict the final outcome of a binary WDNS in an initially circular orbit,
we performed a simulation of a corotating binary pWDNS starting at the Roche limit 
separation. Throughout, we label this case by the letter A. 
Table~\ref{tab:cases} outlines the physical parameters of case A, and Table~\ref{table:GridStructure} 
outlines the adopted AMR grid structure. 

For the simulations performed here, we were able to 
demonstrate 2nd-order convergence for the first quarter of an orbit
monitoring the conservation of angular momentum, and the constraint 
violations. The convergence study showed that angular momentum 
decays linearly with time, but the linear decay rate decreases with
increasing resolution to second order. Moreover, this
decay rate remains roughly constant up until merger.
Furthermore, a resolution study using pWDNSs systems
has been carried out in \cite{WDNS_PAPERII}, where we 
showed that the results were qualitatively insensitive to resolution implying that the 
resolutions used were sufficiently high. The resolution used in our inspiral pWDNS calculations
is twice that used in \cite{WDNS_PAPERII} indicating that
our simulations are well within the convergent regime.


\subsection{Initial configuration}

We prepared valid general relativistic initial data as described in Section ~\ref{sec:Init_data}.
The ADM masses of the compact objects in isolation we consider 
are $0.98\ M_\odot$ and  $1.4\ M_\odot$ for the pWD and NS, respectively. After solving the CTS equations,
we map $\rho_0$, $\Psi$ and $\alpha$, $\beta^i$, and $v^i$, from the
grids used in the elliptic solver code onto the grids used in the
evolution code via second-order polynomial interpolation.
For accuracy, we make sure that the resolution of the initial data grids is always
higher than the resolution of the evolution grids.

\begin{figure}[t]
\centering
\includegraphics[width=0.495\textwidth,angle=0]{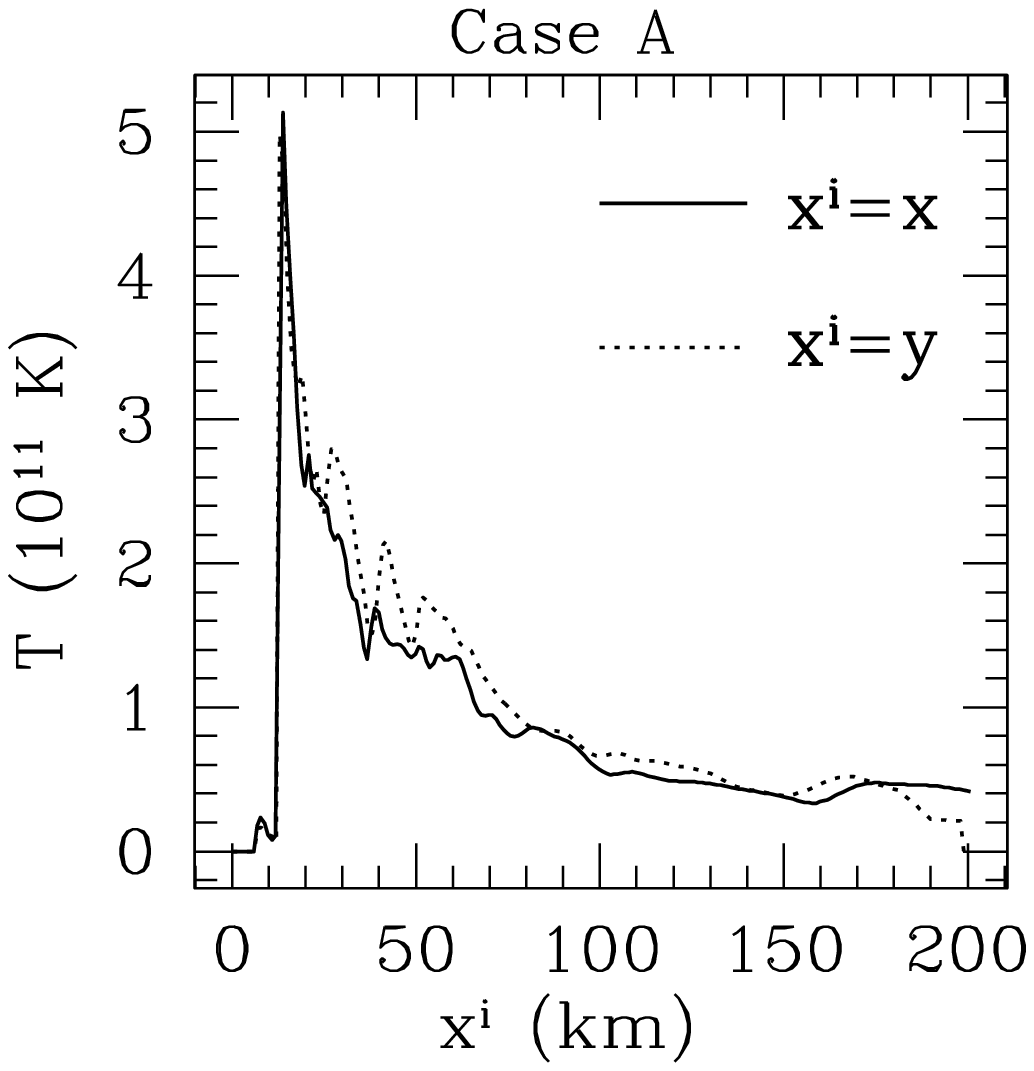}
\caption{Temperature ($\rm T$) profile for case A. The temperature 
is presented in units of $10^{11}\rm\ K$, where K indicates degrees Kelvin
(not to be confused with the EOS entropy parameter $K$).  The solid curve
corresponds to the values of $\rm T$ at the end of the simulations and
along the
$x$-axis for $y=y_{\rm c}$, $x > x_{\rm c}$, where $x_{\rm c}$ ($y_{\rm c}$) is the $x$ ($y$)
position of the center of mass of the remnant. The dotted curve corresponds to
the values of $\rm T$ along the $y$-axis for $y>y_{\rm c}$, $x = x_{\rm c}$.
It is clear that typical temperatures are of order 
$10^{11}\rm\ K$, while the TZlO core is practically at $0 \rm K$. For a realistic massive WDNS merger we expect ${\rm T} \sim 10^{9}$ K 
(see discussion following Eq.~\eqref{Temperature}).
\label{fig:temp}
}
\centering
\end{figure}

\subsection{Dynamics of the WDNS merger}

In \cite{WDNS_PAPERI} we analyzed the stability of corotating binary WDNSs
at the Roche limit, accounting for GR effects on the mass-radius relationship of the WD.
We concluded that if the mass ratio $q=M_{\rm WD}/M_{\rm NS}$ is larger than a critical 
mass ratio $q_{\rm crit}\approx 0.5$, then mass transfer from the WD to the NS will be unstable, 
and the WD will be tidally disrupted. The binary pWDNS system simulated in this work has a 
mass ratio $q=0.7$, so we expect that the system should experience tidal disruption and merge on an
orbital time scale soon after mass transfer has started.

In our simulations, the pWDNS binary completes almost
$2.5$ orbits before the pWD is completely
disrupted. Figure~\ref{fig:Axy} plots rest-mass density contours
in the orbital plane at selected times for case A. The top row,
middle panel shows the binary shortly after completing the first
orbit. At this time, an accretion stream from the pWD to the NS
develops, followed by the formation of an accretion disk around the
NS.  Matter from the accretion stream smashes into the accretion
disk, shock heating the gas at that location. This process continues until
the pWD is completely disrupted. After pWD tidal disruption, a long
tail forms that moves outwards and around the NS. The pWD matter that
orbits the NS at closer separations collides with the tail inducing further strong shocks. 

The bottom row, left panel of Fig.~\ref{fig:Axy}
shows the system after 3 orbits have been completed.  At this point,
the pWD is completely disrupted and a large, rotating, massive mantle
and disk around the NS has begun to form. A tidal tail is also visible. 
This snapshot is followed by a long epoch in which the rotating mantle 
settles onto an extended disk around the central object, composed of a
slowly spinning, cold NS core surrounded by a hot atmosphere and disk
composed of pWD debris. We find that even at this late stage, the NS
core maintains its original spin, and the hot mantle surrounding it
spins and settles into quasiequilibrium.  Non-axisymmetric clumps of
matter inspiraling near the cold NS core launch spiral density
waves into the disk. The remnant of the pWDNS merger may be best
characterized as a spinning TZlO with an extended ($R_{\rm disk}\sim
1000{\rm \ km}$), massive disk.  

We define the radius of the TZlO ($R_{\rm TZlO}$) as the distance between the center of mass
and the ``north'' pole of the remnant. The $z$-radius of the remnant for a cut-off density 
$\sim 10^{-8}\rho_{0,\rm max}$, where $\rho_{0,\rm max}$ is the maximum density of the remnant, 
is $R_{\rm TZlO}\approx300 {\rm \ km}$. We estimate the rest mass of the TZlO ($M_{\rm TZlO}$) as 
the rest mass contained within a sphere of coordinate radius equal to $R_{\rm TZlO}$, and the 
disk mass ($M_{\rm disk}$) by subtracting $M_{\rm TZlO}$ from the total rest mass. 
We find $M_{\rm TZlO}=1.82 \ M_\odot$ and $M_{\rm disk}=0.7 \ M_\odot$. 
The disk is massive and $\gtrsim 50\%$ of the original 
WD rest mass is eventually stored in the disk.

To first order, the TZlO is spherical.
This is evidenced by the XY, XZ, and YZ, density 
contour plots of Fig.~\ref{fig:AKxy}, which focus on the innermost
regions of the remnant at the end of the simulation.

The cutoff density in all the density contour plots we show here is
$10^{-8.5} {{\rm \ km}}^{-2}\approx 10^{-5.3}\rho_{0,\rm max}$, which is
approximately 4 decades above atmosphere density. The
equatorial and polar coordinate radii of these contours is $r_{e}\approx 350\rm
{\rm \ km}$ and $r_{p}\approx 150{\rm \ km}$, respectively. Therefore, the ratio
of these radii is $r_p/r_e\approx 3/7$. Given that the core of the remnant
is approximately spherical, the smallness of the ratio $r_p/r_e$
indicates that the disk stores a large amount of angular momentum.

In the bottom row of Fig.~\ref{fig:AKxy}, we plot contours of $K=P/P_{\rm
  cold}$. These entropy contours show that the neutron 
star core is cold $K\approx 1$ and that $K$ increases with the
distance from the core in the orbital plane and with increasing $z$
in meridional planes. This is reminiscent of the $K$ pattern 
observed in our binary pWDNS head-on collision studies in \cite{WDNS_PAPERII}, where $K\approx 1$
in the core, but increases with distance from the center. Note
the spiral density wave pattern visible in the
bottom row, left panel of Fig.~\ref{fig:AKxy}.

Unlike the head-on collision case, in which the outermost layers of
the NS are shock heated and stripped away when the NS smashes into the
denser parts of the pWD (Fig. 4 insets, \cite{WDNS_PAPERII}), after a
pWDNS binary inspiral and merger, the NS retains its outer layers, and its
structure remains nearly unaffected throughout the simulation
(Fig.~\ref{fig:Axy} insets).  Moreover, in the head-on collision, about
$18\%$ of the total initial rest mass is ejected to infinity, but in the
inspiral case, no ejection of matter to infinity is observed.


As in the case of head-on collisions, we find that the typical temperature
in our inspiraling binary remnant is of order $10^{11}\rm\ K$.
In Fig.~\ref{fig:temp} we show temperature profiles of the remnant. 
To estimate the temperature $\rm T$, we assume that  
the temperature dependence of $\epsilon_{\rm th}$ can be modeled as
\labeq{temp_dep}{
\epsilon_{\rm th} = \frac{3k\rm T}{2m_{\rm n}}+ f\frac{a\rm T^4}{\rho_0},
}
where $m_{\rm n}$ is the mass of a nucleon, $k$ is Boltzmann's constant and $a$ is the radiation constant.
The first term represents the approximate thermal energy of the nucleons, and the second 
term accounts for the thermal energy due to relativistic particles. 
The factor $f$ reflects the number of species of relativistic particles that 
contribute to the thermal energy. When ${\rm T} \ll 2m_{\rm e}/k \sim 10^{10}\rm K$,
where $m_{\rm e}$ is the electron mass, thermal radiation is dominated by photons and $f = 1$. 
When ${\rm T} \gg 2m_{\rm e}/k$, electrons and positrons become relativistic and 
also contribute to radiation, and $f = 1 + 2 \times (7/8) = 11/4$. 
At sufficiently high temperatures and densities
(${\rm T} \gtrsim 10^{11}\ {\rm K}, \rho_0 \gtrsim 10^{12} \rm  g\ cm^{-3}$), neutrinos are generated
copiously and become trapped. So, taking into account three flavors 
of neutrinos and antineutrinos, $f = 11/4+6\times(7/8) = 8$. 
In our temperature estimate, we find $f$ self-consistently in the following sense: 
we first calculate the temperature assuming $f=0$. If the calculated temperature and 
density are inconsistent with our choice of $f$ (which we test based on the above inequalities), 
then we choose a different $f$, until we find the value of $f$ which is consistent. We find that 
$f=11/4$ is consistent with the temperatures and densities in our pWDNS merger. However, 
we expect that in realistic mergers $f=1$, as the expected temperatures are of order $10^9 K$
(see discussion following Eq.~\eqref{Temperature}).

To solve Eq.~\eqref{temp_dep} for $\rm T$ we need to know $\epsilon_{\rm th}$. We calculate 
$\epsilon_{\rm th}$ via
\labeq{}{
\epsilon_{\rm th} = \frac{(K -1 )P_{\rm cold}}{(\Gamma_{\rm th}-1)\rho_0},
}
where Eqs. \eqref{Ptot}, \eqref{Pthermal} and the definition of $K$
were used to obtain this last equation. 


As in our pWDNS
head-on collision studies, we find that the pWDNS binary remnant does
not collapse promptly to a BH.  In contrast to the head-on collision
cases, in which only thermal pressure supported the remnant from
prompt collapse, collapse {\it may} be delayed in the pWDNS binary inspiral 
case by both thermal pressure and centrifugal support. One way to
assess the importance of thermal support in the binary inspiral 
remnant is to apply the same cooling recipe as in the case of TZlOs formed in head-on collisions
(see Section~\ref{cool_TZlOs}), and compare the result to an uncooled
case. 

\begin{figure*}
\centering
%
\subfigure{\includegraphics[width=0.45\textwidth]{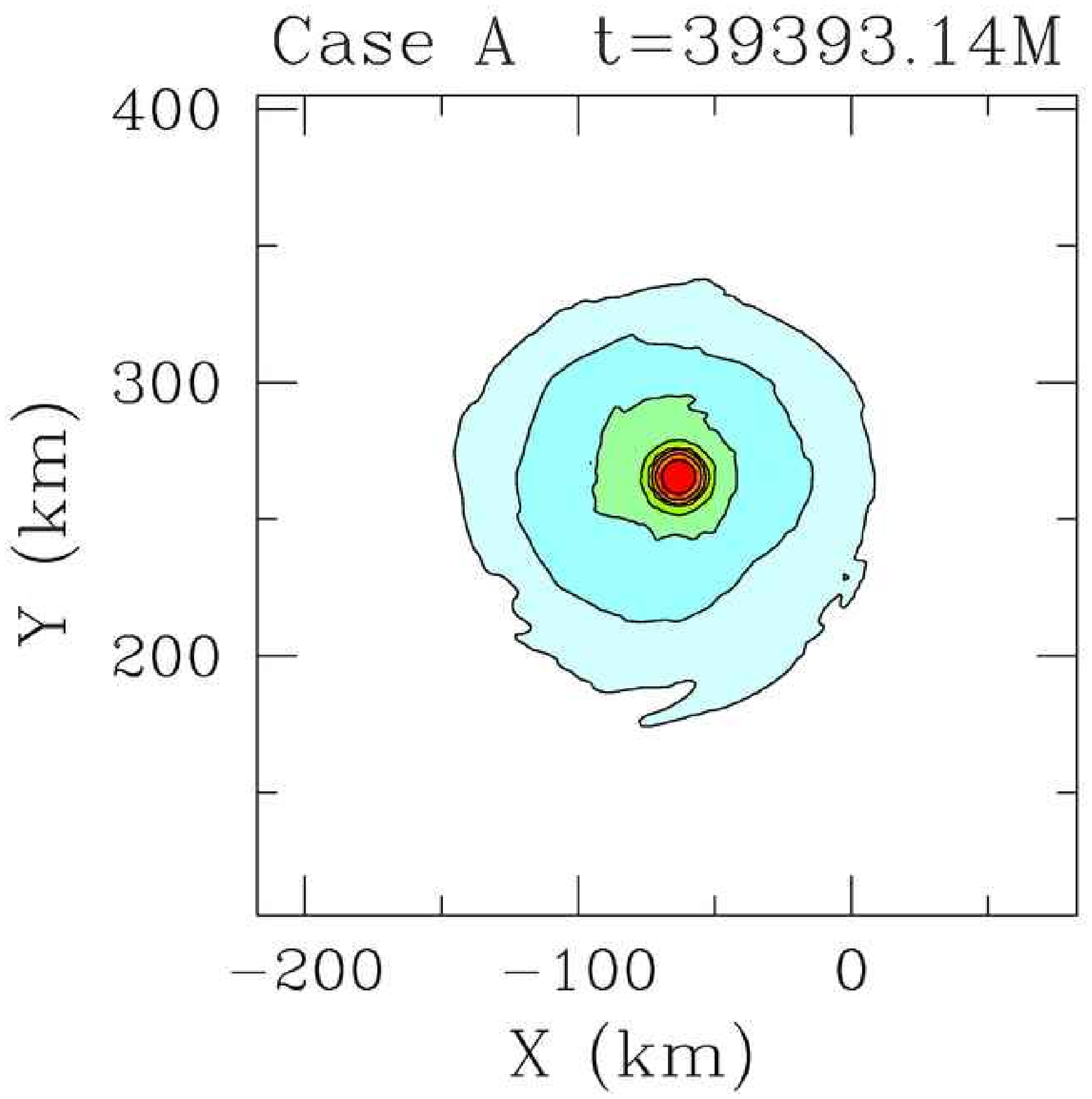}}
\subfigure{\includegraphics[width=0.45\textwidth]{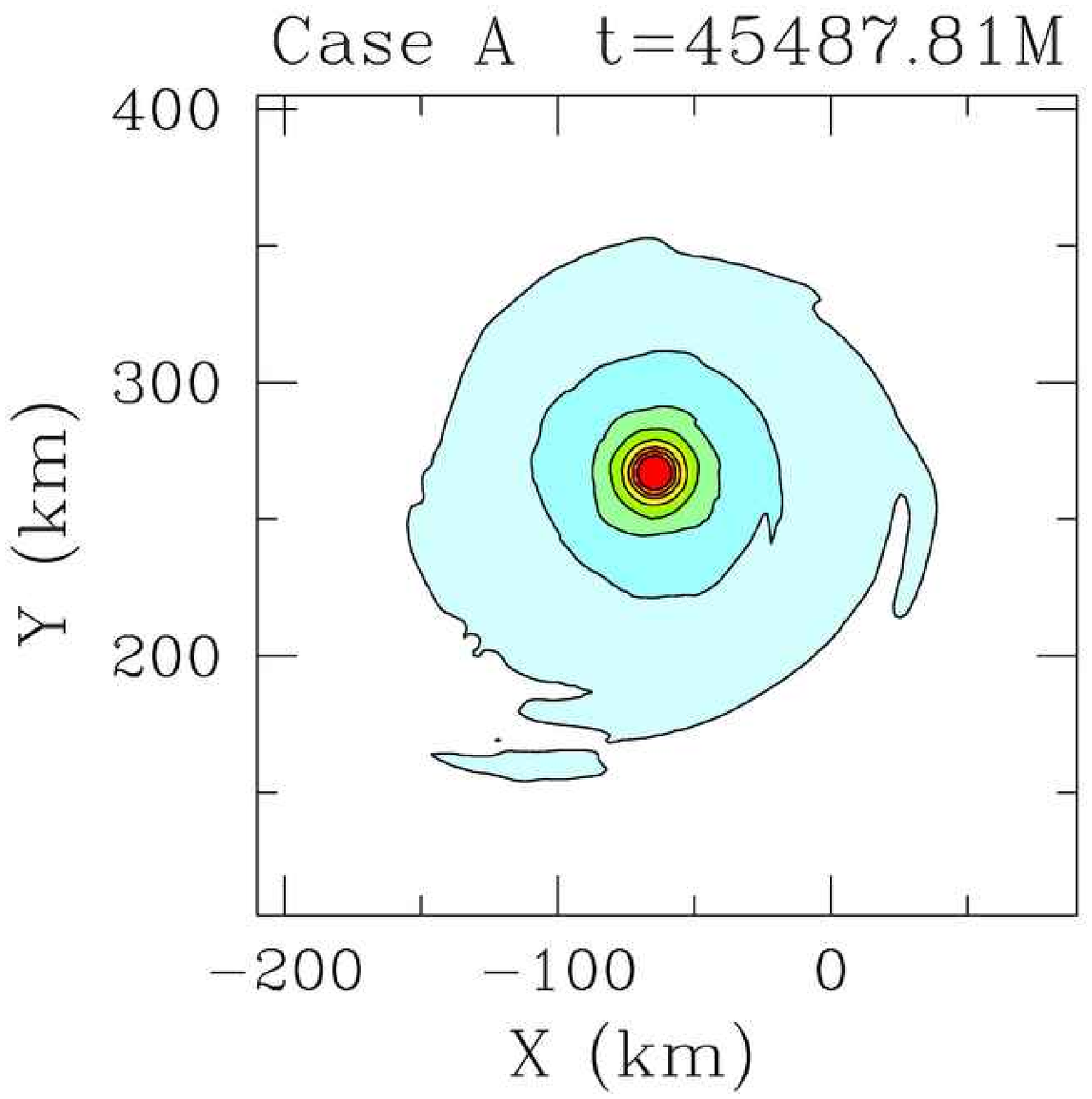}}
\subfigure{\includegraphics[width=0.45\textwidth]{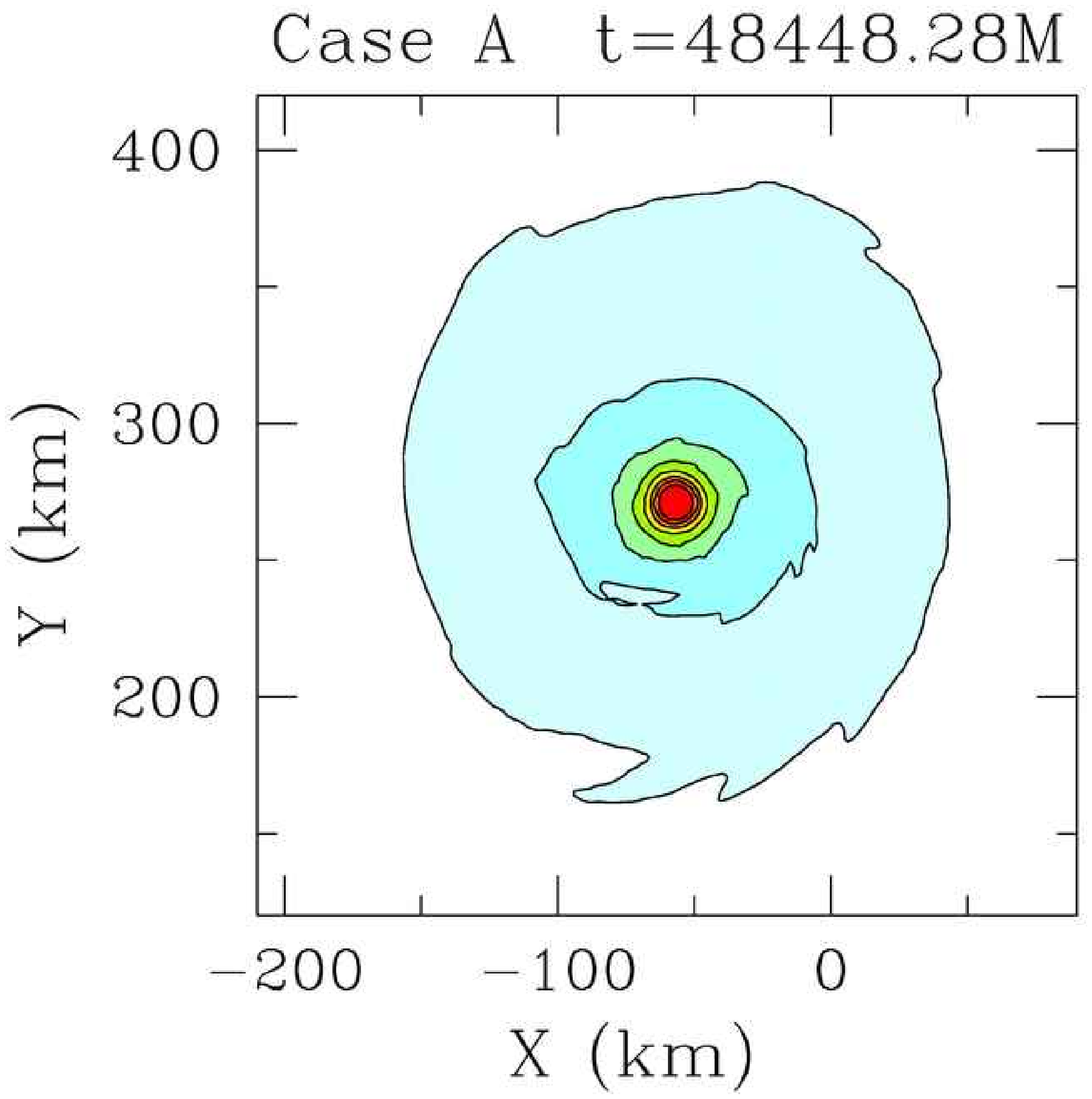}}
\subfigure{\includegraphics[width=0.45\textwidth]{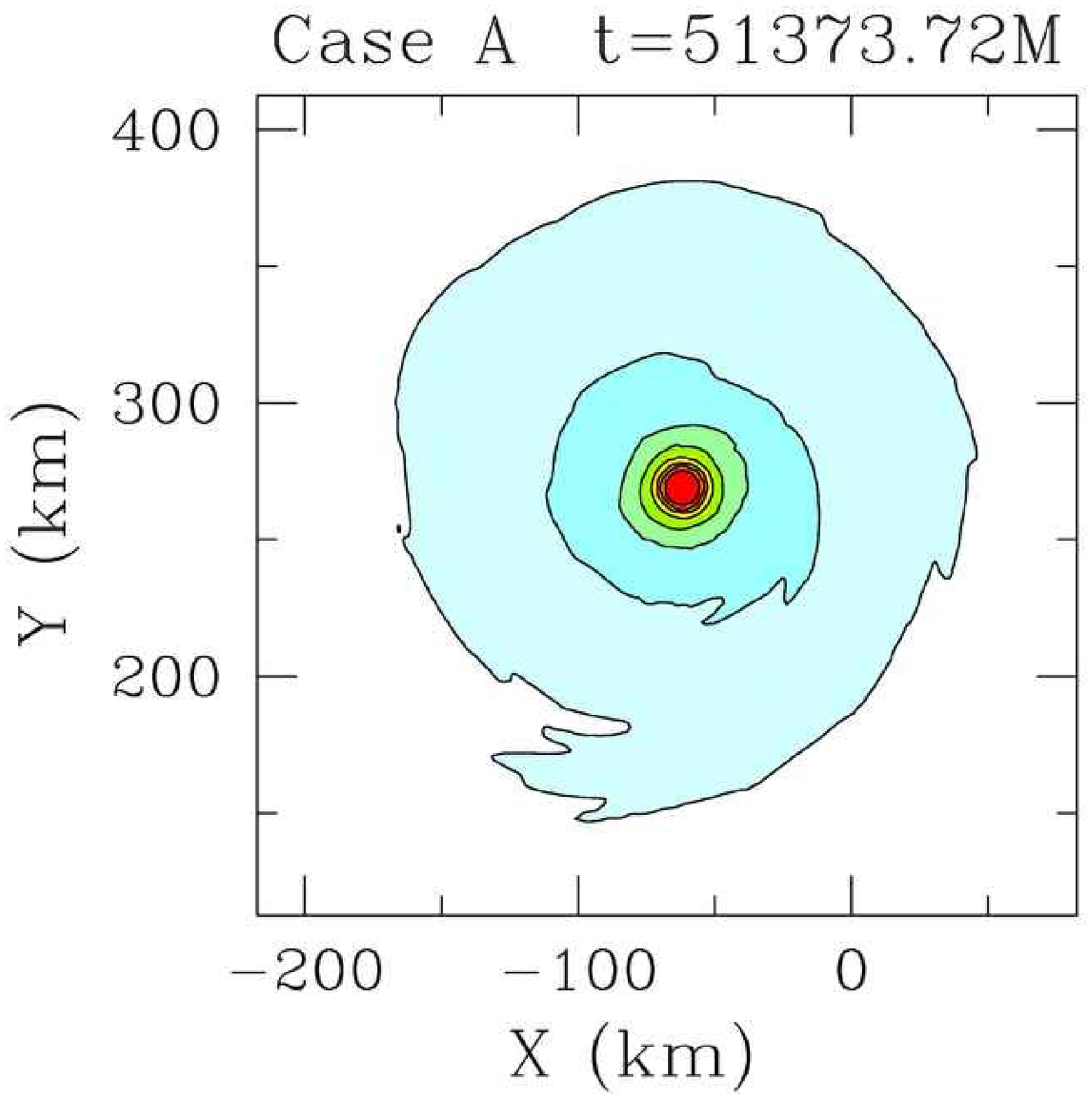}}
\caption{
Snapshots of rest-mass density profiles at selected times for case A with cooling turned on. 
The contours represent the rest-mass density in the orbital plane, plotted 
according to $\rho_0 = \rho_{0,\rm max} 10^{-0.5j-0.16}\ (j=0,1,\ldots, 8)$, where
$\rho_{0,\rm max} = 4.645\rho_{\rm nuc}$, and $\rho_{\rm nuc}=2\times 10^{14}{\rm \ g/cm}^3$.
The last two snapshots show that the contraction practically stops after about 6 cooling time scales. 
The color code used is the same as that defined in Fig.~\ref{fig:A1xy} and
$M = 2.38\ M_\odot = 3.52{\rm \ km} = 1.17\times 10^{-5}\ s $.
\label{fig:coolAxy}
}
\centering
\end{figure*}

\subsection{Cooling of spinning TZ$\rm l$Os formed in WDNS mergers}
\label{sec:mergerTZlOcooling}

To determine whether the spinning TZlO from the WDNS inspiral and merger will collapse to 
a BH following cooling, we apply our cooling technique
setting the cooling time scale to the same value 
used earlier ($\tau_c = 6000{\rm \ km}$) in Sec.~\ref{cool_TZlOs}. 
We turn on cooling after about $4.5$ orbital time scales and follow the subsequent evolution for about
7 cooling time scales.

\begin{figure*}
\centering
\subfigure{\includegraphics[width=0.45\textwidth]{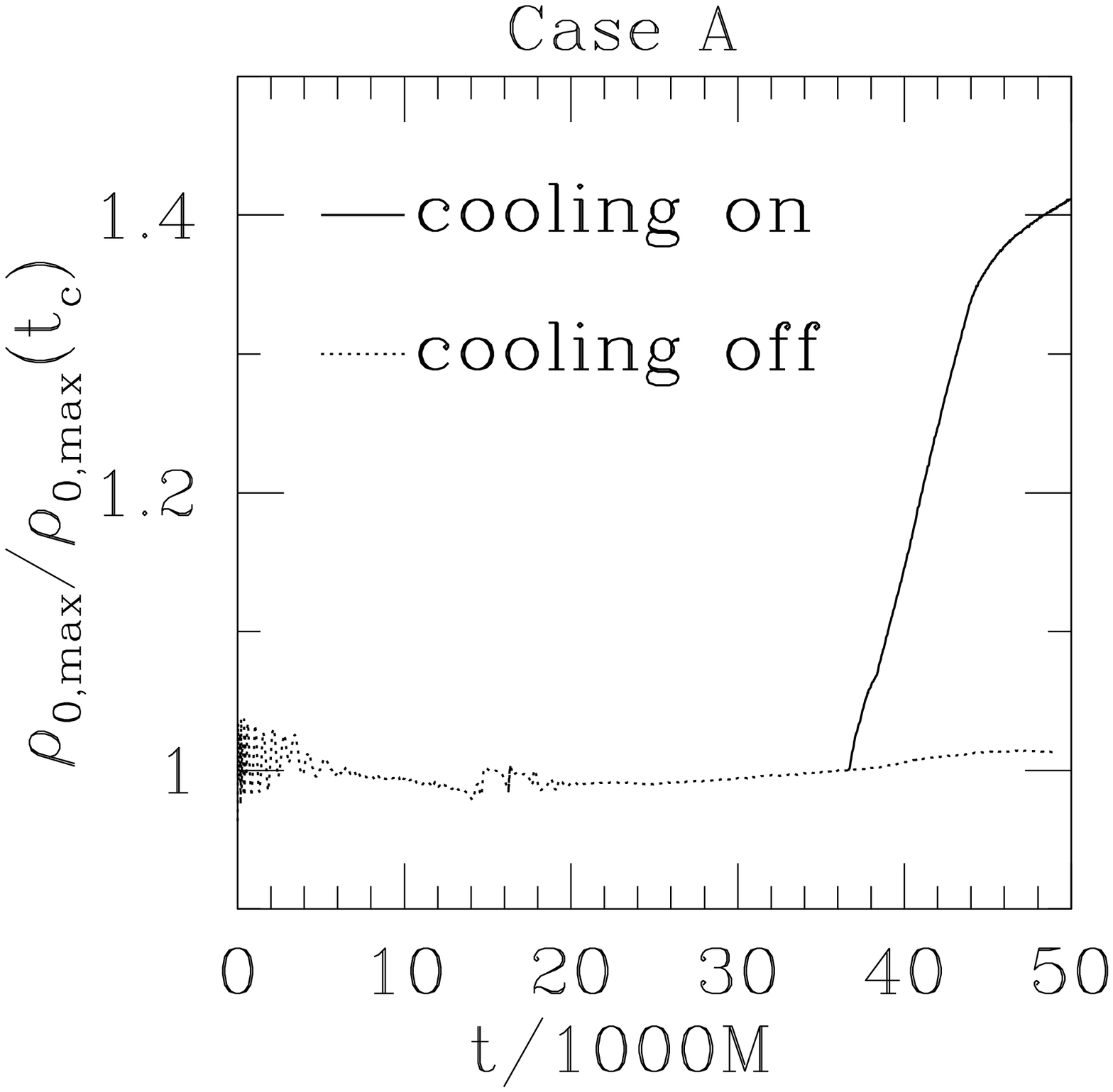}}
\subfigure{\includegraphics[width=0.45\textwidth]{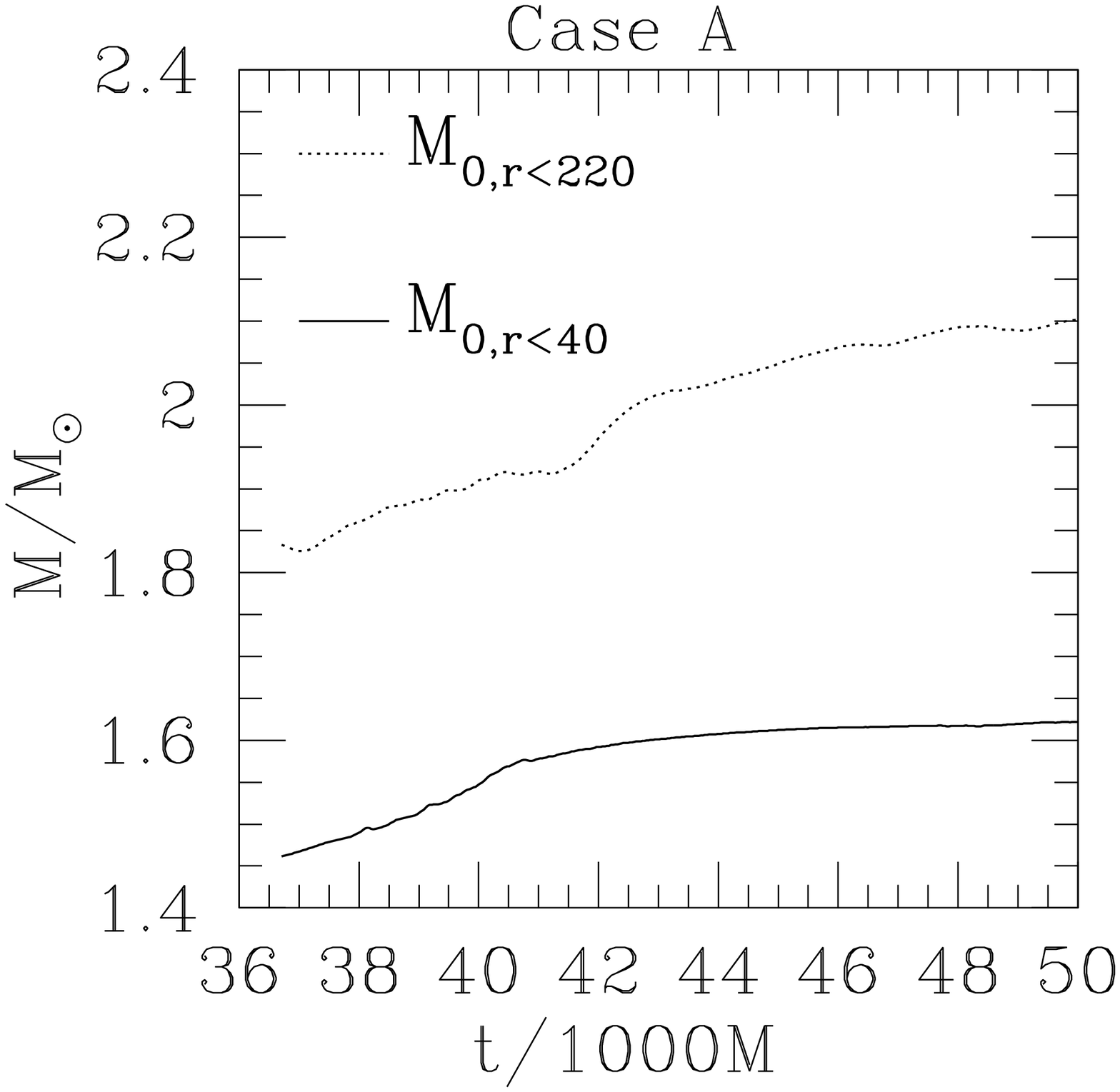}}
\caption{
Left: Evolution of maximum value of rest-mass density with cooling (solid curve) 
and without cooling (dotted curve) for case A. Here $\rho_{0,\rm max}$ is the maximum value of 
the rest-mass density, $\rho_{0,\rm max}({\rm t}_{c})=5.88\times 10^{-4} {\rm \ km}^{-2}=7.92\times 10^{14} {\rm \ g/cm}^3$ 
is the maximum value of rest-mass density at the time ($t_{c}=36705M$) when cooling 
is turned on. 
Right: Evolution of rest-mass within different radii with cooling turned-on.
Here $M_{0,\rm r < r_0}$  stands for the rest mass contained within a coordinate sphere of
radius $r_0$ in units of {\rm \ km}, centered on the remnant's center of mass. These plots demonstrate 
that the TZlO remnant contracts when cooling is turned on. However, the values of $\rho_{0,{\rm max}}$
and  $M_{0,\rm r < r_0}$ begin to plateau after 6 cooling time scales, indicating 
no further contraction proceeds after this time. Here $M = 2.38\ M_\odot = 3.52{\rm \ km} = 1.17\times 10^{-5}\ s $ 
is the sum of the ADM masses of the isolated stars. 
\label{fig:Amass_density}
} 
\centering
\end{figure*}

Figure~\ref{fig:coolAxy} plots density contours in the orbital plane at selected times, and 
Fig.~\ref{fig:Amass_density} shows the evolution of the maximum rest-mass density and the rest 
mass contained within spheres with different coordinate radii. 
These figures
demonstrate that both the maximum density, and the rest mass within
$40${\rm \ km} and $220${\rm \ km} increase as functions of time. 
As the hot outermost layers become cooler they contract and accumulate onto the innermost colder parts. 
For this reason the innermost density contours of like density increase in size (see Fig.~\ref{fig:coolAxy}).
The remnant is contracting with time. The contraction is more rapid in the beginning and 
begins to plateau after about 6 cooling time scales. Therefore, the
spinning TZlO does {\it not} collapse to a BH when thermal support is
removed.

\begin{figure*}
\centering
\subfigure{\includegraphics[width=0.45\textwidth]{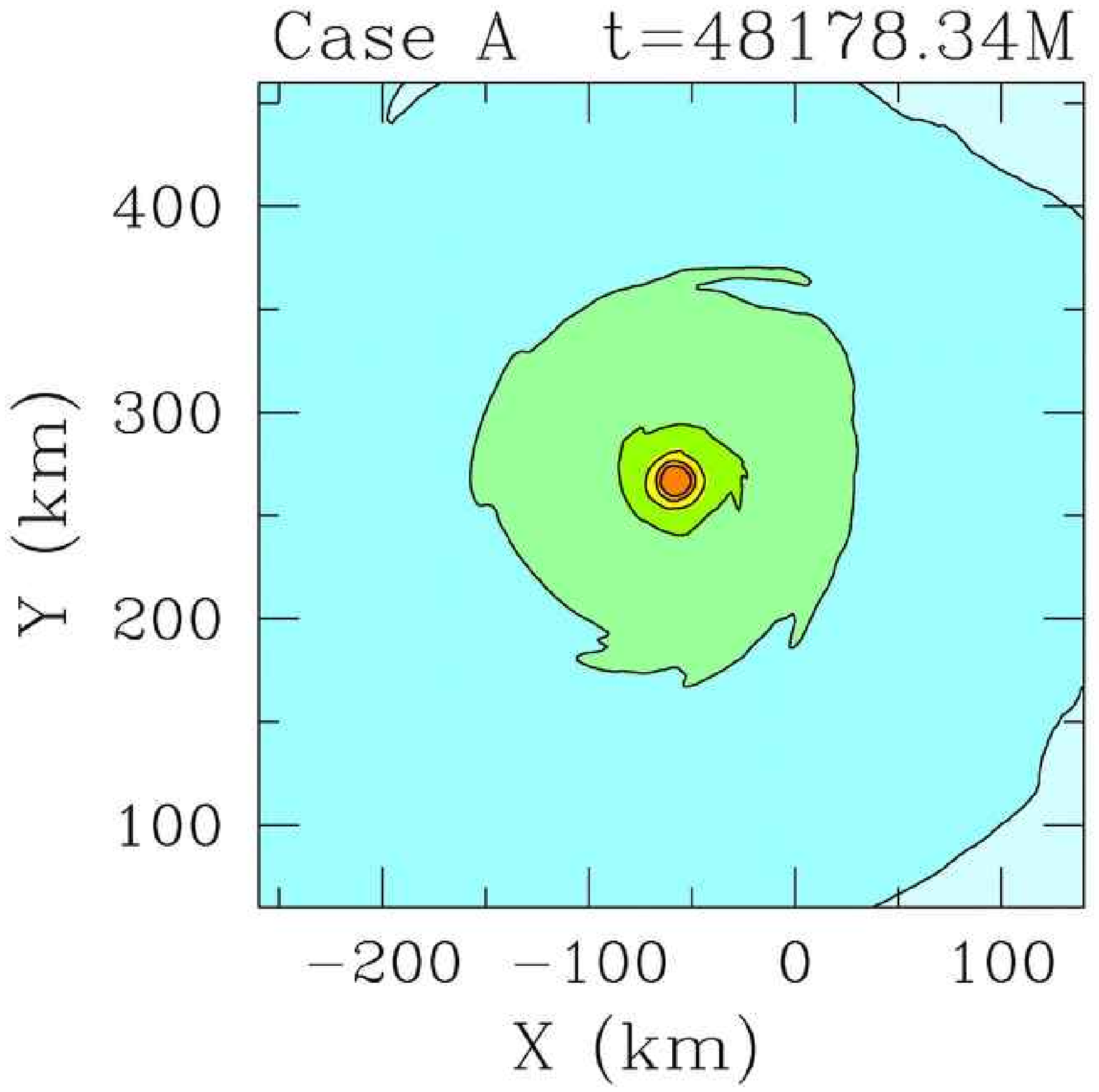}}
\subfigure{\includegraphics[width=0.45\textwidth]{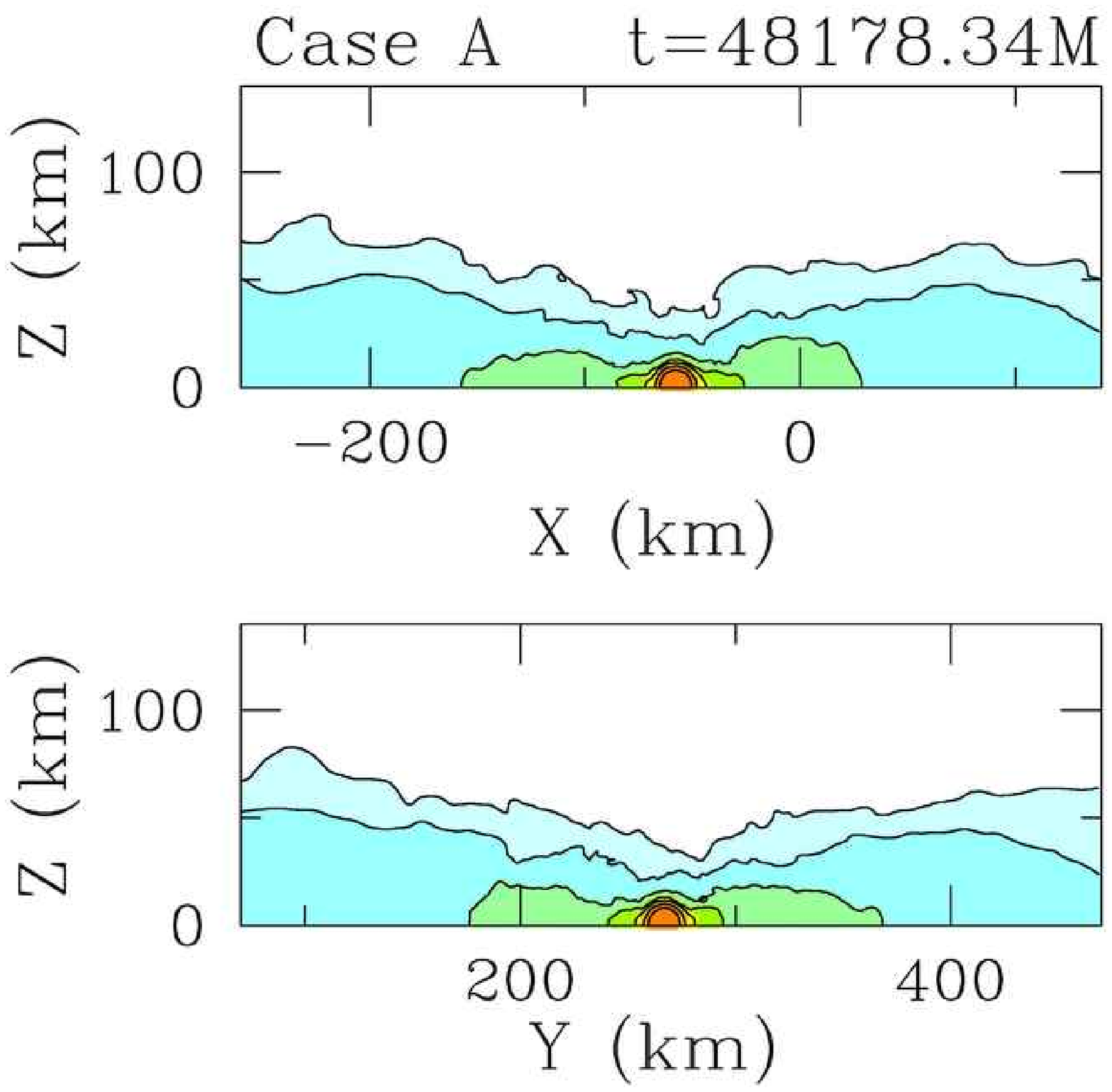}}
\subfigure{\includegraphics[width=0.45\textwidth]{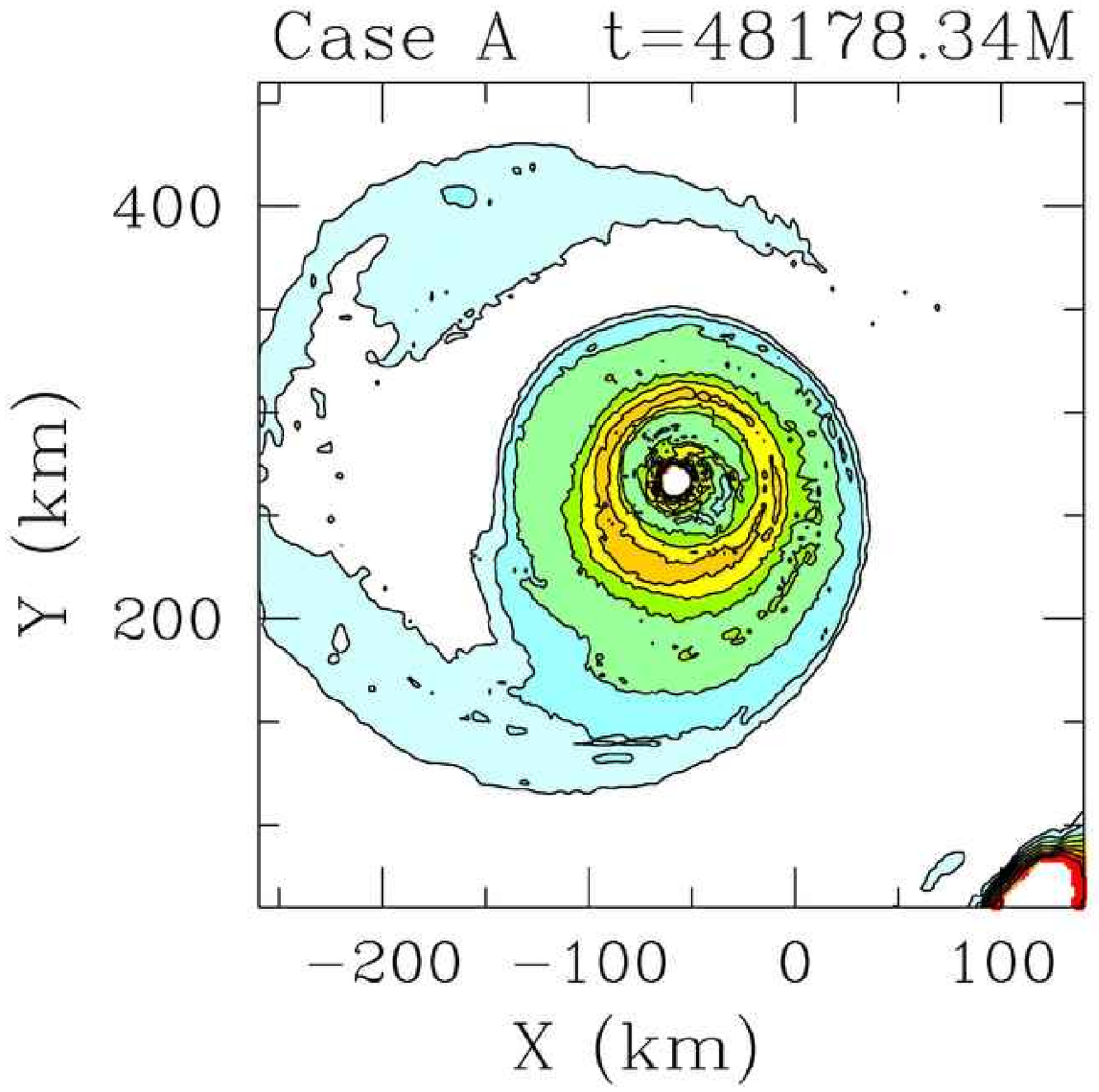}}
\subfigure{\includegraphics[width=0.45\textwidth]{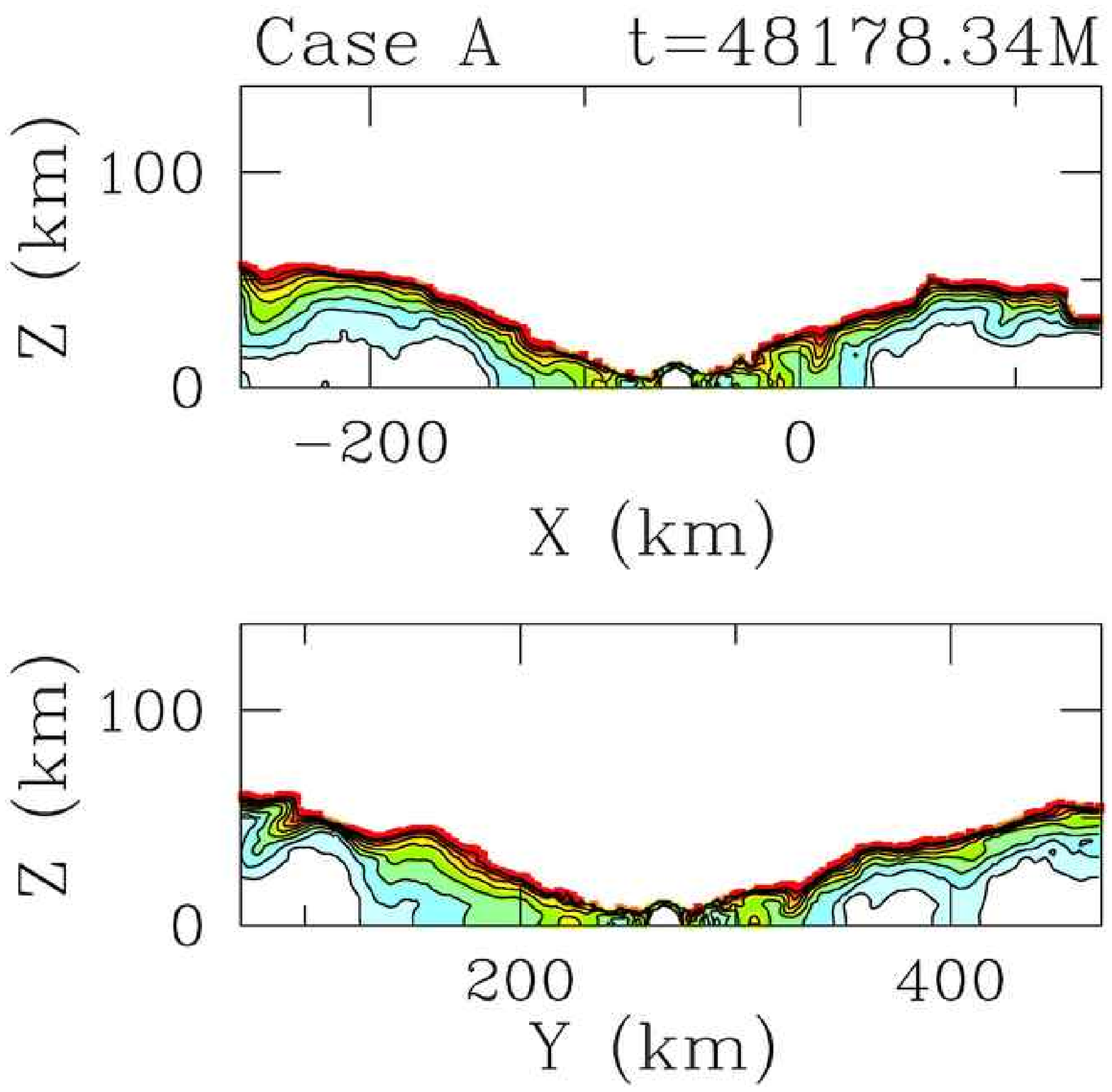}}
\caption{
First row: Snapshots of rest-mass density profiles at selected times for case A with cooling. 
The contours represent the rest-mass density in the orbital plane and the XZ and YZ meridional planes, 
plotted according to $\rho_0 = \rho_{0,\rm max} 10^{-0.69j-0.16}\ (j=0,1,\ldots, 7)$, where
$\rho_{0,\rm max} = 4.645\rho_{\rm nuc}$, and $\rho_{\rm nuc}=2\times 10^{14}{\rm \ g/cm}^3$.
Second row: Snapshots of $K=P/P_{\rm cold}$ profiles at selected times for case A with cooling. 
The contours represent $K$ in the orbital plane and the XZ and YZ meridional planes, 
plotted according to $K = 10^{-0.014j+0.1}\ (j=0,1,\ldots, 7)$. The plots show that 
the remnant NS core is approximately spherical and cold ($K\approx 1$). Far from the core 
the remnant is hotter. $K$ increases as we move away from the core and the orbital plane. In contrast
to the case without cooling (see Fig.~\ref{fig:AKxy}), the maximum value for $K$ here is 
$K_{\rm max} \approx 1.25$. For easy comparison with the case without cooling (Fig.~\ref{fig:AKxy}), 
all plots focus in the innermost $200{\rm \ km}$ from the remnant center of mass.
The color code used is the same as that defined in Fig.~\ref{fig:A1xy}, with white color
in the second row indicating $K\approx 1$. Here $M = 2.38\ M_\odot = 3.52{\rm \ km} = 1.17\times 10^{-5}\ s $.
\label{fig:AKxy_cool}
}
\centering
\end{figure*}

Figure~\ref{fig:AKxy_cool} shows density and $K$ contours for the
innermost regions of the remnant in the XY, XZ, and YZ planes, 6
cooling time scales after cooling was turned on. 
The shape of the NS core remains spherical throughout the evolution. The XZ and YZ 
contours demonstrate that the disk and mantle have become thinner, as expected when cooling 
takes place. Due to this effect, this final configuration is massive accretion disk onto a NS, rather 
than a disk around a TZlO. 

The bottom row of Fig.~\ref{fig:AKxy_cool} shows
contours of $K=P/P_{\rm cold}$. Notice that the neutron 
star core remains cold $K\approx 1$ at the end of the simulation and
that elsewhere $K$ has decreased considerably compared to the run without
cooling. Here $K_{\rm max} \approx 1.25$, while in the run without
cooling $K_{\rm max} \approx 10$. In the innermost region, cold
pressure dominates, with $K\approx1.05$. Given that the rest mass
within $220{\rm \ km}$ of the remnant center of mass is
greater than $2.05\ M_\odot$, which exceeds the maximum
supportable mass by our cold EOS, 
we conclude that the spinning TZlO is {\it centrifugally supported} from
collapse to a BH.

Based on these results and the scalability of our simulations to the realistic scenario,
we are led to the tentative prediction that realistic WDNS mergers with total rest mass
$\lesssim 2.5 M_\odot$, the rest mass in our simulations, 
 will not collapse to a BH following cooling. This conclusion assumes that
angular momentum redistribution takes place on a longer time scale than
cooling.

Given the absence of outflows in our simulations (with the caveat that
we do not model nuclear reactions), the final total
rest mass ($\approx 2.5\ M_\odot$) is larger than the maximum rest mass
supportable by our cold EOS, even allowing for maximal uniform rotation.
Therefore, we expect that after viscosity and/or
magnetic fields redistribute angular momentum, the remnant will
collapse to a black hole. 
This conclusion will be true in the case of realistic WDNS mergers, unless the true nuclear EOS supports 
a uniformly rotating star with a rest mass exceeding the remnant mass. Many viable EOSs do not support 
a uniformly rotating cold configuration with rest mass
as large as $2.5M_\odot$ \cite{2004ApJ...610..941M}, the remnant rest mass in our simulations.

\section{Discussion}
\label{sec:discussion}

To identify the relevant nuclear reaction networks and the dominant cooling mechanisms 
in realistic inspiraling WDNS binaries, we need to estimate the temperatures of realistic TZlOs. 
Moreover, to determine the time scale on which angular momentum redistribution occurs we have 
to consider viscosity and/or magnetic fields. In this section we discuss these issues.

\subsection{Temperature}
\label{sec:temperature}

The characteristic temperature of realistic TZlOs is
expected to be of order $10^9$ K. This is because the energy available for shock heating is of
order the gravitational interaction energy when the two stars first
touch, $M_{\rm NS}M_{\rm WD}/R_{\rm WD}$. Our simulations demonstrate 
that the NS is largely unaffected by shock heating and 
remains cold.  Hence, most of the thermal energy is stored in the
WD debris. The total thermal energy, $E_{\rm th}$, is then 
\labeq{}{
E_{\rm th} \sim \frac{M_{\rm WD}}{m_{\rm n}} k {\rm T} \sim \frac{M_{\rm NS}M_{\rm WD}}{R_{\rm WD}}.
}
From this last equation we can estimate the characteristic temperature as
\labeq{Temperature}{
{\rm T} \sim \frac{C_{\rm WD}m_{\rm n}}{qk},
}
where $C_{\rm WD} = M_{\rm WD}/R_{\rm WD}$ is the WD compaction.
All things being equal (i.e., no mass loss, same mass ratio, etc.),
characteristic TZlO temperatures should be proportional to the WD compaction. 
The compaction of a realistic $1.0\ M_\odot$ WD that obeys the Chandrasekhar EOS is 
$C_{\rm WD}\simeq 10^{-4}$ \cite{Shapiro,WDNS_PAPERI}. If the NS mass is $1.4\ M_\odot$ then
$ q \simeq 0.7$, and Eq.~\eqref{Temperature} predicts
 ${\rm T}\simeq 1.55\times 10^{9}\ {}\rm K$. 

Note that applying Eq.~\eqref{Temperature} to case A, where $C_{\rm pWD}\simeq 10^{-2}$, yields a temperature
 ${\rm T}\simeq 1.55\times 10^{11}\ {}\rm K$, i.e., in good agreement with 
our simulations~\footnote{In \cite{WDNS_PAPERII} we estimated  
the thermal energy of the TZlO as $E_{\rm th} \sim \frac{(M_{\rm NS}+M_{\rm WD})}{m_{\rm n}} k {\rm T}$. 
This estimate does not account for the fact that most of the thermal
energy is stored in the WD debris, so the predicted temperature is
slightly smaller than that found in our simulations. Our current estimate
(Eq.~\eqref{Temperature}) yields results in good agreement not
only with our present simulations, but also with our simulations in
\cite{WDNS_PAPERII}}.

\subsection{Nuclear fusion}

Are realistic WDNS binary remnant densities and temperatures high
enough for nuclear reactions to take place?  The
shock-heated matter is composed of hot (diluted) WD debris, so its
density is of order typical WD densities, i.e., $10^6 {\rm \ g/cm}^3$.  A
$1.0\ M_\odot$ WD is sufficiently massive that its main constituent
elements are carbon and oxygen. While the temperatures and densities
we expect for realistic mergers are probably not high enough for
oxygen burning to become important, they are sufficiently high for
carbon fusion to become dynamically relevant.

Non-explosive nuclear reactions in the context of WDNS mergers 
were recently considered in \cite{Metzger11}.
A 1D steady state model of accretion onto a NS was introduced,
allowing for disk wind outflows that do not exert any torque on the disk. 
It was found that heating from nuclear burning is so important that a
disk wind eventually unbinds $50\%-80\%$ of the original WD mass. It
was suggested that these ejecta may include small quantities of
radioactive ${}^{56}Ni$. In such scenarios, detectable EM signals will
likely follow a WDNS merger. Although this 1D steady-state model
includes much of the important physics (albeit in parametrized form),
it is simplified and does not apply to the large mass-ratio 
mergers simulated here. However, as in the 1D model we do
expect that nuclear burning will also be non-explosive in a realistic WDNS merger, as we now
explain.

In a {\it head-on} collision of a WDNS binary with
companions of comparable mass that collide
at free-fall velocity, the kinetic energy
of motion is converted by shocks into thermal energy
in the WD remnant. This shock heating at merger guarantees 
that a degenerate WD initially in hydrostatic equilibrium will acquire 
shock-induced thermal pressure comparable in magnitude to
its original equilibrium degeneracy pressure, thereby lifting the
degeneracy, i.e.,
\labeq{}{
\frac{P_{\rm th}}{\rho_0} \sim \frac{k{\rm T}}{m_n} \sim v_{\rm ff}^2 \sim \frac{G M}{R_{\rm WD}} \sim \frac{P_{\rm eq,WD}}{\rho_0} \sim 
\frac{P_{\rm cold}}{\rho_0}.
}
The net effect should be to reduce the likelihood of explosive
carbon burning, since a carbon flash requires a degenerate environment. 
The reason for this is that if gas pressure becomes a significant
component of the total pressure, then the pressure will be sensitive to the temperature.
Therefore, if carbon fusion takes place, the released heat will 
increase the gas temperature which will, in turn, increase the pressure. 
As a result the gas will expand, decreasing its density and temperature, and
eventually carbon fusion will be turned off. Such a process 
is self-regulated, a well-known fact.

Shock heating plays a similar role in the merger of an
inspiraling binary, only it is not as strong and
the fraction of the thermal pressure generated will be smaller,
due to the role of angular momentum in lessening the impact
and contributing to the support of the remnant. 
Using our estimated temperature ${\rm T}\approx 10^9\ {\rm K}$ and
characteristic density $10^6  {\rm \ g/cm}^3$ for realistic TZlOs, the
ratio of thermal gas pressure to the electron degenerate pressure is 4. This
implies that the WD debris would be non-degenerate.
Under these conditions a carbon flash is likely suppressed, but further
simulations would be useful to confirm this.

\subsection{Neutrino cooling}



For the estimated characteristic temperature ${\rm T}=10^9\rm\ K$ and
density $\rho_0 = 10^{6} {\rm \ g/cm}^3$, the
dominant cooling mechanism likely will involve neutrino emission. At
these densities and temperatures, thermal neutrino processes (pair
neutrinos, photoneutrinos, plasmon decay, and bremsstrahlung
\cite{Shapiro}) are important, with pair annihilation
($e^-+e^+\rightarrow \nu+\bar\nu$) being slightly more important than
the other processes (see Fig. 1 in \cite{BARKAT75} and Fig. 3(a) in
\cite{BPSCooling}). The pair neutrino cooling rate can be estimated as
\cite{Kippenhahn}
\labeq{epsnunu}{
\epsilon_{\nu}^{{\rm pair}}\approx 4.45\times 10^{9}\frac{{\rm T}_{9}{}^{9}}{\rho_6}\ \rm [erg/g/s],
}
where ${\rm T}_9={\rm T}/10^9\rm\ K$, $\rho_6=\rho_0/10^6\rm g\ cm^{-3}$ and the high-temperature and
non-degenerate limit has been assumed.

The specific thermal energy is approximately given by $\epsilon_{\rm th} = 3k{\rm T}/m_n$.
Based on this, the cooling time scale can be estimated as
\labeq{}{
  \tau_{\rm cooling} = \frac{\epsilon_{\rm th}}{\epsilon_\nu^{\rm pair}} \approx 1.76{\rm \ yr}\ \frac{\rho_6}{{\rm T}_9{}^{8}}.
}
Notice that for ${\rm T_9=0.1}$, $\tau_{\rm cooling}\approx 10^8 {\rm \ yr}$.
Thus the object cools fast when it is very hot, but when the temperature drops to $10^8\rm K$ it takes 
hundreds of millions of years for cooling to take place.

Based on these considerations and Eq.~\eqref{epsnunu}, the net
conclusion is that the neutrino cooling time scale is highly
temperature sensitive, and our pWDNS inspiral simulation may only
provide a crude estimate of temperature.  Therefore, simulations with
more physics are necessary to precisely calculate realistic TZlO
temperatures, so that the relevant cooling time scales may be better
estimated.

Adopting the cooling rate \eqref{epsnunu} we can estimate whether neutrinos from WDNS mergers are detectable.
The number of detectable neutrinos ($N_d$) are approximately given by
\labeq{}{
N_d \approx \frac{L_{\nu}\sigma_\nu \Delta T}{4\pi D^2 \bar \epsilon_\nu},
}
where $L_\nu$ is the total neutrino luminosity, $\sigma_\nu$ the neutrino detection cross section, 
$\Delta T$ the time interval over which neutrinos are emitted, $D$ the distance to the binary,
and $\bar \epsilon_\nu$ the average neutrino energy. Given that $\sigma_\nu \approx 10^{-44}\rm cm^2$, 
and the expected neutrino energy from pair annihilation is $\epsilon_\nu \approx 0.5 \rm MeV$, we estimate
\begin{widetext}
\labeq{}{
N_d \approx 10^{-32}\frac{{\rm T}_{9}{}^{9}}{\rho_6}
            \bigg(\frac{M}{M_{\odot}}\bigg)
            \bigg(\frac{\sigma_{\nu}}{10^{-44}\rm cm^2}\bigg)
            \bigg(\frac{\Delta T}{1 \rm yr}\bigg)
            \bigg(\frac{\bar\epsilon_\nu}{0.5 \rm MeV}\bigg)^{-1}
            \bigg(\frac{D}{1\rm kpc}\bigg)^{-2},
}
\end{widetext}
where we have assumed that the entire TZlO mantle emits neutrinos at the same rate for a year.

Given this result, we conclude that neutrinos emitted in WDNS mergers are unlikely  
to be detectable. However, simulations with detailed microphysics \cite{1996ApJS..102..411I}
would be useful to confirm this.

\subsection{Angular momentum redistribution}

Our inspiraling WDNS merger simulation with cooling turned on shows that the remnant 
does not collapse to a BH following cooling, because it is centrifugally supported. 
Given that the mass of the remnant is larger than the maximum mass supportable by our cold
EOS, it is likely that delayed collapse will take place after angular momentum is redistributed.

Angular momentum redistribution will occur on the viscous or Alfv\'en time scale. 
Assuming an $\alpha$-disk, the viscous time scale (neglecting the disk self-gravity)
is given by
\labeq{tvis}{
t_{\rm vis} \simeq \alpha^{-1}\bigg(\frac{H}{R}\bigg)^{-2} \sqrt{\frac{R^3}{M_{\rm TZlO}}},
}
where $H$ is the disk scale height, $R$ the characteristic disk radius, and $\alpha$ the turbulent
viscosity parameter. Using the values for $H/R$, and $M_{\rm TZlO}$ found in our simulations we estimate that 
in realistic WDNS mergers the viscous time scale is
\begin{widetext}
\labeq{tvis2}{
t_{\rm vis} \simeq 20 s \bigg(\frac{\alpha}{0.1}\bigg)^{-1}\bigg(\frac{H/R}{1.0}\bigg)^{-2} \bigg(\frac{R}{10^4 {\rm \ km}}\bigg)^{3/2}\bigg(\frac{M_{\rm TZlO}}{1.8 \ M_\odot}\bigg)^{1/2},
}
\end{widetext}
where $R \approx 2R_{\rm WD}$, i.e., near the Roche limit for a $1.0M_\odot$ WD with a $1.4M_\odot$ NS.

The Alfv\'en time scale ($t_A=R/v_A$, where $v_A$ is the Alfv\'en speed) is given by
\labeq{tA2}{
t_A \simeq \beta^{-1/2}\bigg(\frac{H}{R}\bigg)^{-1} \sqrt{\frac{R^3}{M_{\rm TZlO}}},
}
where the $\beta$ parameter
\labeq{beta}{
\beta \equiv \frac{B^2}{8\pi P}
}
\\
was introduced to obtain the last expression. If we use the same values for $R$, $H/R$ and
$M_{\rm TZlO}$ as in Eq.~\eqref{tvis2}, the Alfv\'en time scale becomes
\begin{widetext}
\labeq{tA3}{
t_A \simeq 6.5 s \bigg(\frac{\beta}{0.1}\bigg)^{-1/2}\bigg(\frac{H/R}{1.0}\bigg)^{-1}\bigg(\frac{R}{10^4{\rm \ km}}\bigg)^{3/2}\bigg(\frac{M_{\rm TZlO}}{1.8\ M_\odot}\bigg)^{-1/2}.
}
\end{widetext}
We emphasize that the dimensionless parameters $\alpha$ and $\beta$ above are unknown and may both 
be $\ll 1$, in which case the angular momentum redistribution
time scale may be as long as the cooling time scale.
For example, observations of magnetic WDs indicate that surface magnetic field strengths are
$B \sim 10^4-10^9\ G$ \cite{MWDsReview2000} or
$\beta \sim 10^{-17}-10^{-7} \ll 1$, where we calculated the thermal pressure as 
$P = \rho_0 k {\rm T}/m_n$ with $\rho_0 = 10^6 {\rm \ g/cm}^3$, ${\rm T}= 10^9\ \rm K$.
If $\beta \lesssim 10^{-15}$, then the Alfv\'en time scale is longer than $1{\rm \ yr}$,
i.e., the cooling time scale for ${\rm T}_9=1$. However, field amplification via winding and instabilities 
(e.g. magnetorotational instability) is always possible. Hence, we must await detailed 
calculations for reliable estimates of the angular momentum redistribution time scale.

%
%

\section{Summary and Conclusions}
\label{sec:summary}

This work is a follow-up to our study of binary WDNS head-on collisions \cite{WDNS_PAPERII},
focusing on the dynamics of an initially circular, quasiequilibrium WDNS binary through
inspiral and merger.
In particular, we begin with a circular binary in which the WD
has just filled its Roche lobe (the Roche limit) and with systems
whose total mass exceeds the maximum mass that a cold EOS can
support. The goal is to determine whether a WDNS
merger leads to either prompt collapse to a BH
or a spinning quasiequilibrium configuration consisting of a cold NS
surrounded by a hot gaseous mantle of WD debris, or something else.

Due to the vast range of dynamical time and
length scales, hydrodynamic simulations in full GR
of realistic WDNS mergers (head-on or otherwise) are computationally
prohibitive. For this reason, we tackle the problem using the same approach as
in our investigation of binary WDNS head-on collisions.
In particular, we adopt the pseudo-white dwarf (pWD) approximation
with the 10:1 EOS constructed in \cite{WDNS_PAPERII}.
This EOS captures the main physical features of NSs,
but scales down the size of WDs so that the ratio of
the isotropic radius of a TOV $0.98\ M_\odot$ pWD to that of a TOV
$1.5\ M_\odot$ NS is 10:1 (hence the name of the EOS), rather than the more
realistic ratio 500:1. These pWDs enable
us to reduce the range of length and time scales involved while
maintaining all length and time-scale inequalities, rendering the
computations tractable and the results scalable. 

If the pWDNS merger does not result in prompt collapse to a
black hole, it is unlikely that the corresponding WDNS merger will
collapse promptly.

The reason for this expectation is that the pWD approximation is based on scaling.
In particular, both the collision velocity and the pre-shocked WD sound speed 
scale as $\sim (M/R_{\rm WD})^{1/2}$. This implies that the Mach number is 
invariant  under scaling of $R_{\rm WD}$ and so is the degree of shock heating. 
So the thermal energy, as well as the rotational kinetic energy ($T$) 
and the gravitational potential energy ($W$) all scale 
as $\sim M^2/R_{\rm WD}$, when the binary merges. Thus
$T/|W|$ is also invariant under scaling of $R_{\rm WD}$. 
These considerations simply mean that with respect to gravity
the relative importance of thermal and rotational support in a WDNS merger remnant 
is approximately invariant, when the masses of the binary components are fixed and the only 
quantity that changes is the WD radius. As a consequence, the results obtained
when adopting pWDNS systems can be scaled up to realistic WDNS systems.

To predict whether a TZlO, which does not collapse to a BH promptly,
will collapse following cooling, we introduced an artificial cooling 
mechanism (see Sec.~\ref{sec:cooling}). 
If following cooling the remnant collapses, we expect that delayed 
collapse in the corresponding WDNS case
likely will take place on a cooling time scale.

To test our cooling prescription, we applied it to the TZlOs formed in the WDNS 
head-on collision simulations we performed in \cite{WDNS_PAPERII}. We demonstrated that these 
remnants collapse to a black hole when the excess thermal energy is radiated away, as expected.

Finally, we simulated the merger of an initially quasiequilibrium, corotational pWDNS
system in circular orbit at the Roche limit, comprised of a
1.4$M_\odot$ NS and a 0.98$M_\odot$ pWD.  We find that the remnant of
the pWDNS inspiral is a spinning TZlO which is surrounded by an extended, hot 
disk. The coordinate radius of the TZlO remnant and disk is approximately $300{\rm \ km}$
and $1000{\rm \ km}$, respectively.
We estimated the disk mass to be $\gtrsim 50\%$ of the initial original WD rest mass. 
In contrast to our binary WDNS head-on collision investigations, 
no outflows were observed in the circular case. The final total ADM mass ($\sim 2.4\ M_\odot$) 
is greater than the maximum mass supportable by a cold, degenerate star
with our adopted NS EOS. However, the remnant does
not collapse promptly to a black hole. 
This is because the remnant is both 
thermally and centrifugally supported. To determine whether
centrifugal support by itself supports the remnant from collapse, we enabled our radiative cooling mechanism 
and found that the object does not collapse to a black hole following cooling. Therefore, the extra support
provided by rotation is sufficient for holding the collapse.

Although the TZlO does not collapse following cooling, ultimate
collapse to a BH is almost certain, since the final total mass is
larger than the maximum possible mass supportable by our cold EOS (and
many nuclear EOSs), even allowing for maximal uniform rotation.  Therefore, delayed collapse likely will take
place after viscosity or magnetic fields redistribute the angular
momentum and/or following cooling. 
This conclusion will be true in the case of realistic WDNS mergers, unless the true nuclear EOS supports 
a uniformly rotating star with a rest mass exceeding the remnant mass. Many viable EOSs do not support 
rest masses as large as $2.5M_\odot$ \cite{2004ApJ...610..941M}, the remnant rest mass in our simulations.

Our results hold true provided that nuclear burning remains unimportant
in the post-merger event. We estimated that typical realistic TZlO temperatures will be
of order $10^9\rm\ K$.  For typical WD densities of order $10^6{\rm \ g/cm}^3$
carbon is ignited and can become an important source of heating.
Though nuclear burning likely will play some role in the
post-merger evolution of a massive WDNS system,
we do not expect a carbon flash to occur.
The reason for this is that shocks at merger lift the
degeneracy of the WD matter. Given that a carbon flash requires a
cold degenerate environment, the net effect should be to reduce the
likelihood of explosive carbon burning. Nevertheless, further
simulations would be useful. 

The neutrino cooling time scale is highly
temperature sensitive, and our pWDNS inspiral simulation may only
provide a crude estimate of temperature.  Therefore, simulations with
more physics are necessary to precisely calculate realistic TZlO
temperatures, so that the relevant cooling time scales may be better
determined.  Finally, while our simulations indicate that prompt collapse 
to a black-hole is not possible for WDNS systems with total rest mass 
$\lesssim 2.5 M_\odot$, it is likely that systems with greater mass can collapse promptly. 
Therefore, more simulations in full GR are necessary before a definitive solution to 
the problem can be given. We plan to address these issues in a future work.

\acknowledgments

We would like to thank Brian D. Farris and Thomas W. Baumgarte for helpful discussions. 
We are also grateful to Morgan MacLeod for providing the Newtonian binary pWDNS equilibrium 
configurations, which we used to generate our CTS initial data. 
This paper was supported in part by NSF Grants 
PHY06-50377 and PHY09-63136 as well as NASA Grants NNX07AG96G and NNX10AI73G to the 
University of Illinois at Urbana-Champaign. Z. Etienne gratefully acknowledges 
support from NSF Astronomy and Astrophysics Postodoctoral Fellowship AST-1002667.

\bibliography{paper}

\end{document}